\documentclass[a4paper,11pt]{article}
\pdfoutput=1

\usepackage[utf8]{inputenc}
\usepackage{jheppub}
\usepackage{amsmath,amssymb,mathtools}
\usepackage{enumerate}
\usepackage{booktabs}
\usepackage{multirow}
\usepackage{graphicx}
\usepackage{xspace}
\usepackage{subcaption}

\renewcommand{\tabcolsep}{2em}

\allowdisplaybreaks

\DeclareRobustCommand{\ensuremathrm}[1]{\ensuremath{\mathrm{#1}}\xspace}

\DeclareRobustCommand{\ensuremathcal}[1]{\ensuremath{\mathcal{#1}}\xspace}
\DeclareRobustCommand{\rd}{\ensuremathrm{d}} 
\DeclareRobustCommand{\ord}{\ensuremathcal{O}} 

\DeclareRobustCommand{\rs}{\ensuremathrm{s}}
\DeclareRobustCommand{\rT}{\ensuremathrm{T}}
\DeclareRobustCommand{\rR}{\ensuremathrm{R}}
\DeclareRobustCommand{\rF}{\ensuremathrm{F}}
\DeclareRobustCommand{\jet}{\ensuremathrm{jet}\xspace}

\DeclareRobustCommand{\alphas}{\ensuremath{\alpha_\rs}\xspace}
\DeclareRobustCommand{\mur}{\ensuremath{\mu_\rR}\xspace}
\DeclareRobustCommand{\muR}{\mur}
\DeclareRobustCommand{\muf}{\ensuremath{\mu_\rF}\xspace}
\DeclareRobustCommand{\muF}{\muf}

\DeclareRobustCommand{\PZ}{{\ensuremathrm{Z}}\xspace}

\DeclareRobustCommand{\Pp}{{\ensuremathrm{p}}\xspace}

\DeclareRobustCommand{\GeV}{\ensuremathrm{GeV}\xspace}
\DeclareRobustCommand{\TeV}{\ensuremathrm{TeV}\xspace}

\DeclareRobustCommand{\CF}{\ensuremath{C_\rF}\xspace}
\DeclareRobustCommand{\CA}{\ensuremath{C_\mathrm{A}}\xspace}

\DeclareRobustCommand{\TR}{\ensuremath{T_\rR}\xspace}
\DeclareRobustCommand{\NF}{\ensuremath{N_\rF}\xspace}
\DeclareRobustCommand{\pt}{\ensuremath{p_\rT}\xspace}
\DeclareRobustCommand{\ptj}[1]{\ensuremath{p_{\rT,#1}}\xspace}
\DeclareRobustCommand{\htj}{\ensuremath{H_\rT}\xspace} 
\DeclareRobustCommand{\htp}{\ensuremath{\hat{H}_\rT}\xspace} 

\preprint{{\raggedleft%
  IPPP/18/38 \\
  ZU-TH 24/18 \\
  CFTP/18-010 \\
  CERN-TH-2018-159 \\
}}

\title{Infrared sensitivity of single jet inclusive production at hadron colliders}

\author[a]{James Currie,}
\author[b,c]{Aude Gehrmann--De Ridder,}
\author[c]{Thomas Gehrmann,}
\author[a]{Nigel Glover,}
\author[d]{Alexander Huss,}
\author[e]{Jo\~{a}o Pires}
\affiliation[a]{Institute for Particle Physics Phenomenology, Durham University, Durham, DH1 3LE, UK}
\affiliation[b]{Institute for Theoretical Physics, ETH, CH-8093 Z\"urich, Switzerland}
\affiliation[c]{Physik-Institut, Universit\"at Z\"urich, Winterthurerstrasse 190, CH-8057 Z\"urich, Switzerland}
\affiliation[d]{Theoretical Physics Department, CERN, 1211 Geneva 23, Switzerland}
\affiliation[e]{Centro de F\'{\i}sica Te\'{o}rica de Part\'{\i}culas - CFTP, 
Instituto Superior T\'ecnico IST,
Universidade de Lisboa, Av.\ Rovisco Pais,
P-1049-001 Lisboa, Portugal}
\emailAdd{james.currie@durham.ac.uk}
\emailAdd{gehra@phys.ethz.ch}
\emailAdd{thomas.gehrmann@uzh.ch}
\emailAdd{e.w.n.glover@durham.ac.uk}
\emailAdd{alexander.huss@cern.ch}
\emailAdd{joao.ramalho.pires@tecnico.ulisboa.pt}

\abstract{%
  Jet production at hadron colliders is a benchmark process to probe the dynamics of the strong interaction and the structure of the colliding hadrons. 
  One of the most basic jet production observables is the single jet inclusive cross section, which is obtained by summing all jets that are observed in an event. 
  Our recent computation of next-to-next-to-leading order (NNLO) QCD contributions to single jet inclusive observables uncovered large corrections in certain kinematical regions, which also resulted in a sizeable ambiguity on the appropriate choice of renormalization and factorization scales. 
  We now perform a detailed investigation of the infrared sensitivity of the different ingredients to the single jet inclusive cross section. We show that the contribution from the second jet, ordered in transverse momentum \pt, in the event is particularly sensitive to higher order effects due to implicit restrictions on its kinematics. 
   By investigating  the second-jet transverse momentum distribution, we identify large-scale 
   cancellations between different kinematical event configurations, which are aggravated by certain types of scale  choice. 
   Taking perturbative convergence and stability as selection criteria enables us to single out the total 
   partonic transverse energy \htp and twice the individual jet transverse momentum 2\,\pt (with which \htp coincides in Born kinematics) as the
   most appropriate scales in the perturbative description of single jet inclusive production. 
   }

\begin{document}
\maketitle

\section{Introduction}
\label{Sec:intro}

At hadron colliders, the factorised form of the inclusive cross section is given by,
\begin{equation}
  \label{eq:hadroncross}
  \rd\sigma = 
  \sum_{i,j} \int_0^1\rd x_1 \int_0^1\rd x_2 \; f_i(x_1,\muF) \; f_j(x_2,\muF)
  \;\rd\hat{\sigma}_{ij}(\alphas(\muR),\muR^2,\muF^2)\;,
\end{equation}
where $\rd \hat\sigma_{ij}$ is the parton-level scattering cross section for parton $i$ to scatter off parton $j$ and the sum runs over the possible parton types $i$ and $j$. 
The probability of finding a parton of type $i$ in the proton carrying a momentum fraction $x$ is described by the parton distribution function (PDF) $f_i(x)\rd x$.  
By applying suitable cuts, one can study more exclusive observables such as the transverse momentum distribution or the rapidity distribution of the hard objects (jets or vector bosons, Higgs bosons or other new particles) produced in the hard scattering.  
In Eq.~\eqref{eq:hadroncross}, one has to fix the renormalization scale $\muR$ for the strong coupling $\alphas(\muR)$, and the mass factorization scale $\muF$ for the parton distribution functions $f_i(x,\muF)$. 

In this paper we study jet production at hadron colliders, in particular the single jet inclusive cross section in proton-proton collisions, $\rd \sigma(\Pp+\Pp \to \jet + X)$, which is obtained by summing over all jets in the event. 
The observable is inclusive over all additional radiation as no further kinematical constraints are imposed on the final state particles beyond the requirement of observing at least a single jet.
In this way, the full event can contain multiple jets and all jets that lie in a given range of rapidity $y$ and transverse momentum $\pt$  are taken into account in determining the single jet inclusive cross section for that bin. 

Large-$\pt$ jet production at hadron colliders has been studied in particle accelerators over a period of many years by the UA1~\cite{Arnison:1986vk} and UA2~\cite{Banner:1982kt} experiments at the Sp$\bar{\text{p}}$S collider ($\sqrt{s}=546~\GeV$ and $630~\GeV$) and by the CDF~\cite{Aaltonen:2008eq} and D0~\cite{Abazov:2011vi} experiments at the Tevatron ($\sqrt{s}=1.96~\TeV$). 
At the Large Hadron Collider (LHC) at CERN, the ALICE, ATLAS and CMS collaborations have measured inclusive jet cross sections in proton-proton collisions at centre-of-mass energies of $\sqrt{s}=2.76~\TeV$~\cite{Abelev:2013fn,Aad:2013lpa,Khachatryan:2015luy}, $7~\TeV$~\cite{Aad:2010ad,Chatrchyan:2012bja}, $8~\TeV$~\cite{Aaboud:2017dvo,Khachatryan:2016mlc} and $13~\TeV$~\cite{Khachatryan:2016wdh,Aaboud:2017wsi}. 
These precise measurements are crucial for understanding physics at hadron colliders as jet cross sections provide valuable information about the strong coupling constant $\alphas$, the non-perturbative structure of the proton encoded in the PDFs, and probe the shortest distance scales that are experimentally attainable. 
More recently, jet substructure techniques have been applied to understand the internal dynamics of QCD jets in order to identify discriminator variables which can more easily disentangle jets originating in 
the QCD parton scattering process from those produced by the hadronic decay of new heavy beyond-Standard-Model particles~\cite{Larkoski:2017jix}.

Hadron collider jet observables can be computed at a given fixed order in $\alphas$ in perturbative QCD, by retaining the corresponding terms in the series expansion in $\alphas$ for the parton-level cross sections and the PDFs, as presented in Eq.~\eqref{eq:hadroncross}.
Next-to-leading-order (NLO) QCD corrections to jet production at hadron colliders were computed in~\cite{Ellis:1992en,Giele:1994gf,Nagy:2001fj} and later combined with a parton shower in~\cite{Alioli:2010xa,Hoche:2012wh}. 
First-order corrections in the electroweak (EW) coupling have been derived in~\cite{Dittmaier:2012kx,Campbell:2016dks}, and the combination of NLO QCD and EW corrections was studied in~\cite{Frederix:2016ost}. A study of joint jet radius and threshold resummation has been presented in~\cite{Liu:2018ktv}.
Progress in next-to-next-to-leading-order (NNLO) QCD calculations has been made over the past several years~\cite{Glover:2010im,GehrmannDeRidder:2011aa,Ridder:2012dg,Currie:2013vh}. 
After the completion of the first calculations of the gluons-only subprocess~\cite{Ridder:2013mf,Currie:2013dwa}, the complete leading-colour and leading-$\NF$ NNLO QCD corrections to the single jet inclusive production~\cite{Currie:2016bfm} and to di-jet production~\cite{Currie:2017eqf} were obtained recently.

The recent NNLO calculation provides new opportunities for QCD studies at hadron colliders, it enables precise theoretical predictions for jet observables to be compared with the wealth of experimental jet data which have similar precision. 
More formally, the knowledge of three orders in the perturbative expansion in $\alphas$ for these jet observables provides a testing ground for the impact of the higher order corrections through the notion of perturbative convergence and the reduction of theoretical scale uncertainties. 

One issue which requires particular attention is the role of the renormalization and factorization scales in the theoretical predictions. 
At a formal level, the parameters $\muR$ and $\muF$ are introduced as auxiliary quantities which allow meaningful predictions to be calculated at each order in perturbation theory. 
As auxiliary quantities, an all-order prediction would be independent of these parameters. 
However, truncating at fixed order yields a residual dependence, formally of one order higher in the strong coupling. Varying the numerical value of the scale (usually in an interval around a pre-defined central scale choice) is frequently used to quantify the uncertainty on the theory prediction due to the uncomputed higher orders. 
The huge dynamical range of jet production at the LHC and the three available perturbative orders  in the theoretical prediction provide the opportunity to test thoroughly the commonly used arguments about scale dependence in perturbative calculations. 

For these reasons we have recently provided jet cross section predictions for the LHC at NNLO using both the leading jet transverse momentum in an event $\muR=\muF=\ptj1$~\cite{Currie:2016bfm} or each individual jet transverse momentum $\muR=\muF=\pt$~\cite{Currie:2017ctp} as a central scale choice. 
We have observed an overall reduction in the scale dependence of the prediction at NNLO with respect to the NLO result with either of these scale choices. 
However, comparing the two predictions (which are both based on well-motivated central scale choices) against each other, we noticed a substantial difference in their quantitative behaviour~\cite{Currie:2017ctp}, which can be viewed as a further uncertainty on the theory prediction. 
It is therefore important to arrive at a sensible central scale choice, which covers the range of jet kinematics accessible at the LHC. 

In this paper, we perform a detailed study of the perturbative behaviour of the individual contributions to the single jet inclusive production cross section for a given set of sensibly chosen dynamical scales. 
In Section~\ref{Sec:scaledep} we present the structure and the scale dependence of the single jet inclusive cross section computed through to  NNLO in QCD. 
We discuss possible functional forms for the scale choice in terms of kinematical variables, thereby carefully distinguishing scales which are based on individual jet kinematics (jet-based) or on full event kinematics (event-based). 
Particular attention is paid to the effects of the jet clustering algorithm and the jet resolution parameter on the kinematical variables used in the different scale choices. 

In Section~\ref{Sec:scalecomp} we subsequently perform a detailed investigation of the infrared sensitivity of the different ingredients to the single jet inclusive cross section for the jet-based scale choice $\muR=\muF=\pt$ and the event-based scale choice $\muR=\muF=\ptj1$. 
It is the aim of this section to identify the source of the different quantitative behaviour in the NNLO predictions between the $\muR=\muF=\pt$ and $\muR=\muF=\ptj1$ scale choices. 

In Section~\ref{Sec:kfactors13TeV} we analyse the behaviour of the 
 perturbative expansion of the single jet inclusive observable, for the different functional forms for the central 
 scale choice established in Section~\ref{Sec:scaledep}. 
This allows us to assess the perturbative stability and convergence properties for each scale choice up to NNLO, thereby identifying the most appropriate 
candidates.  
Subsequently we compare our predictions at NLO and NNLO to the available CMS $13~\TeV$ jet data for the first time in Section~\ref{Sec:numerics13TeV}. 
Finally in Section~\ref{Sec:conclusions} we present our conclusions.

\section{Renormalization and factorization scales in the single jet inclusive cross section}
\label{Sec:scaledep}

When calculating jet cross sections to fixed order in perturbation theory~\eqref{eq:hadroncross}, one has to fix the renormalization scale $\muR$ for the strong coupling $\alphas(\muR)$, and the mass factorization scale $\muF$ for the parton distribution functions $f_i(x,\muF)$. 

The behaviour of the coupling constant and parton distributions under scale variations is determined by evolution equations. 
After fixing a central reference scale, all scale-dependent terms of hard scattering cross sections can be inferred by expanding the solutions of the evolution equations in powers of the strong coupling constant. 
These are collected in Section~\ref{subsec:scales} below. 
For processes involving massive particles (vector bosons, Higgs bosons or top quarks), the particle mass provides a natural candidate for the central reference scale. 
In contrast, no natural fixed scale is present in jet production processes, which involve only massless objects at parton level. 
Consequently, the central scale for jet production must be chosen dynamically, based on the kinematics of the final-state objects (jets or full events) under consideration. Section~\ref{subsec:scalechoice} discusses different prescriptions for the central scale in single jet inclusive production, based on the kinematics of each individual jet, or of the whole event. These kinematical variables depend on the jet algorithm and the jet resolution parameter. In Section~\ref{subsec:indiv} we define the individual jet contributions to the single jet inclusive production cross section 
which are individually infrared safe only if they are inclusive in the jet rapidity.
Analysing the different possible final state configurations up to NNLO, we finally discuss the impact of the jet resolution on the event properties and on the scale choices in Section~\ref{subsec:rdep}.

\subsection{Scale dependent terms up to NNLO}
\label{subsec:scales}

\subsubsection{Renormalization scale dependence}

The renormalization group equation describing the 
running of $\alphas$ as a function of the renormalization scale $\muR$
reads:
\begin{eqnarray}
\muR^2 \frac{\rd \alphas(\muR)}{\rd \muR^2} &=& -\alphas(\muR) 
\left[\beta_0 \left(\frac{\alphas(\muR)}{2\pi}\right) 
+ \beta_1 \left(\frac{\alphas(\muR)}{2\pi}\right)^2 
+ \beta_2 \left(\frac{\alphas(\muR)}{2\pi}\right)^3 
+ \ord(\alphas^4) \right]\,,\nonumber \\
\label{eq:running}
\end{eqnarray}
with the $\overline{\text{MS}}$-scheme coefficients~\cite{Tarasov:1980au,Larin:1993tp}
\begin{eqnarray}
\beta_0 &=& \frac{11 \CA - 4 \TR  \NF}{6}\;,\nonumber  \\
\beta_1 &=& \frac{17 \CA^2 - 10 \CA  \TR  \NF- 6\CF  \TR  \NF}{6}\;, \nonumber \\
\beta_2 &=&\frac{1}{432}
\big( 2857 \CA^3 + 108 \CF^2 \TR  \NF  -1230 \CF \CA  \TR  \NF 
-2830 \CA^2\TR \NF  \nonumber \\ &&
+ 264 \CF \TR^2 \NF^2 + 316 \CA \TR^2\NF^2\big)\;,
\end{eqnarray}
where $\CA = 3$, $\CF=4/3$, $\TR=1/2$ and $\NF$ is the number of light quark flavours.

Using the solution of this 
equation, the coupling at a fixed scale $\mu_{\rR_0}$ can be truncated
in terms of the coupling at $\muR$ by introducing 
\begin{equation}
L_\rR  = \log \left(\frac{\muR^2}{\mu_{\rR_0}^2}\right)\,
\end{equation}
as
\begin{equation}
\alphas(\mu_{\rR_0}) = \alphas(\muR) \left[1 + \beta_0 L_\rR 
\frac{\alphas(\muR)}{2\pi} + \left[\beta_0^2L^2_R + \beta_1L_\rR  \right] 
\left(\frac{\alphas(\muR)}{2\pi}\right)^2 + \ord (\alphas^3) 
\right] \,.
\label{eq:asfix}
\end{equation}

The perturbative expansion of the single jet inclusive cross section starts at order 
$\alphas^2$. In evaluating the expansion coefficients 
$\sigma^{(n)}=\sigma^{(n)}(\mu_{\rR_0})$, 
the renormalization scale is fixed to a value $\mu_{\rR_0}$ (which can 
be dynamically evaluated event-by-event). Rescalings can then be made 
for a fixed ratio $\muR/\mu_{\rR_0}$ for all events; e.g.\
if $\mu_{\rR_0}=\ptj1$, we can rescale to $\muR=2\,\ptj1$ or 
$\muR=\ptj1/2$, but not to $\muR=M_Z$ or $\muR=H_T$ ).

\subsubsection{Factorization scale dependence}

The evolution of parton distributions associated to a variation of the factorization scale $\muF$ is determined by the Altarelli-Parisi
equations~\cite{Altarelli:1977zs}
which  read (omitting for simplicity the dependence on the Bjorken scaling variable $x$): 
\begin{equation}
\muF^2 \frac{\rd}{\rd \muF^2} f_i(\muF,\muR) 
= \sum_j P_{ij}(\alphas(\muR),\muF,\muR) \otimes f_j(\muF,\muR)\,.
\label{eq:apmaster}
\end{equation}
The expansion to the third order of the 
splitting functions $P^{(n)}_{ij}$ computed for $\muF=\muR$ is~\cite{Moch:2004pa,Vogt:2004mw}:
\begin{eqnarray}
P_{ij}(\alphas(\muR),\muF,\muR) & = &
\frac{\alphas(\muR)}{2\pi} P_{ij}^{(0)}  +
\left(\frac{\alphas(\muR)}{2\pi}\right)^2 
\left[ P_{ij}^{(1)} +\beta_0 l  P_{ij}^{(0)} 
\right]  \nonumber \\ && 
+
\left(\frac{\alphas(\muR)}{2\pi}\right)^3 
\left[ P_{ij}^{(2)} +\left(\beta_1   P_{ij}^{(0)} 
+ 2  \beta_0  P_{ij}^{(0)}\right) l + \beta_0^2 l^2  P_{ij}^{(0)}  
\right] + \ord (\alphas^{4})\,,\nonumber \\
\label{eq:apsplit}
\end{eqnarray}
 where we introduced
\begin{equation}
l = \log \left(\frac{\muR^2}{\muF^2}\right).
\end{equation}

Note that (\ref{eq:apsplit}) can be rewritten as
\begin{eqnarray}
P_{ij}(\alphas(\muR),\muF,\muR) & = &
\frac{\alphas(\muF)}{2\pi} P_{ij}^{(0)}  +
\left(\frac{\alphas(\muF)}{2\pi}\right)^2 
P_{ij}^{(1)} 
+
\left(\frac{\alphas(\muF)}{2\pi}\right)^3 
P_{ij}^{(2)} + \ord (\alphas^{4})\,,\nonumber \\
\label{eq:apsplit1}
\end{eqnarray}
which implies that $f_i(\muF,\muR)$ and $f_i(\muF,\muF)$ fulfil the 
same evolution equation to all perturbative orders. The finite scheme 
transformation between both possible choices for $\muF$, equal or different from $\muR$, is thus vanishing 
to all orders and both functions can at most vary in their non-perturbative boundary 
conditions. 
For all perturbative purposes, we thus have
\begin{equation}
f_i(\muF,\muR) = f_i(\muF,\muF) \equiv f_i(\muF)\,,
\end{equation}
which we will normally use in what follows (except if the 
scale transformation of the parton distribution is not expanded in 
$\alphas(\muF)$, but in $\alphas(\muR)$). 

The parton distribution at a fixed scale $\mu_{\rF_0}$ can be expressed in terms of parton 
distributions at $\muF$ by expanding the solution of (\ref{eq:apmaster}). We 
distinguish the expansion in powers of $\alphas(\muR)$ and in powers of $\alphas(\muF)$ and 
introduce
\begin{equation}
L_\rF  = \log \left(\frac{\muF^2}{\mu_{\rF_0}^2}\right)\;. 
\end{equation}
The expansion in $\alphas(\muR)$ of the parton distribution at  $\mu_{\rF_0}$ reads then:
\begin{eqnarray}
f_i(\mu_{\rF_0}) & = & f_i(\muF) - \frac{\alphas(\muR)}{2\pi} P_{ij}^{(0)}\otimes
f_j(\muF) L_\rF  \nonumber \\
&& - \left( \frac{\alphas(\muR)}{2\pi}\right)^2 \bigg[ P_{ij}^{(1)}\otimes
f_j(\muF) L_\rF  - \frac{1}{2} P_{ij}^{(0)}\otimes P_{jk}^{(0)}\otimes f_k(\muF) L_\rF ^2 
\nonumber \\ && \hspace{3cm}
+ P_{ij}^{(0)} \otimes f_j(\muF) \beta_0 L_\rF \left(l + \frac{1}{2} L_\rF  \right)
\bigg] + \ord (\alphas^{3})\;.
\label{eq:pdffix}
\end{eqnarray}
The expansion in powers of  $\alphas(\muF)$ is obtained from the above by 
setting $\muR=\muF$ in $\alphas$ to yield 
\begin{eqnarray}
f_i(\mu_{\rF_0}) & = & f_i(\muF) - \frac{\alphas(\muF)}{2\pi} P_{ij}^{(0)}\otimes
f_j(\muF) L_\rF  \nonumber \\
&& - \left( \frac{\alphas(\muF)}{2\pi}\right)^2 \bigg[ P_{ij}^{(1)}\otimes
f_j(\muF) L_\rF  - \frac{1}{2} P_{ij}^{(0)}\otimes P_{jk}^{(0)}\otimes f_k(\muF) L_\rF ^2 
\nonumber \\ && \hspace{3cm}
+\frac{1}{2} P_{ij}^{(0)} \otimes f_j(\muF) \beta_0 L_\rF ^2\bigg] 
+ \ord (\alphas^{3})\;.
\end{eqnarray}
In both expressions, a summation over indices appearing twice is implicit.

\subsubsection{Hadron collider jet cross section}

Using the results presented above, 
one can compute the perturbative coefficients of the 
hadron collider cross section with default values of $\mu_{\rF_0}=\mu_{\rR_0}=\mu_0$. 
The perturbative expansion to NNLO reads:
\begin{eqnarray}
\sigma(\mu_{\rR_0},\mu_{\rF_0},\alphas(\mu_{\rR_0})) 
&=& \left(\frac{\alphas(\mu_{\rR_0})}{2\pi}\right)^2 \hat{\sigma}_{ij}^{(0)}\otimes f_i(\mu_{\rF_0}) 
\otimes f_j(\mu_{\rF_0}) \nonumber \\ &&
+\left(\frac{\alphas(\mu_{\rR_0})}{2\pi}\right)^{3} \hat{\sigma}_{ij}^{(1)}\otimes f_i(\mu_{\rF_0}) 
\otimes f_j(\mu_{\rF_0})  \nonumber \\ &&
+\left(\frac{\alphas(\mu_{\rR_0})}{2\pi}\right)^{4} \hat{\sigma}_{ij}^{(2)}\otimes f_i(\mu_{\rF_0}) 
\otimes f_j(\mu_{\rF_0}) 
+\ord (\alphas^{5}) \,.
\end{eqnarray}

The full scale dependence of this expression, for $\muF$ and $\muR$ different from each other,
 can be recovered by inserting (\ref{eq:asfix}) and (\ref{eq:pdffix}) into the above equation. It yields
\begin{eqnarray}
\lefteqn{\sigma(\muR,\muF,\alphas(\muR),L_\rR ,L_\rF ) 
=}\nonumber \\ &&
 \left(\frac{\alphas(\muR)}{2\pi}\right)^2 \hat{\sigma}_{ij}^{(0)}\otimes f_i(\muF)  \otimes f_j(\muF) \nonumber \\ &&
+\left(\frac{\alphas(\muR)}{2\pi}\right)^{3} \hat{\sigma}_{ij}^{(1)}\otimes f_i(\muF) \otimes f_j(\muF)  \nonumber \\ &&
\hspace{5mm} +L_\rR  \, \left(\frac{\alphas(\muR)}{2\pi}\right)^{3}  2\beta_0 \, \hat{\sigma}_{ij}^{(0)}\otimes f_i(\muF) \otimes f_j(\muF)  \nonumber \\ &&
\hspace{5mm} +L_\rF \, \left(\frac{\alphas(\muR)}{2\pi}\right)^{3} \Big[- \hat{\sigma}_{ij}^{(0)}\otimes f_i(\muF) \otimes \left( P_{jk}^{(0)}\otimes f_k(\muF) \right)\nonumber \\ && 
\hspace{5mm}\phantom{+L_\rF \, \left(\frac{\alphas(\muR)}{2\pi}\right)^{3} \Big[}-  \hat{\sigma}_{ij}^{(0)}\otimes  \left( P_{ik}^{(0)}\otimes f_k(\muF)\right) 
\otimes f_j(\muF) 
\Big] \nonumber \\ &&
+\left(\frac{\alphas(\muR)}{2\pi}\right)^{4} \hat{\sigma}_{ij}^{(2)}\otimes f_i(\muF) \otimes f_j(\muF) \nonumber \\ &&
\hspace{5mm} +L_\rR  \, \left(\frac{\alphas(\muR)}{2\pi}\right)^{4}  \left(3\, \beta_0 \, \hat{\sigma}_{ij}^{(1)} + 2\, \beta_1 \, \hat{\sigma}_{ij}^{(0)}\right) \otimes f_i(\muF) \otimes f_j(\muF)  \nonumber \\ &&
\hspace{5mm} +L_\rR ^2 \, \left(\frac{\alphas(\muR)}{2\pi}\right)^{4}  3\, \beta^2_0 \,\hat{\sigma}_{ij}^{(0)}\otimes f_i(\muF) \otimes f_j(\muF)  \nonumber \\ &&
\hspace{5mm} +L_\rF \, \left(\frac{\alphas(\muR)}{2\pi}\right)^{4}\Big[- \hat{\sigma}_{ij}^{(1)}\otimes f_i(\muF) \otimes \left( P_{jk}^{(0)}\otimes f_k(\muF) \right)\nonumber \\ && 
\hspace{5mm}\phantom{+L_\rF \, \left(\frac{\alphas(\muR)}{2\pi}\right)^{4}\Big[}-  \hat{\sigma}_{ij}^{(1)}\otimes  \left( P_{ik}^{(0)}\otimes f_k(\muF)\right) \otimes f_j(\muF)   \nonumber \\ &&
\hspace{5mm}\phantom{+L_\rF \, \left(\frac{\alphas(\muR)}{2\pi}\right)^{4}\Big[}- \hat{\sigma}_{ij}^{(0)}\otimes f_i(\muF) \otimes \left( P_{jk}^{(1)}\otimes f_k(\muF) \right)\nonumber \\ && 
\hspace{5mm}\phantom{+L_\rF \, \left(\frac{\alphas(\muR)}{2\pi}\right)^{4}\Big[}-  \hat{\sigma}_{ij}^{(0)}\otimes  \left( P_{ik}^{(1)}\otimes f_k(\muF)\right) \otimes f_j(\muF) \Big] \nonumber \\ && 
\hspace{5mm} +L_\rF ^2\, \left(\frac{\alphas(\muR)}{2\pi}\right)^{4}
\Big[
 \hat{\sigma}_{ij}^{(0)}\otimes \left( P_{ik}^{(0)}\otimes f_k(\muF)\right) 
\otimes \left( P_{jl}^{(0)}\otimes f_l(\muF) \right)\nonumber \\ && \hspace{4cm}
+ \frac{1}{2} \hat{\sigma}_{ij}^{(0)}\otimes f_i(\muF)\otimes \left( 
P_{jk}^{(0)}\otimes P_{kl}^{(0)}\otimes f_l(\muF)\right)\nonumber \\ && \hspace{4cm}
+ \frac{1}{2} \hat{\sigma}_{ij}^{(0)}\otimes \left( 
P_{ik}^{(0)}\otimes P_{kl}^{(0)}\otimes f_l(\muF)\right)
\otimes f_j(\muF) \nonumber \\ && \hspace{4cm}
+\frac{1}{2} \beta_0 \,\hat{\sigma}_{ij}^{(0)}\otimes f_i(\muF) 
\otimes \left( P_{jk}^{(0)}\otimes f_k(\muF) \right)\nonumber \\ && \hspace{4cm}
+\frac{1}{2} \beta_0 \, \hat{\sigma}_{ij}^{(0)}\otimes  \left( P_{ik}^{(0)}\otimes f_k(\muF)\right) 
\otimes f_j(\muF)  \Big] \nonumber \\ &&
\hspace{5mm} +L_\rF  L_\rR \, \left(\frac{\alphas(\muR)}{2\pi}\right)^{4}\Big[- 3\, \beta_0\,\hat{\sigma}_{ij}^{(0)}\otimes f_i(\muF) \otimes \left( P_{jk}^{(0)}\otimes f_k(\muF) \right)\nonumber \\ && 
\hspace{5mm}\phantom{+L_\rF  L_\rR \, \left(\frac{\alphas(\muR)}{2\pi}\right)^{4}\Big[}- 3\, \beta_0\, \hat{\sigma}_{ij}^{(0)}\otimes  \left( P_{ik}^{(0)}\otimes f_k(\muF)\right) 
\otimes f_j(\muF) \Big]  \nonumber \\ &&
+\ord (\alphas^{5}) \,.
\end{eqnarray}

\subsection{Scale choices}
\label{subsec:scalechoice}

Inclusive jet observables accumulate each reconstructed jet in the event to the same kinematic distribution, resulting in multiple bookings of the event into a given histogram. 
The set of possible scale choices is consequently large and we shall distinguish two generic types: event-based and jet-based scales.
A \emph{jet-based scale} only uses kinematic information from the individual jet to determine the scale associated with the contribution from this jet to the cross section. 
In a given event, the event weight is thus evaluated at several different scales, one scale for each jet. 
In contrast to this, an \emph{event-based scale} uses information from the full final state of the event to set a common scale for all binnings of the jets that are contained in this event. 

In this paper we will consider the following set of functional forms for the scale choice (and multiples thereof):
\begin{description}
  \item[the individual jet transverse momentum \pt:] 
  When this jet-based scale choice is used for the inclusive \pt distribution, the observable is directly aligned with the scale itself, making it a convenient choice for PDF fits.
  It mimics kinematical hierarchies in an event, where multiple jets can be 
  reconstructed with very different \pt.  
  However, this can lead to the scale being set to values that are not at all representative of the underlying hard scattering process.
  \item[the leading-jet transverse momentum \ptj1:]
  This event-based scale uses the $\pt$ of the hardest jet in the event, which is a better proxy for the scale of the hard interaction compared to the $\mu=\pt$ choice.
  For multi-jet events comprising many hard resolved jets, \ptj1 can still underestimate the scale of the hard interaction. Moreover, \ptj1 does not take account of scale hierarchies in an event. 
  \item[the scalar sum of the transverse momenta of all reconstructed jets \htj:]
  \  \ With\\  this event-based scale one incorporates the kinematics of all individual jets by summing up their 
  respective transverse momenta, $\htj = \sum_{i \in \text{jets}} \ptj{i}$.
  As such, it constitutes the hardest scale discussed so far and for the Born-level $2\to2$ process, it is related to the \pt scales as $\htj=2\,\pt=2\,\ptj1$.
  It however suffers from a discontinuous behaviour, when the number of reconstructed jets changes:
  \begin{align}
    n_\text{jets} = 1 &\Rightarrow \htj = \ptj1 \,, &
    n_\text{jets} = 2 &\Rightarrow \htj = \ptj1+\ptj2 \,, &
    \ldots \,.
  \end{align}
  For this reason, it displays a large displacement 
 at the phase-space boundaries where $(n+1)$-jet events migrate to $n$-jet events. As a consequence, higher order corrections for values of $\pt$ close to
 the minimum jet acceptance $p_{\rT,\min}$ become unstable and we will no longer consider this scale in the remainder of this paper.
   \item[the scalar sum of the transverse momenta of all partons \htp:]
  The undesirable discontinuous behaviour of \htj can be alleviated if the transverse momentum sum 
  is not based on the reconstructed jets, but instead obtained as  
 the transverse momentum sum of all partons in the event: $\htp = \sum_{i \in \text{partons}} \ptj{i}$.
  This event-based 
  scale choice also has the advantage of being insensitive to the jet reconstruction applied in the analysis and 
 is an infrared-safe event shape variable.
\end{description}

\begin{table}
  \centering
  \begin{tabular}{l@{\;}l c c c}
    \toprule
    \multicolumn{2}{c}{scale} & 
    type &
    $n_\text{jets} \to n_\text{jets}+1$ &
    \shortstack{analysis \\ dependence} \\
    \midrule
    \pt      & ($2\,\pt$)   & jet based   & continuous    & yes \\
    \ptj1    & ($2\,\ptj1$) & event based & continuous    & yes \\
    $\htj/2$ & (\htj)       & event based & discontinuous & yes \\
    $\htp/2$ & (\htp)       & event based & continuous    & no \\
    \bottomrule
  \end{tabular}
  \caption{
    Possible scale choices in inclusive jet production and their properties.
    \label{tab:scales}
  }
\end{table}

Any scale choice that is based on the kinematics of the reconstructed jets, i.e.\ \pt, \ptj1, and \htj from the list above, inherits a dependence on the jet cuts and the details of the clustering employed in the analysis~\cite{Dasgupta:2016bnd}. 
This means that for a given partonic configuration and the same scale definition, the determined value for the scale depends on the details of the jet algorithm, the allowed rapidity range and the rapidity and $\pt$ range probed by the experiment.
In particular, the scale choice introduces an indirect dependence on the cone size $R$ of the jet algorithm and on the jet cuts:
\begin{itemize}
  \item A sensitivity of event-based scales on the jet cuts induces the unwanted property that a variation of 
  rapidity cuts can impact the predictions in the other rapidity regions well away from the variation. 
  \item The dependence of the scale on the jet-clustering algorithm can introduce an indirect sensitivity on the cone size $R$.
  Such an effect becomes hard to disentangle from the purely \emph{kinematical} dependence on $R$, which is discussed in Sect.~\ref{subsec:rdep} and which induces potentially enhanced corrections of the form $\log(R)$.
\end{itemize}

As an example of the above, consider the event-based scale $\mu=\ptj1$ and a configuration in which the leading jet is relatively forward and thus does not contribute in a central rapidity slice of the single jet inclusive cross section. 
If the detector rapidity coverage includes the jet, the scale will be the $\pt$ of this forward jet.  
On the other hand, if the forward jet lies outside of the detector coverage it will not be identified as the leading jet, and the event-based scale will be different.
As a consequence, predictions for the jet cross section in the central region of the detector will depend on the rapidity coverage of the detector when the event-based scale $\ptj1$ is used. 

In contrast, the $\mu=\pt$ scale choice always uses the transverse momentum of the jet in the rapidity slice where the jet is observed and therefore its predictions are not sensitive to the jet-defining cuts.
However, as will be detailed in Section~\ref{subsec:rdep}, the scales $\mu=\ptj1$ and $\mu=\pt$ show a different sensitivity to the jet cone size.

Both of these issues are avoided for $\mu=\htp$ that is defined on the basis of the parton kinematics. While 
not being directly accessible in the experimental measurement, \htp is infrared-safe and theoretically well-defined. Its use 
for scale settings is not problematic, since the renormalization and factorization scales are simply auxiliary quantities in the theoretical prediction. 

Table~\ref{tab:scales} summarises the different scale choices together with their respective properties discussed in this section.

\subsection{Individual jet contributions to inclusive jet production}
\label{subsec:indiv}

To illustrate the difference between an event-based and a jet-based scale choice, we consider two of the most common scale choices in studies of
jet production at hadron colliders, i.e. $\mu = \ptj1$ and $\mu=\pt$. To this end,  it is instructive to look at the 
composition of the single jet inclusive cross section in terms of contributions from individual jets in an event (ordered 
in decreasing transverse momentum).  With the event-based scale choice $\mu_{\rR_0} = \mu_{\rF_0} = \mu =\ptj1$, then through to $\ord(\alphas^4)$ we have:
\begin{eqnarray}
\frac{\rd\sigma}{\rd\pt}(\mu=\ptj1)&=&\frac{\rd\sigma}{\rd\ptj1}(\mu=\ptj1)+\frac{\rd\sigma}{\rd\ptj2}(\mu=\ptj1)+\frac{\rd\sigma}{\rd \ptj3}(\mu=\ptj1)+\frac{\rd\sigma}{\rd\ptj4}(\mu=\ptj1)\;.\nonumber
\end{eqnarray}

Predictions for the jet-based scale choice  $\mu=\pt$ can subsequently be obtained in the following way,
\begin{eqnarray}
\label{eq:mupT}
\frac{\rd\sigma}{\rd\pt}(\mu=\pt)&=&\frac{\rd\sigma}{\rd\ptj1}(\mu=\ptj1)+\frac{\rd\sigma}{\rd\ptj2}(\mu=\ptj2)+\frac{\rd\sigma}{\rd\ptj3}(\mu=\ptj3)+\frac{\rd\sigma}{\rd\ptj4}(\mu=\ptj4)\nonumber\\
&=&\frac{\rd\sigma}{\rd\pt}(\mu=\ptj1)\nonumber\\
&+&\frac{\rd\sigma}{\rd\ptj2}(\mu=\ptj2)-\frac{\rd\sigma}{\rd\ptj2}(\mu=\ptj1)\nonumber\\
&+&\frac{\rd\sigma}{\rd\ptj3}(\mu=\ptj3)-\frac{\rd\sigma}{\rd\ptj3}(\mu=\ptj1)\nonumber\\
&+&\frac{\rd\sigma}{\rd\ptj4}(\mu=\ptj4)-\frac{\rd\sigma}{\rd\ptj4}(\mu=\ptj1),
\end{eqnarray}
such that the difference between the $\mu=\ptj1$ and $\mu=\pt$ results can be identified in the last three lines in equation~\eqref{eq:mupT}.
It will therefore be important to numerically study the individual sub-leading jet contributions to the inclusive jet sample and in particular the effects that can arise from 
changing the scale from an event-based scale to a jet-based scale. 

When decomposing the inclusive jet cross section in terms of the contributions from leading and subleading jets, 
the individual jet distributions are well-defined and infrared-safe 
only if they are inclusive in the jet rapidity (with the same global rapidity 
cuts applied to all jets). 
Since the notion of leading and sub-leading jet is not well defined at leading order ($\ptj1 = \ptj2$ at LO), the 
rapidity assignment to the leading and subleading jet is ambiguous for leading-order kinematics. When 
computing higher-order corrections, the rapidity of the leading and subleading jet may thus be interchanged between 
event and counter-event, 
causing them to end up in different rapidity bins, thereby obstructing their cancellation in infrared-divergent 
limits. On the other hand, in the inclusive jet transverse momentum distribution (which sums over all jets in the event)
IR-safety is restored in differential distributions in rapidity $y$, since leading and subleading jet contributions are 
treated equally.

\subsection{Dependence on the jet resolution parameter \texorpdfstring{$R$}{R}}
\label{subsec:rdep}

In this subsection we discuss the effects stemming from the jet definition itself, in particular the jet resolution. For 
the sake of illustration, we represent the jets by cones of radius $R$
in rapidity and azimuthal angle, as obtained~\cite{Cacciari:2008gn} by either a cone algorithm 
or the commonly used anti-$k_\rT$ clustering/recombination algorithm~\cite{Cacciari:2008gp}. 
\begin{figure}[b]
  (a)~\includegraphics[width=2.90cm]{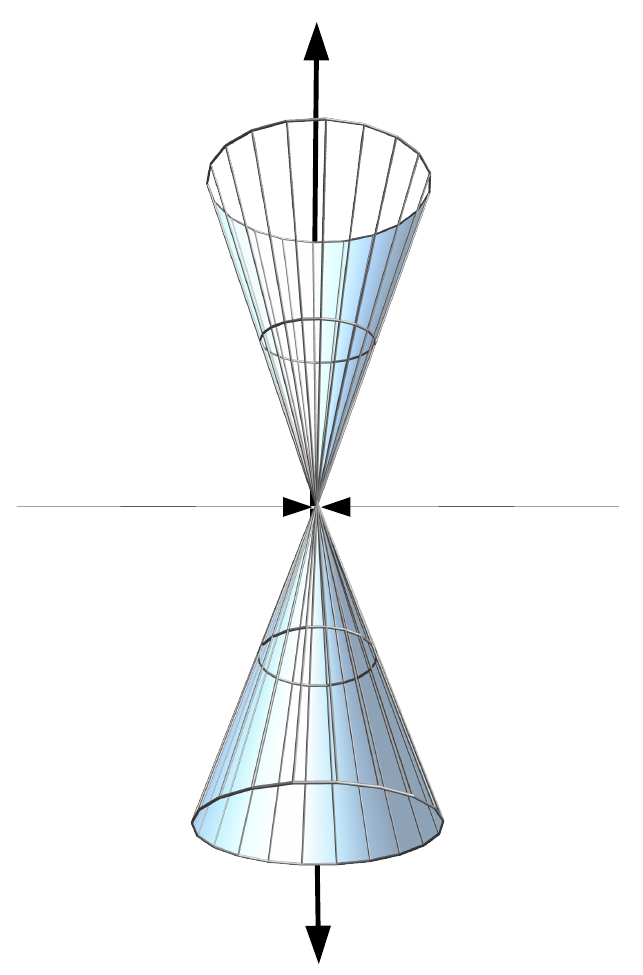} \hfill
  (b)~\includegraphics[width=2.85cm]{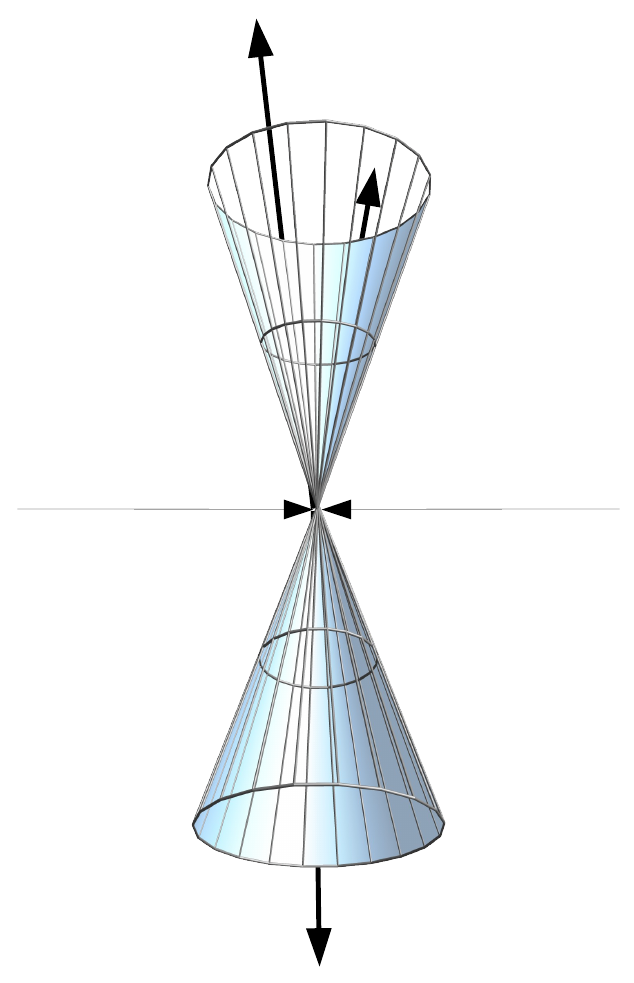} \hfill
  (c)~\includegraphics[width=2.55cm]{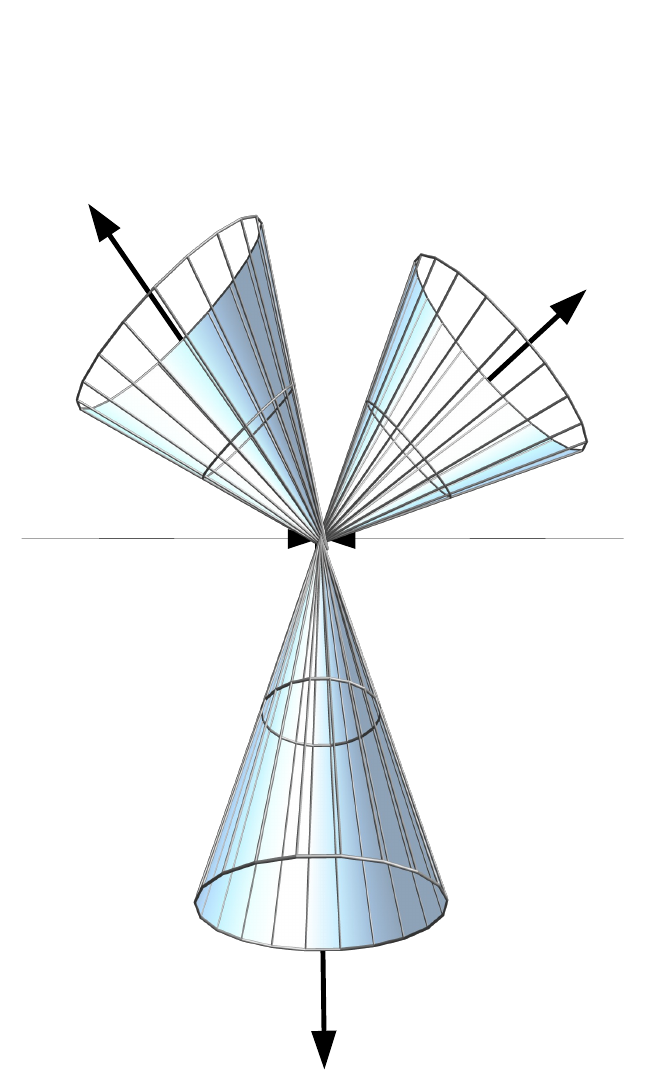} \hfill
  (d)~\includegraphics[width=2.85cm]{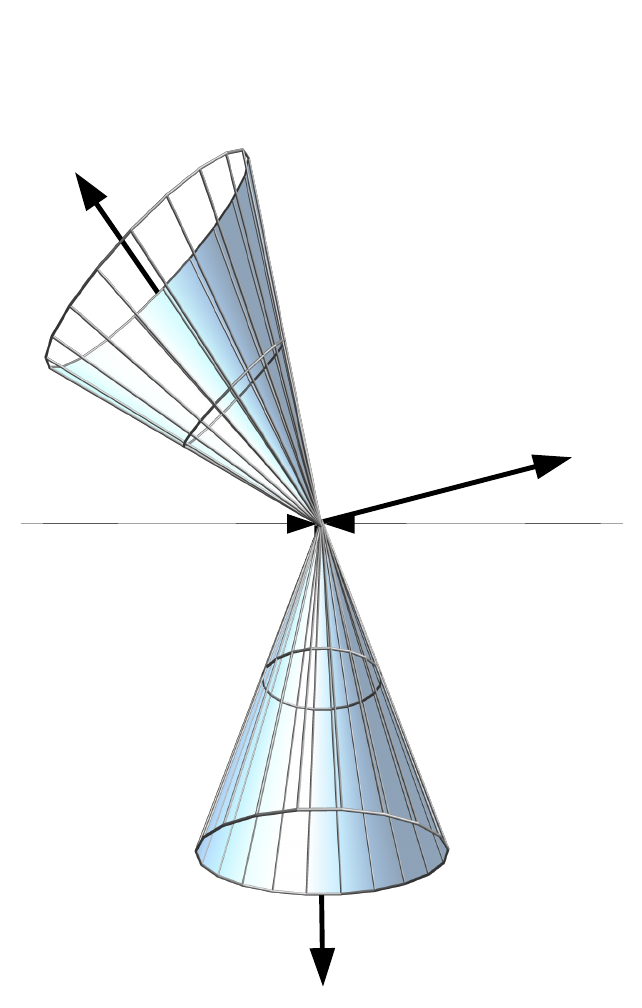} 
  \caption{Illustrations of jet events at LO and NLO with arrows representing partons and cones representing jets.}
  \label{fig:jet1}
\end{figure}

Figure~\ref{fig:jet1} shows some illustrations of various jet configurations at LO and NLO where solid arrows 
represent partons and cones represent jets resulting from the jet algorithm. Fig.~\ref{fig:jet1}(a) shows a
 dijet event at leading order where two back-to-back partons form two jets and $\ptj1= \ptj2$. In this
 case there is no difference in scale choice between \ptj1 and \pt. Fig.~\ref{fig:jet1}(b) shows 
  a dijet event where three partons are clustered by the jet algorithm into two jets such that the jets are still 
  balanced in \pt and the scale choice is identical. Fig.\ref{fig:jet1}(c) shows a   
  trijet event where three partons are sufficiently hard and separated to form three distinct jets. In this configuration
  $\ptj1\ne \ptj2\ne \ptj3$ and so the scale choice does make a difference, although the three-jet contribution 
  makes up only 
 a very small fraction of the inclusive jet cross section as we will observe in Section~\ref{Sec:scalecomp}.
  Fig.~\ref{fig:jet1}(d) depicts a dijet event where the third parton falls outside the jet radius and is not clustered
  but also is not sufficiently hard to form a jet on its own; such configurations typically lead to a small imbalance in the leading
  and subleading jet \pt and their description is sensitive to the scale parameterization. 
\begin{figure}[t!]
\centering
\begin{subfigure}{.5\textwidth}
  \centering
  \includegraphics[width=3.0cm]{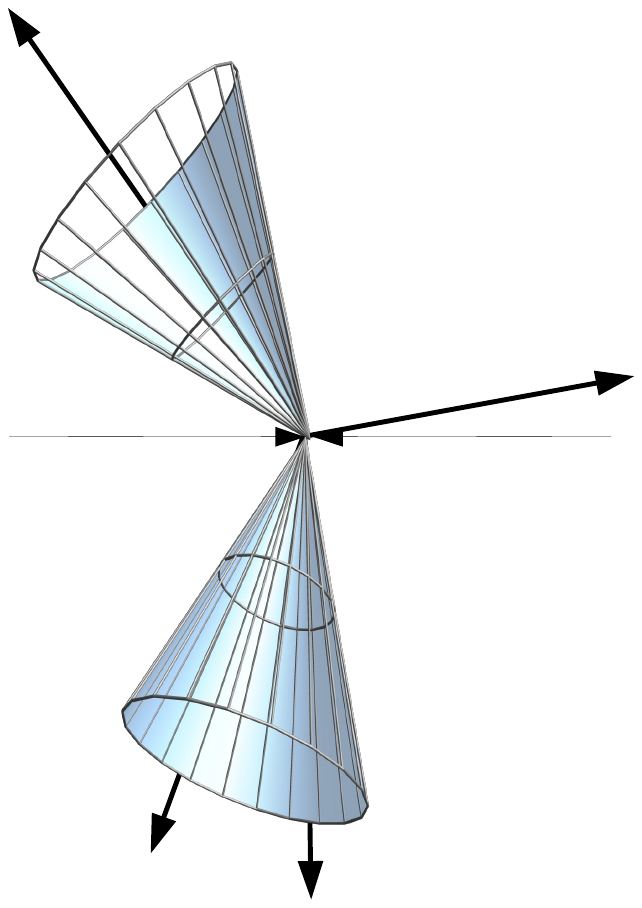}
  \caption{}
\end{subfigure}%
\begin{subfigure}{.5\textwidth}
  \centering
  \includegraphics[width=3.0cm]{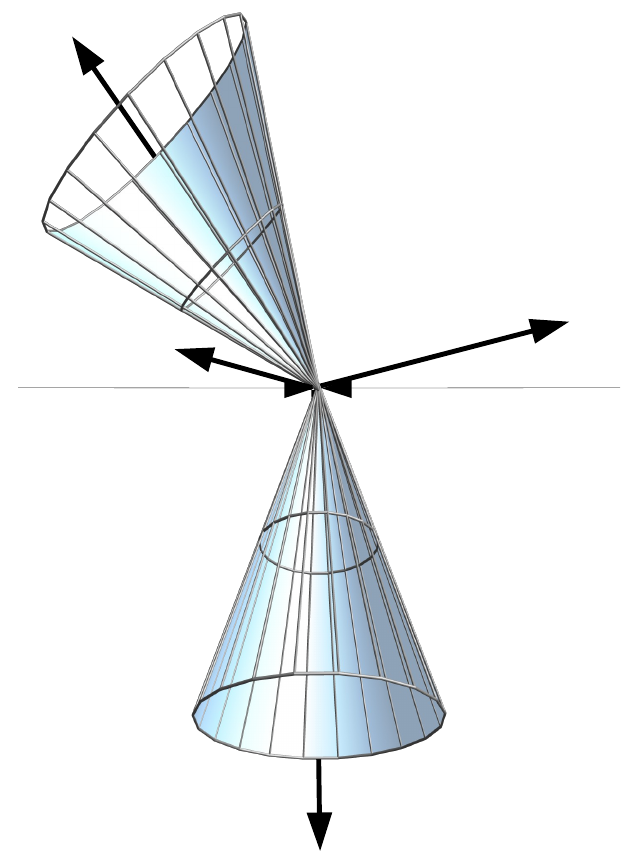}
  \caption{}
\end{subfigure}
\caption{An illustration of 2-jet NNLO configurations that contribute to the $\pt$ imbalance between the leading and subleading jet with arrows representing partons and cones representing jets.}
\label{fig:jetNNLO}
\bigbreak
\centering
\begin{subfigure}{.5\textwidth}
  \centering
  \includegraphics[width=3.0cm]{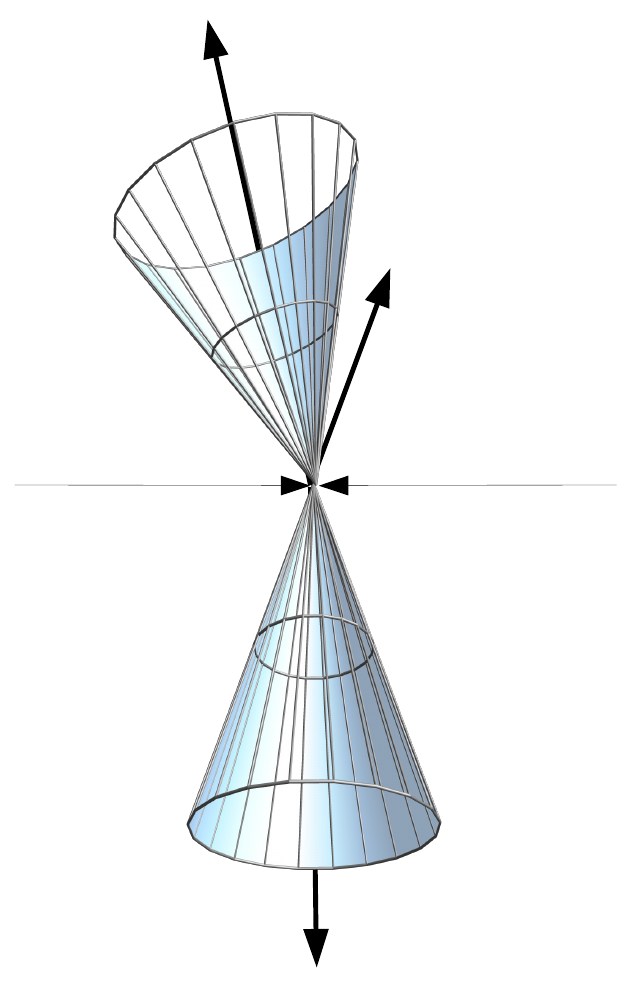}
  \caption{}
\end{subfigure}%
\begin{subfigure}{.5\textwidth}
  \centering
  \includegraphics[width=3.0cm]{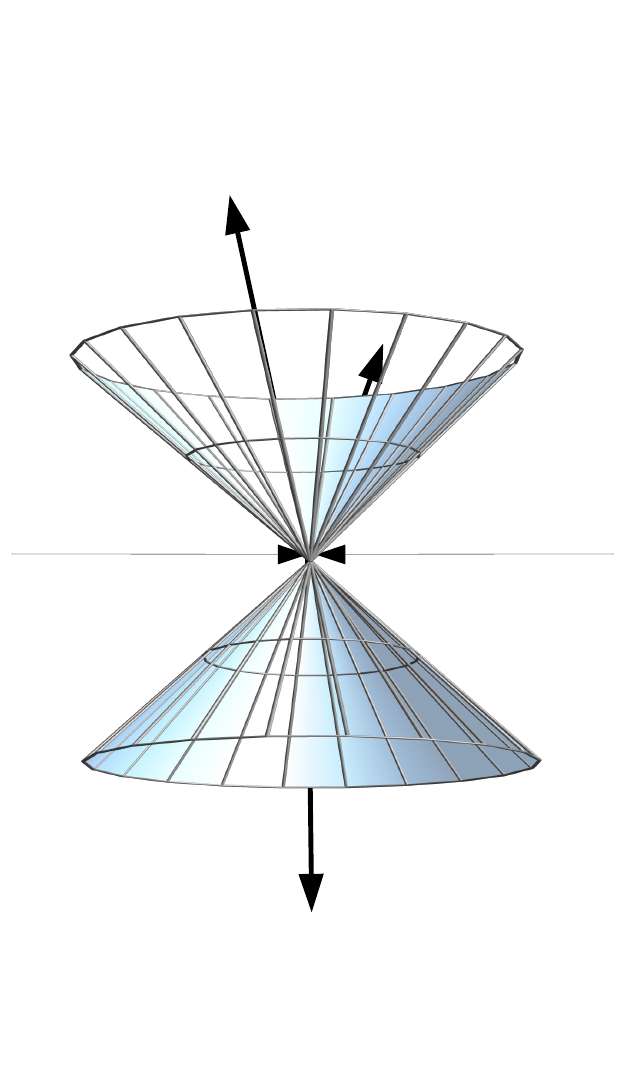}
  \caption{}
\end{subfigure}
\caption{An illustration of clustering the same three-parton event with smaller (a) and larger (b) values of $R$.}
\label{fig:jet2}
\end{figure}

At NNLO there are more configurations to consider due to the presence of four final-state partons in the double real
contribution forming either two, three or four jets. Once again, many configurations do not contribute to the 
difference in scale parameterization. Whenever the jet algorithm clusters two, three or four partons into two jets then
the jets are balanced in \pt and there is no difference between the $\mu=\ptj1$ and $\mu=\pt$ scale choices. 
The only NNLO configurations that can contribute to the difference are: three- or four-jet events (for which the cross section is very small) 
or two jet events where additional radiation falls outside of the jet radius, see Fig.~\ref{fig:jetNNLO}.

As illustrated in Fig.~\ref{fig:jet2} the choice of the $R$ parameter in the jet algorithm can have an effect on how
the partons are clustered into jets. We can take the same three-parton configuration and consider the 
clustering for different values of $R$, Fig.~\ref{fig:jet2}(a), and a larger value of $R$, Fig.~\ref{fig:jet2}(b). For the smaller $R$ value
the most subleading parton is more likely to fall outside the jet radius of the two leading jets and so generate a difference between the
$\mu=\ptj1$ and $\mu=\pt$ scale parameterizations. Therefore, when using the $\mu=\pt$  
scale choice, the value of 
the scale can vary with $R$ for a fixed event. On the other hand, with the choice $\mu=\ptj1$, the scale for the event is $R$-independent at NLO, where the leading jet
is not sensitive to radiation outside the cone, and becomes $R$-dependent only at NNLO.

This difference between the two scale choices grows significantly
for small $R$, decreases for large $R$, and is moderate for the phenomenologically 
relevant values used at the LHC, for $R=0.4$ (0.7) as we will observe in Section~\ref{Sec:scalecomp}.

\section{The scale choices \texorpdfstring{$\mu=\ptj1$}{ptj1} and \texorpdfstring{$\mu=\pt$}{pt}}
\label{Sec:scalecomp}

As observed in Ref.~\cite{Currie:2017ctp}, the spread in the NNLO predictions for single jet inclusive production between using the dynamical scales $\mu=\ptj1$ and $\mu=\pt$ can be comparable or even larger in size than the respective uncertainties estimated through scale variations.
The significant effect of this scale ambiguity on the NNLO predictions, and the lack of a theoretically well-motivated preference motivates us to revisit these results and to further study this issue. 

For the leading jet in the event, the scale $\mu=\pt$ is identical to $\mu=\ptj1$ and its contribution is therefore insensitive to the scale choice between $\pt$ and $\ptj1$. 
Furthermore, two-jet events where the jets are balanced in $\pt$ cannot generate any difference as $\pt$ = $\ptj1 = \ptj2$. 
Away from these jet configurations, the subleading jets will have a smaller $\pt$ than the leading jet in the event so that $\ptj2, ~\ptj3, ~\ldots < \ptj1$. 

For these reasons, at LO the two scale choices generate the same prediction and similarly, for all events at higher order that have LO kinematics there is no difference between the two scale choices. 
In particular at high $\pt$ the scale choices once again converge as is to be expected for the largely back-to-back configurations encountered at high $\pt$.
Kinematical configurations where the scale choices do not coincide are events with three or more hard jets and events with hard
emissions outside the jet fiducial cuts that generate an imbalance in $\pt$ between the leading and subleading jets in the event. For this reason, we can expect also that for larger jet cone sizes the difference in the predictions using $\mu=\pt$ or $\mu=\ptj1$ will be smaller, since the increased number of parton clusterings driven by a larger cone size promotes final state jets balanced in $\pt$.

It is the aim of this section to scrutinise how the contributions to the single jet inclusive transverse momentum distribution behave according to the choice of the functional form of the scale. 
To this end, we use the two central scale choices $\mu=\pt$ and $\mu=\ptj1$ as representatives for jet-based and event-based scale settings. 
After describing our calculational set up in Section~\ref{subsec:setup}, in Sections~\ref{subsec:corrtoptdist} and \ref{sec:jetfrac} we study 
the impact on the transverse momentum distribution of the individual jet fractions at LO, NLO and NNLO level for 
the two central scale choices $\mu=\pt$ and $\mu=\ptj1$ and for the two cone sizes $R=0.7$ and $R=0.4$. 
Having identified the crucial role of the second jet distribution 
from this analysis, in Section~\ref{subsec:secondjet} we focus our attention to this particular contribution and present 
how it behaves at a given perturbative order.

\subsection{Calculational setup}
\label{subsec:setup}

In order to investigate the differences between the scale choices $\mu=\pt$ and $\mu=\ptj1$ and their origin, we perform a numerical study for the single jet inclusive cross section at a center of mass energy $\sqrt{s}=13~\TeV$.
The jets are identified using the anti-$k_\rT$ algorithm~\cite{Cacciari:2008gp} and results are presented for both $R=0.4$ and $0.7$ to further allow to inspect the dependence on the jet cone size.

Jets are accepted within the fiducial volume defined through the cuts
\begin{align}
  \lvert y^j \rvert &< 4.7 ,&
  \pt^j &> 114~\GeV , 
\end{align}
covering jet-$\pt$ values up to $2~\TeV$, and ordered in transverse momentum.%
\footnote{Unless otherwise stated, we use the $\pt$ binning and rapidity bin widths used by the CMS collaboration in their 13~TeV jet measurement~\cite{Khachatryan:2016wdh}.} 
As explained in Section~\ref{subsec:indiv}, we can not apply a rapidity binning to leading and subleading jet distributions. 

We systematically use the \verb|PDF4LHC15_nnlo_100| PDF set~\cite{Butterworth:2015oua} 
for the evaluation of the LO, NLO and NNLO contributions. 
This choice of a fixed PDF across the different perturbative orders allows us to quantify the effects of the two scale choices at the partonic cross section level rendering our conclusions independent of the PDF set used. 
The value for the strong coupling constant is given by $\alphas(M_\PZ) = 0.118$, as provided by the PDF set.

\subsection{Corrections to the transverse momentum distribution}
\label{subsec:corrtoptdist}

\begin{figure}[t]
\begin{subfigure}{.5\textwidth}
  \centering
  \includegraphics[width=\linewidth]{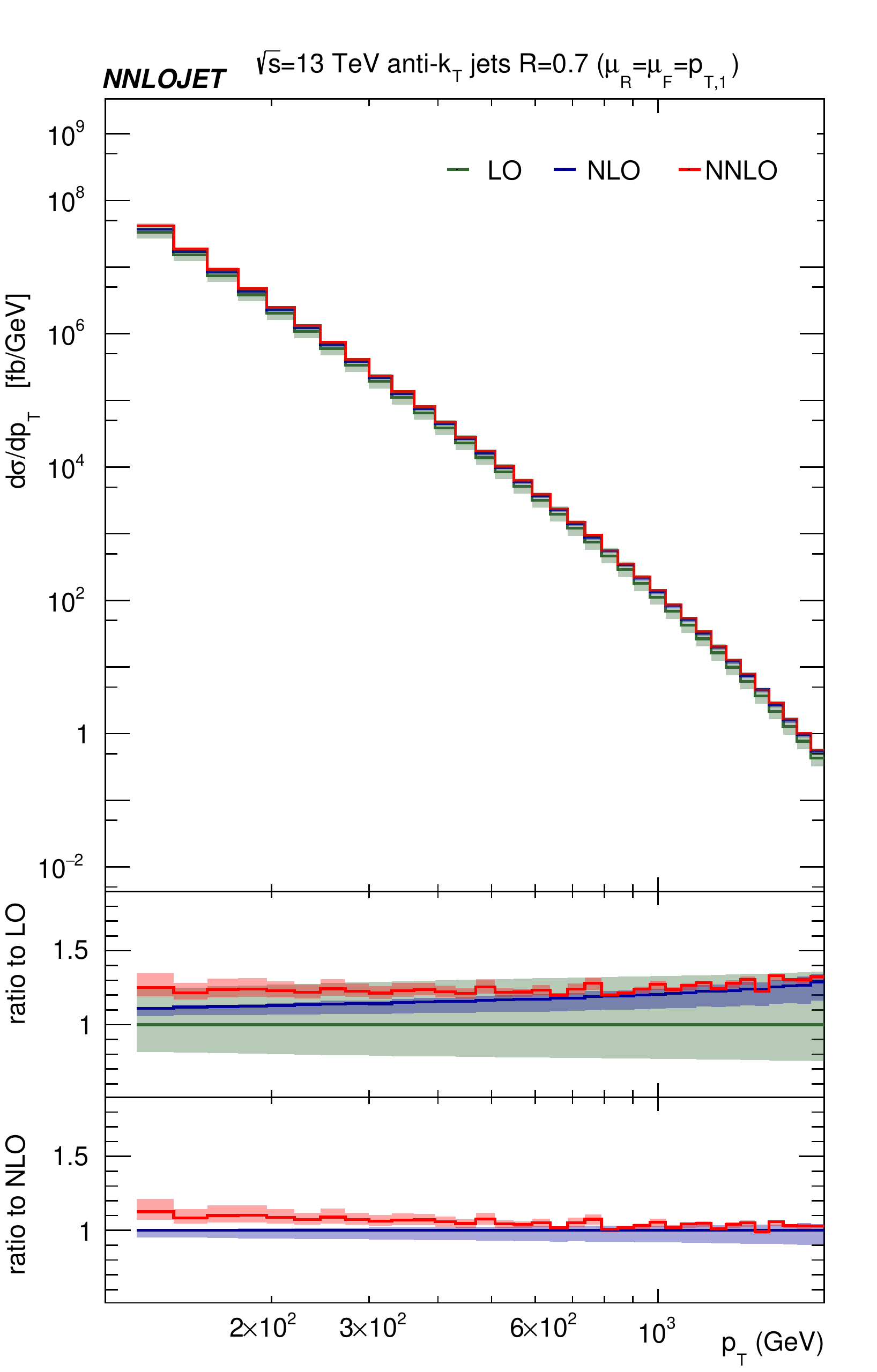}
  \caption{}
  \label{fig:PT0muPT1R07}
\end{subfigure}%
\begin{subfigure}{.5\textwidth}
  \centering
  \includegraphics[width=\linewidth]{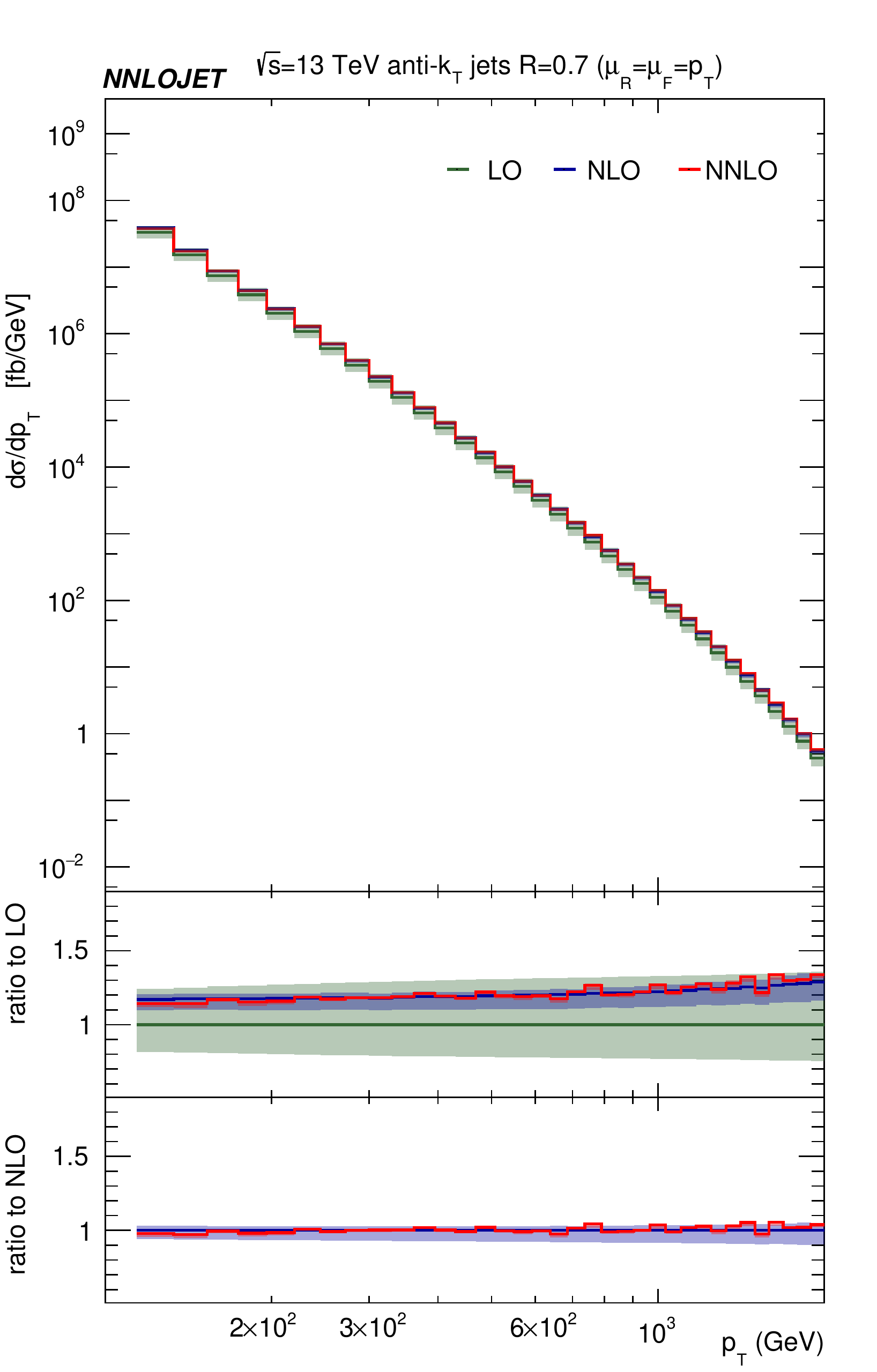}
  \caption{}
  \label{fig:PT0muPTR07}
\end{subfigure}
\caption{Perturbative corrections to the single jet inclusive distribution at 13 TeV (CMS cuts, $|y| < 4.7$, $R=0.7$), integrated over rapidity and normalised to 
lower order predictions. Central scale choice: (a) $\mu=\ptj1$, (b) $\mu=\pt$. }
\label{fig:PT0R07}
\end{figure}

As a first step, we investigate the impact of including NLO and NNLO corrections to the single jet inclusive transverse momentum distribution. 
Figure~\ref{fig:PT0R07} shows the size of the higher order corrections to the single jet inclusive cross section obtained with the scale choice $\mu=\ptj1$ (left) and $\mu=\pt$ (right) for a fixed cone size $R=0.7$. 
The top and bottom panels show respectively ratios of perturbative predictions
to the LO and NLO results and the shaded bands represent the theoretical uncertainty estimated by varying the central scale choices 
by factors of 2 and 1/2 and taking the envelope of the resulting cross sections. For all event-based scales in the remainder of this paper the variation includes doubling and halving the central value of the scale independently for $\mu_{R}$ and $\mu_{F}$, with the constraint $1/2\le\mu_{R}/\mu_{F}\le2$.  For jet-based scales the reevaluation of the event at several different scales is increasingly expensive to compute. For this reason we restrict the scale variation for event-based scales to 3-point symmetric $\mu_{R}$, $\mu_{F}$ scale variations noting that the bulk of the scale dependence comes from $\mu_{R}$ variations and that no significant differences are observed with respect to a 7-point scale variation.

As expected, we can observe that at high $\pt$ the NLO and NNLO effects are small and similar using either of the two central scale choices while more pronounced and different effects can be observed at low $\pt$. In the low $\pt$ region we can observe
larger NLO corrections with the scale $\mu=\pt$ than with the scale $\mu=\ptj1$, while one observes smaller NNLO corrections with the $\mu=\pt$ scale than with the scale $\mu=\ptj1$. As a result we see a faster convergence of the perturbative expansion when using the scale  $\mu=\pt$, where in particular the NNLO result lies inside the NLO scale uncertainty band, which itself lies inside the LO scale band. Furthermore, the scale uncertainty at NNLO displays a greater reduction for the scale choice $\mu=\pt$.

\begin{figure}[t!]
\begin{subfigure}{.5\textwidth}
  \centering
  \includegraphics[width=\linewidth]{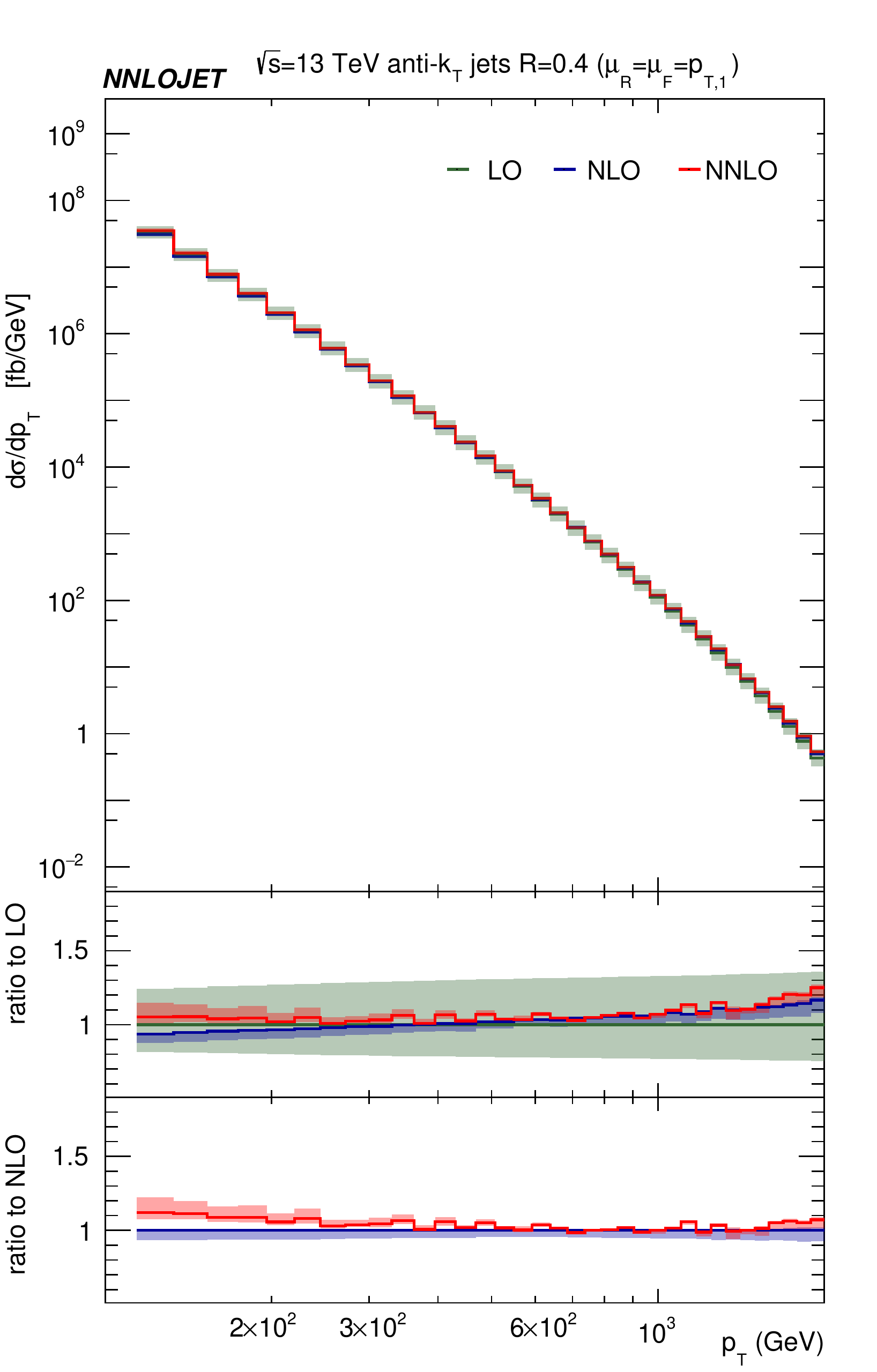}
  \caption{}
  \label{fig:PT0muPT1R04}
\end{subfigure}%
\begin{subfigure}{.5\textwidth}
  \centering
  \includegraphics[width=\linewidth]{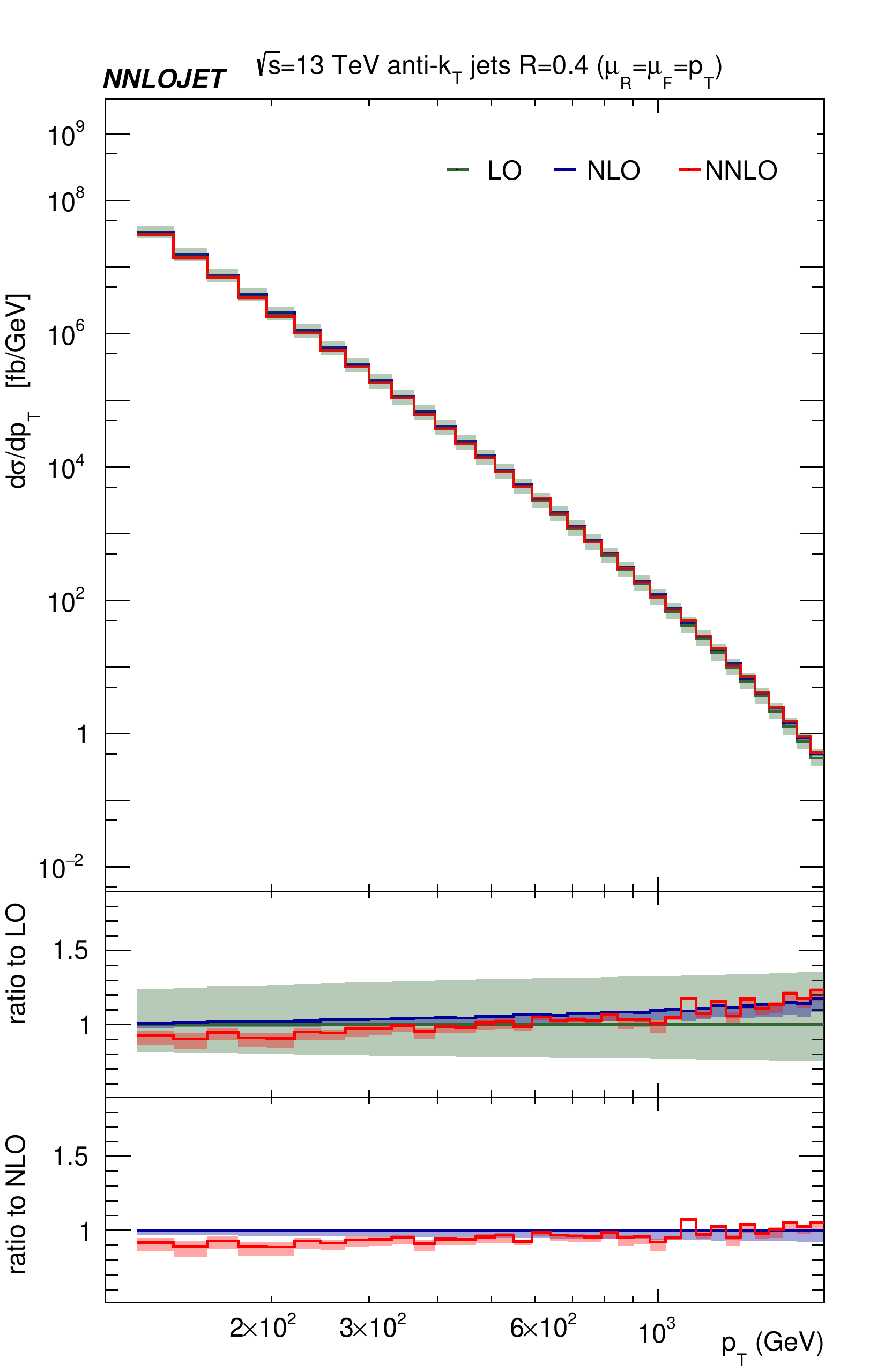}
  \caption{}
\end{subfigure}
\caption{Perturbative corrections to the single jet inclusive distribution at 13 TeV (CMS cuts, $|y| < 4.7$, $R=0.4$), integrated over rapidity and normalised to 
lower order predictions. Central scale choice: (a) $\mu=\ptj1$ and (b) $\mu=\pt$. }
\label{fig:PT0R04}
\end{figure}

It is instructive to compare what happens when using a smaller jet cone size. 
In this case, we fix the jet cone size to $R=0.4$ and present the
results in Fig.~\ref{fig:PT0R04}. Similarly to the $R=0.7$ case, we observe identical higher order effects at high $\pt$ between the two scale choices while more pronounced effects can be seen at low $\pt$. 
In this case, we observe that the NLO corrections using the central scale $\mu=\pt$ 
are smaller than those corrections obtained for $R=0.7$. The NLO scale uncertainty band is artificially small with the central scale choice sitting at the top of the band and the overlap between the NNLO result and the NLO scale band is no longer observed. Looking at the $\mu=\ptj1$ results, we observe
an almost identical NLO scale band as for the results obtained with $R=0.7$ 
and again non-overlapping NNLO and NLO scale bands. 
By comparing the NNLO/NLO
$K$-factors for the two scale choices we observe that the NNLO cross section decreases (increases) with respect the NLO 
result with $\mu=\pt$ ($\mu=\ptj1$).
This is not unexpected since we have anticipated that for smaller jet cone sizes the effects of changing the scale from  $\mu=\pt$ 
to  $\mu=\ptj1$ would be more pronounced. In particular, by comparing the NNLO/LO curves for $\mu=\ptj1$ and $\mu=\pt$ (remembering that the LO result is identical for the two scale choices) 
we see that in the $R=0.7$ case the NNLO predictions lie significantly closer to each other than for $R=0.4$.

\begin{figure}[t]
\centering
\begin{subfigure}{.5\textwidth}
  \centering
  \includegraphics[width=\linewidth]{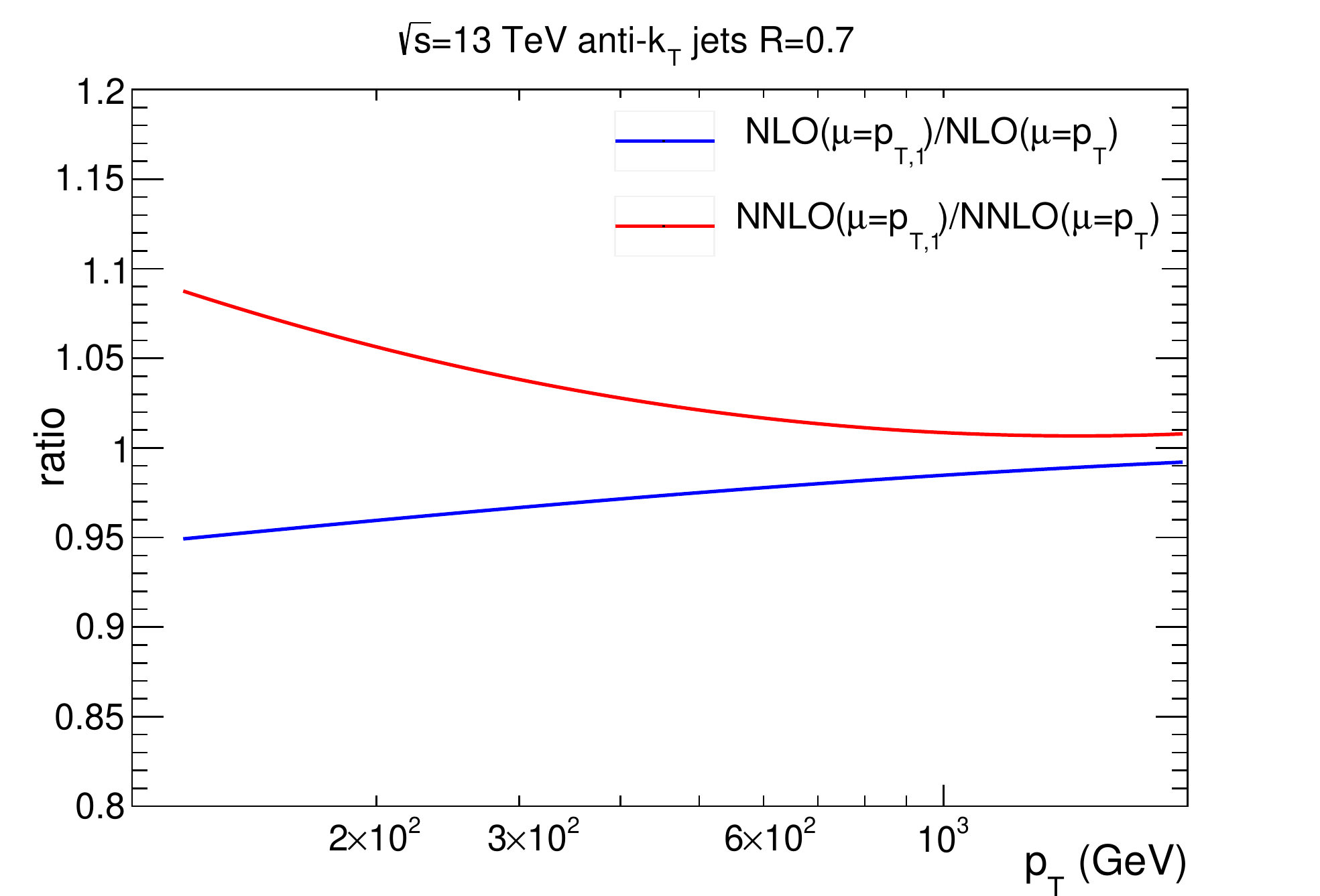}
  \caption{}
  \label{fig:XsecRatio07}
\end{subfigure}%
\begin{subfigure}{.5\textwidth}
  \centering
  \includegraphics[width=\linewidth]{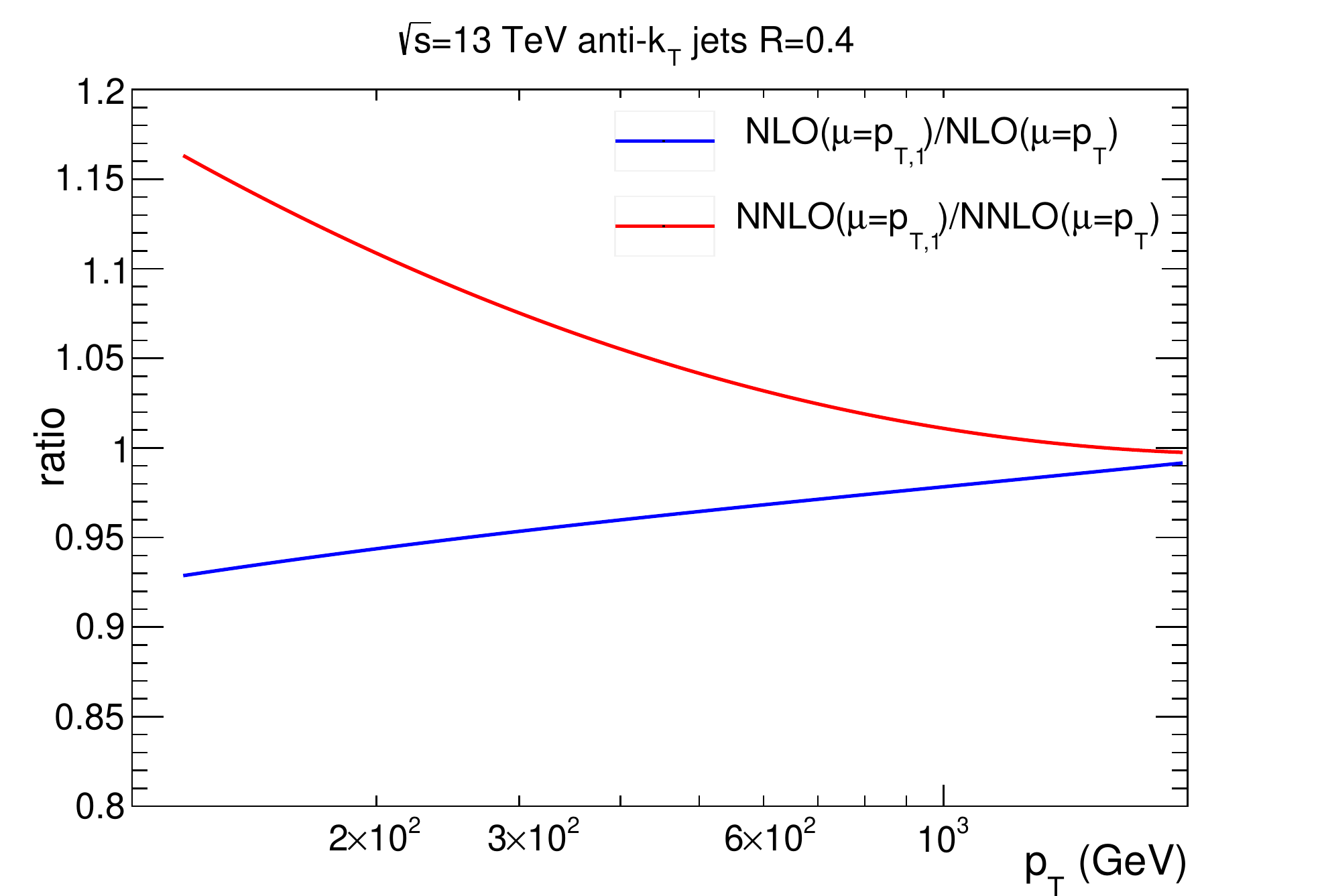}
  \caption{}
  \label{fig:XsecRatio04}
\end{subfigure}
\caption{Ratio between predictions for the single jet inclusive distribution integrated over rapidity obtained with $\mu=\ptj1$ and $\mu=\pt$ at NLO (blue) and 
NNLO (red) for different jet resolution: (a) $R=0.7$; (b) $R=0.4$.}
\label{fig:XsecRatio}
\end{figure}

In order to demonstrate the last effect more clearly, Fig.~\ref{fig:XsecRatio} shows the ratios of the predictions at NLO (in blue) and NNLO (in red) for the two scale choices. We see that at NLO and NNLO, the impact of changing the scale from $\mu=\ptj1$ to $\mu=\pt$ is more pronounced for the smaller jet size $R=0.4$ (right) than it is for the larger jet size $R=0.7$ (left). Interestingly, we also observe that for the two jet sizes,  the impact of this change is bigger at NNLO than it is at NLO, which contradicts our expectation that the higher order corrections should lead to a smaller scale dependence.

It is worth noting that we do not necessarily expect that a change in the form of the central scale choice from $\mu=\ptj1$ to $\mu=\pt$ can be captured by varying the values taken for renormalization and factorization scales in the predictions computed at a given fixed order. When the scale variation is performed, all the events are shifted simultaneously by a rescaling of the $\muR$ and $\muF$ scales.
On the other hand,
when we change the central scale from $\mu=\ptj1$ to $\mu=\pt$, events with LO kinematics are unchanged while events with higher order kinematics can change significantly. 

The renormalization group equations (see Section~\ref{Sec:scaledep}) can be used to predict a change in the cross section due to a multiplication of the scales by a constant shift factor, but are otherwise unable to predict the behaviour of the cross section with another functional form for the central scale choice.  For this reason we can expect the potentially different behaviour of the two scales used to compute IR sensitive observables (which are subject to delicate cancellations between real and virtual corrections) to be the underlying cause of the discrepancy in the results at NNLO between $\mu=\ptj1$ and $\mu=\pt$.

\begin{figure}
\centering
(i)\phantom{ii}\begin{subfigure}{.5\textwidth}
  \centering
  \includegraphics[width=\linewidth,page=1]{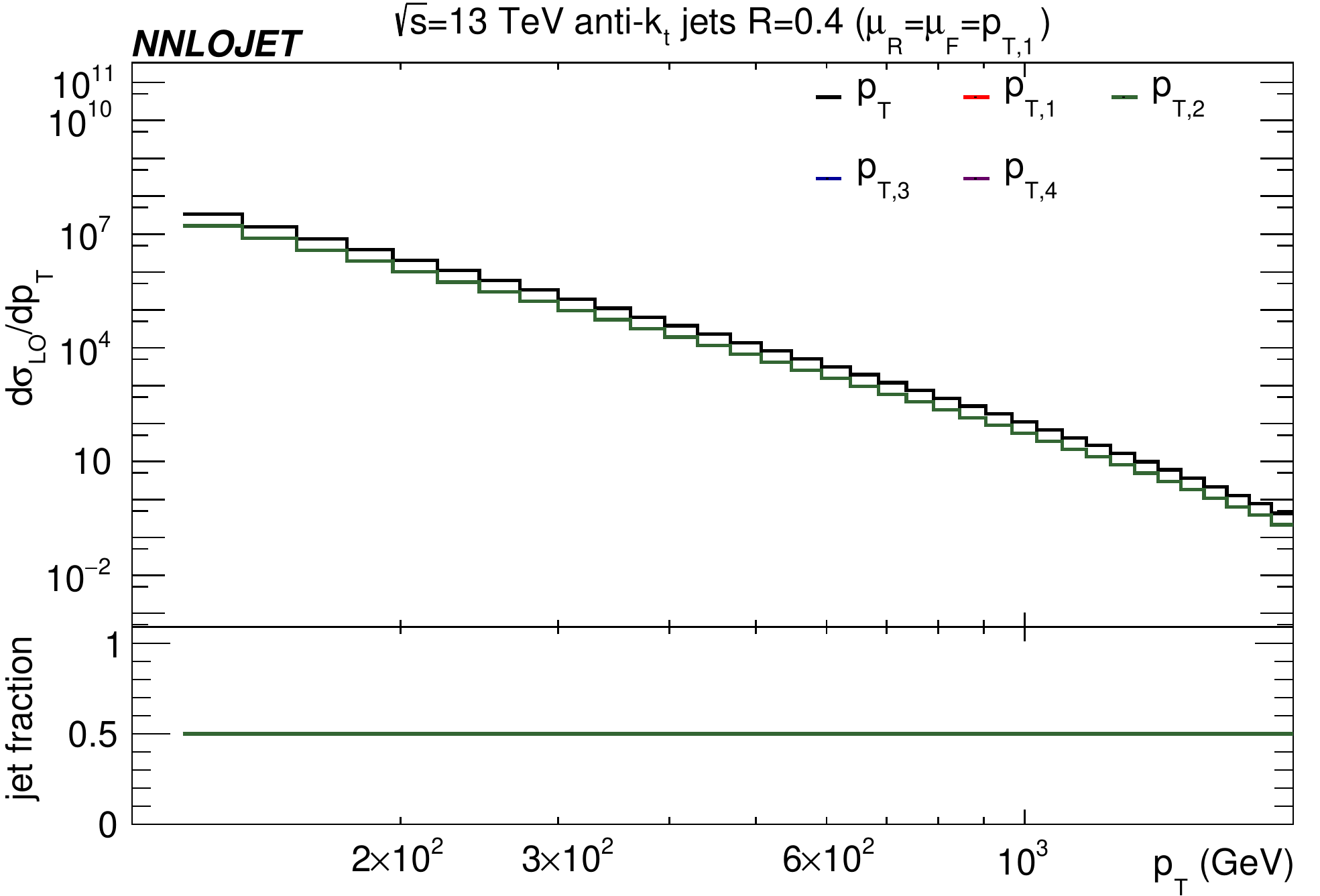}
\end{subfigure}%
\begin{subfigure}{.5\textwidth}
  \centering
  \includegraphics[width=\linewidth,page=1]{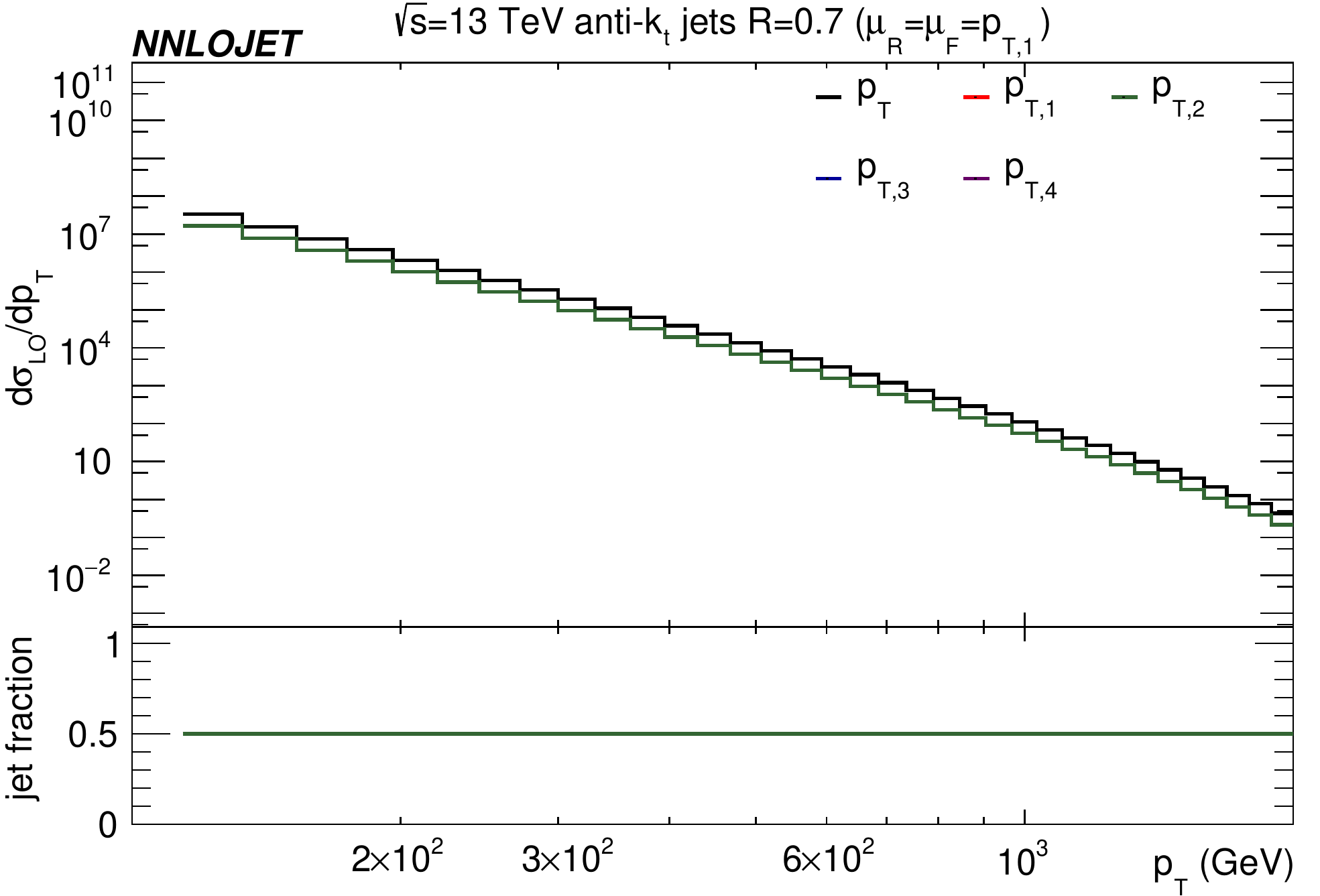}
\end{subfigure}
(ii)\phantom{i}\begin{subfigure}{.5\textwidth}
  \centering
  \includegraphics[width=\linewidth,page=2]{figures/jetfracMUpt1-R04.pdf}
\end{subfigure}%
\begin{subfigure}{.5\textwidth}
  \centering
  \includegraphics[width=\linewidth,page=2]{figures/jetfracMUpt1-R07.pdf}
\end{subfigure}
(iii)\begin{subfigure}{.5\textwidth}
  \centering
  \includegraphics[width=\linewidth,page=3]{figures/jetfracMUpt1-R04.pdf}
  \caption{}
\end{subfigure}%
\begin{subfigure}{.5\textwidth}
  \centering
  \includegraphics[width=\linewidth,page=3]{figures/jetfracMUpt1-R07.pdf}
  \caption{}
\end{subfigure}
\caption{Breakdown of single jet inclusive cross section integrated over rapidity into contributions from first, second, third 
and fourth jet at (i) LO, (ii) NLO and (iii) NNLO evaluated for $\mu=\ptj1$ for jet cone sizes (a) $R=0.4$ and (b) $R=0.7$.}
\label{fig:jetfracMUpt1}
\end{figure}

\begin{figure}
\centering
(i)\phantom{ii}\begin{subfigure}{.5\textwidth}
  \centering
  \includegraphics[width=\linewidth,page=1]{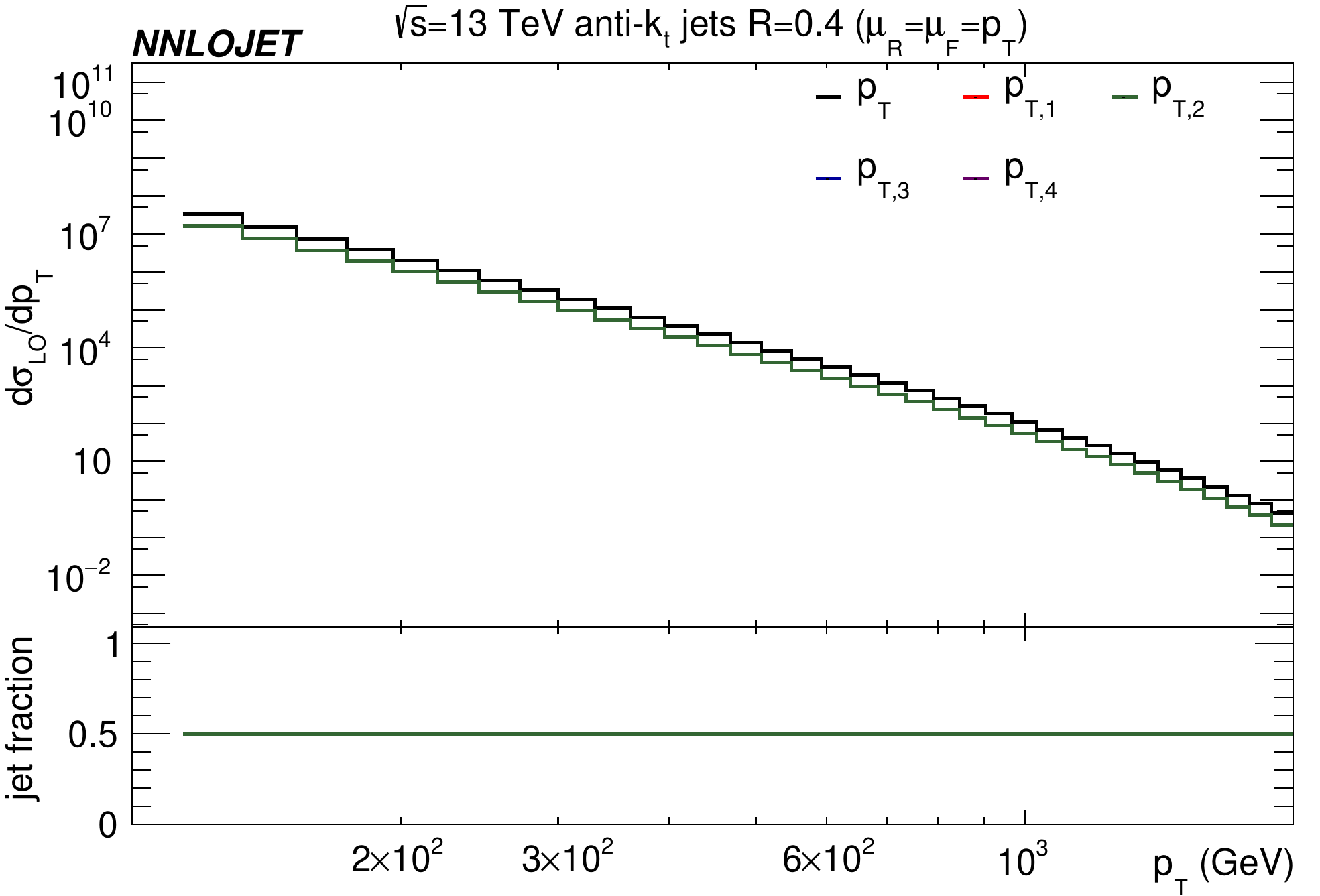}
\end{subfigure}%
\begin{subfigure}{.5\textwidth}
  \centering
  \includegraphics[width=\linewidth,page=1]{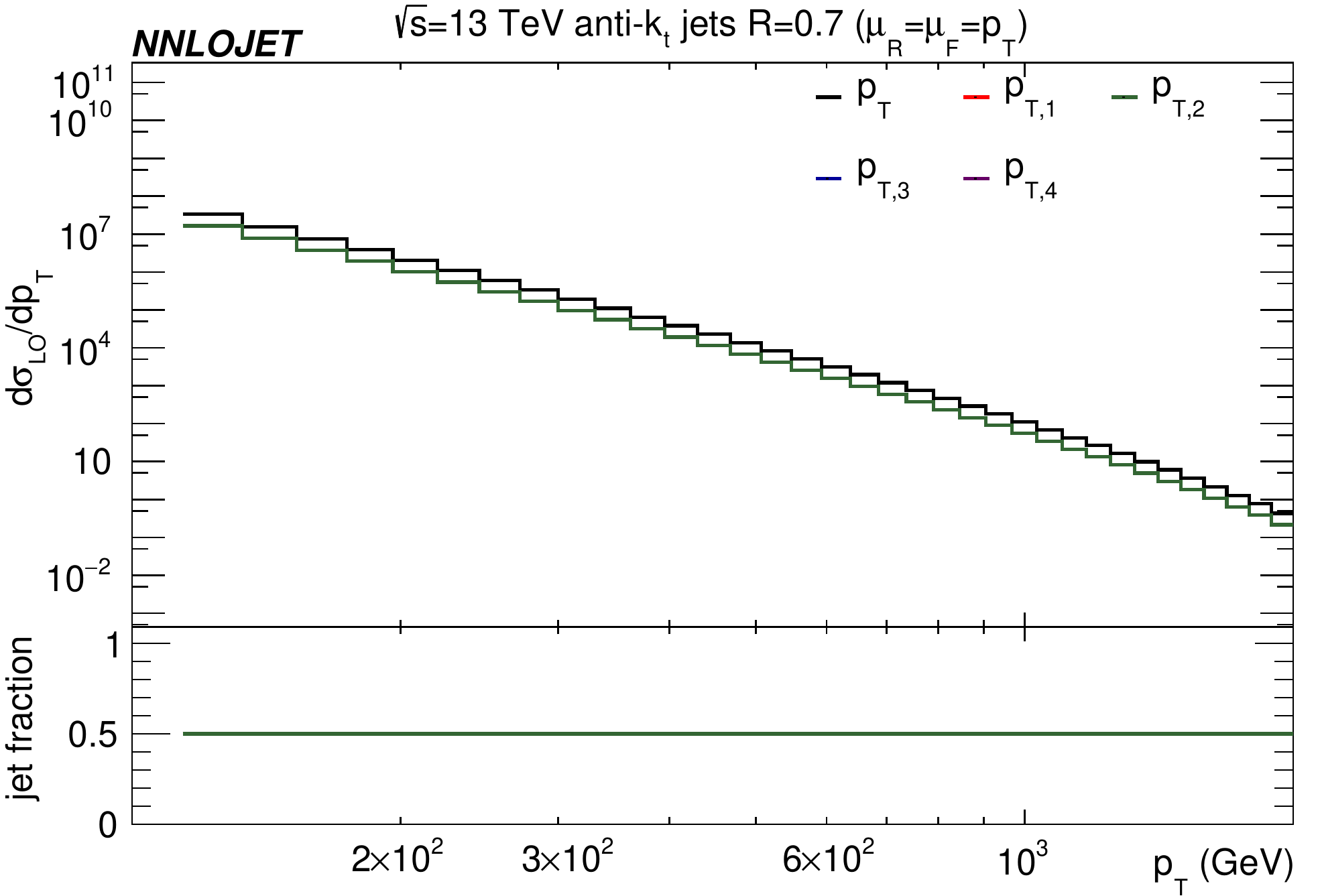}
\end{subfigure}
(ii)\phantom{i}\begin{subfigure}{.5\textwidth}
  \centering
  \includegraphics[width=\linewidth,page=2]{figures/jetfracMUpt-R04.pdf}
\end{subfigure}%
\begin{subfigure}{.5\textwidth}
  \centering
  \includegraphics[width=\linewidth,page=2]{figures/jetfracMUpt-R07.pdf}
\end{subfigure}
(iii)\begin{subfigure}{.5\textwidth}
  \centering
  \includegraphics[width=\linewidth,page=3]{figures/jetfracMUpt-R04.pdf}
  \caption{}
\end{subfigure}%
\begin{subfigure}{.5\textwidth}
  \centering
  \includegraphics[width=\linewidth,page=3]{figures/jetfracMUpt-R07.pdf}
  \caption{}
\end{subfigure}
\caption{Breakdown of single jet inclusive cross section integrated over rapidity into contributions from first, second, third 
and fourth jet at (i) LO, (ii) NLO and (iii) NNLO evaluated for $\mu=\pt$ for jet cone sizes (a) $R=0.4$ and (b) $R=0.7$.}
\label{fig:jetfracMUpt}
\end{figure}

\subsection{Jet fractions in the single jet inclusive distribution}
\label{sec:jetfrac}

In order to explore this idea further, it is instructive to observe the breakdown of the single jet inclusive transverse momentum 
distribution into leading and subleading jet fractions, 
which is shown in Fig.~\ref{fig:jetfracMUpt1} for the scale $\mu=\ptj1$ and jet sizes $R=0.4$ (left) and $R=0.7$ (right) at LO (top), NLO (middle) and NNLO (bottom). 
Beyond
the trivial LO result, which as expected shows an equality between the first and second jet transverse momentum distributions, we observe interesting effects at higher
orders. In particular, at NLO, we find that the leading jet contribution dominates the inclusive jet $\pt$ spectrum 
for both jet sizes, while the contribution from the third jet is negligible. As expected for the larger cone size we produce more events with  jets that are  balanced in $\pt$ and the jet fractions for the first and second jet are closer to the 
symmetric LO result. We can also identify a significant depletion of the second jet contribution in the NLO result for the jet cone size $R=0.4$ at low $\pt$ with the scale choice $\mu=\ptj1$. 
Finally, the NNLO results show a substantial increase in the second jet fraction for both jet sizes with respect to the NLO case, thereby coming closer to the LO result of similar-size first and second jet fractions. 

With these results in mind, we can conclude that a small change in the second jet $\pt$ distribution can have a potentially larger impact
on the inclusive jet transverse momentum distribution at NNLO than at NLO, since the second jet contributes significantly more to the inclusive jet sample at NNLO than it does at NLO. 
It is therefore plausible that a change in scale from $\mu=\ptj1$ to $\mu=\pt$ which affects the second jet $\pt$ distribution produces a larger shift in the prediction of the inclusive jet $\pt$ distribution at NNLO than it does at NLO (as shown in Fig.~\ref{fig:XsecRatio}). 

For comparison, Fig.~\ref{fig:jetfracMUpt} shows the corresponding jet fractions for the 
$\mu=\pt$ scale choice and jet sizes $R=0.4$ (left) and $R=0.7$ (right) at LO (top), NLO (middle) and NNLO (bottom). 
As expected,
when we compare with the results obtained with the scale $\mu=\ptj1$ we do not see significant differences in the jet fractions for the larger jet size of $R=0.7$. On the other hand, for $R=0.4$ we observe an increase in the second jet contribution at low $\pt$ at NLO and a reduction in the same region at NNLO with respect to the results for $\mu=\ptj1$.

\subsection {The second jet transverse momentum distribution}
\label{subsec:secondjet} 

\begin{figure}[t!]
\centering
\begin{subfigure}{.5\textwidth}
  \centering
  \includegraphics[width=\linewidth]{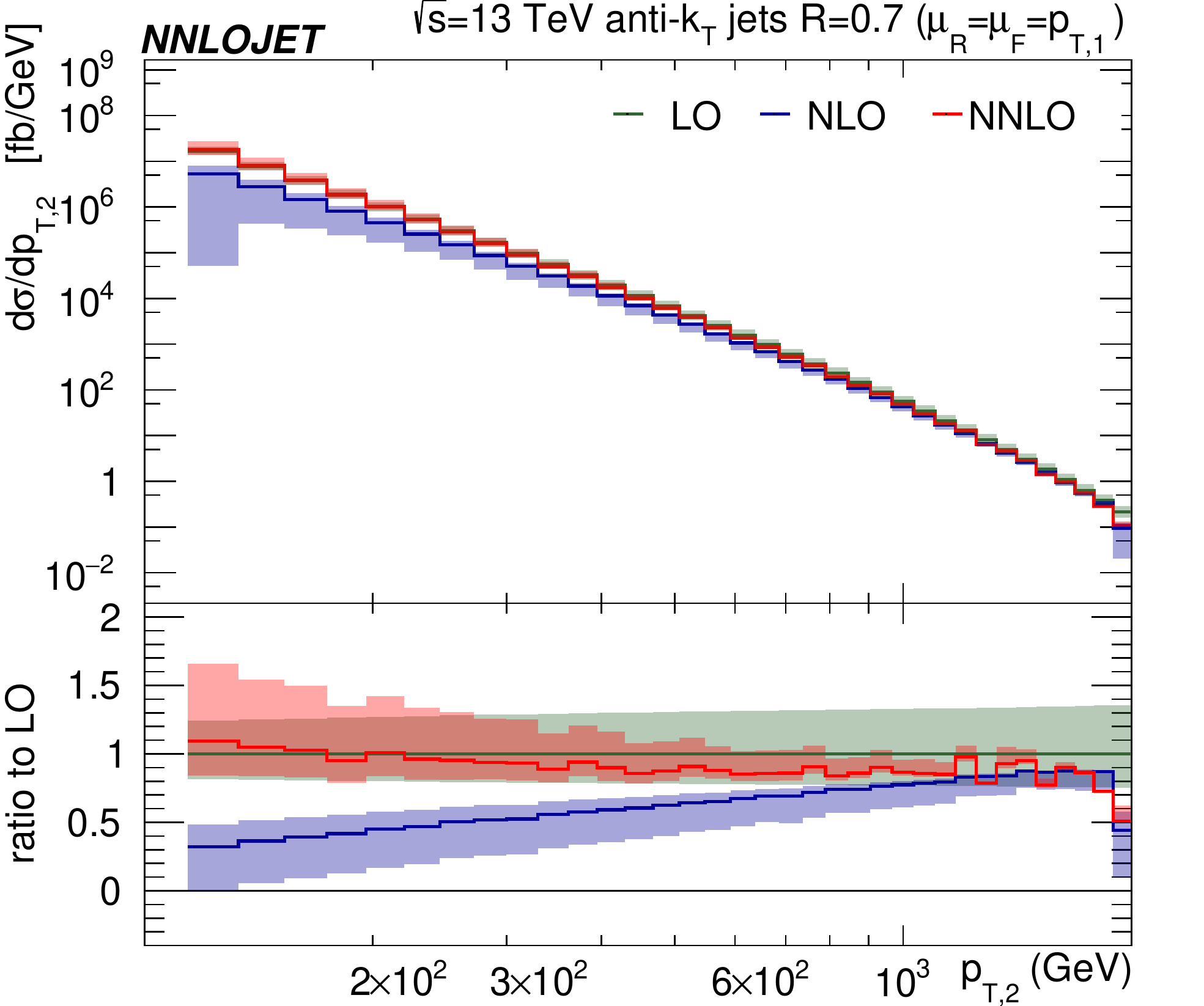}
  \caption{}
  \label{fig:PT2muPT1R07}
\end{subfigure}%
\begin{subfigure}{.5\textwidth}
  \centering
  \includegraphics[width=\linewidth]{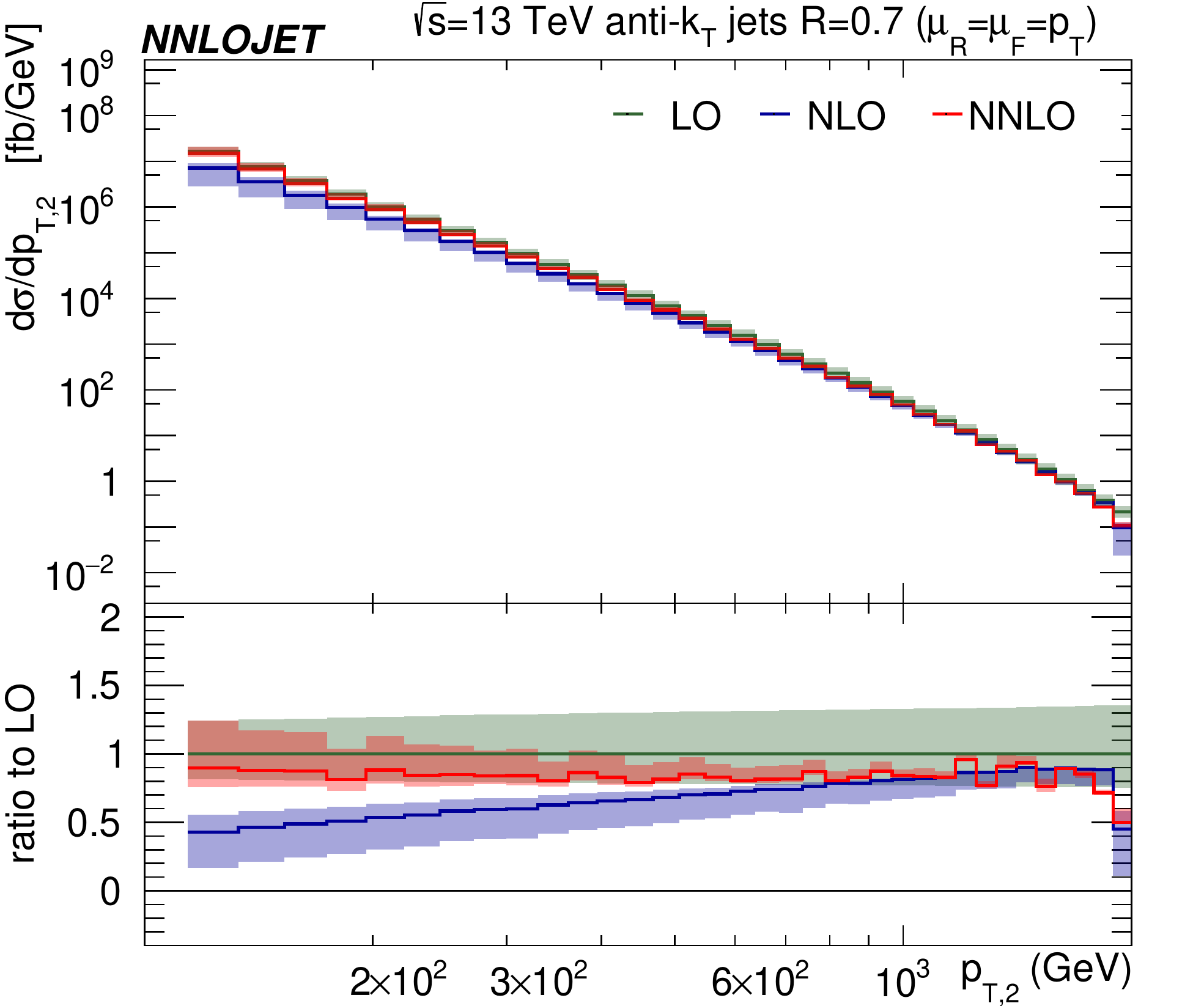}
  \caption{}
  \label{fig:PT2muPTR07}
\end{subfigure}
\caption{Perturbative corrections to the transverse momentum  distribution of the second jet 
at 13 TeV (CMS cuts, $|y| < 4.7$, $R=0.7$),  integrated over rapidity and normalised to 
the LO prediction. Central scale choice: (a) $\mu=\ptj1$ and (b) $\mu=\pt$. 
Shaded bands represent the theory uncertainty due to the variation of the factorization and renormalization scales.}
\label{fig:PT2R07}
\end{figure}

\begin{figure}[t!]
\centering
\begin{subfigure}{.5\textwidth}
  \centering
  \includegraphics[width=\linewidth]{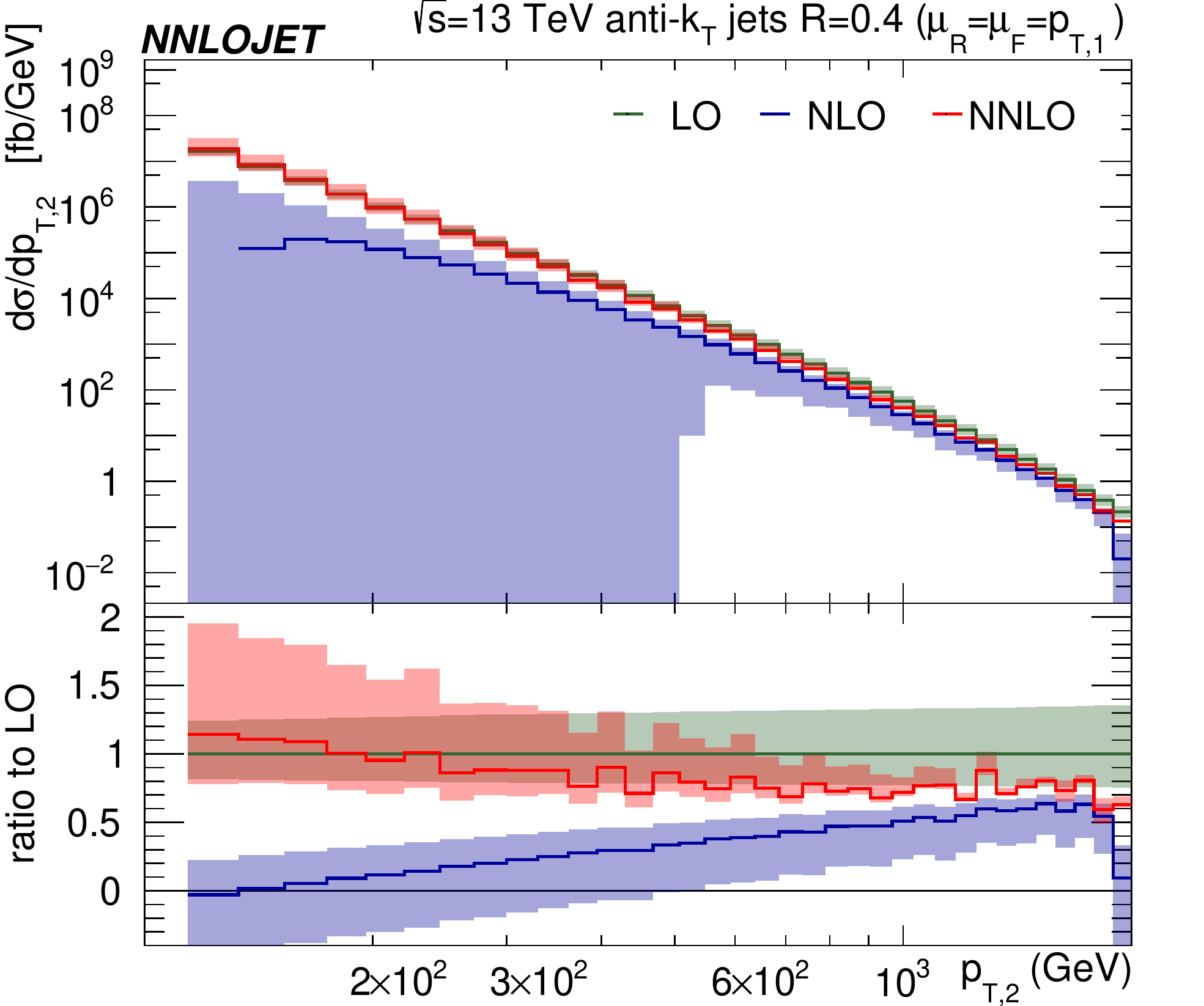}
  \caption{}
  \label{fig:PT2muPT1R04}
\end{subfigure}%
\begin{subfigure}{.5\textwidth}
  \centering
  \includegraphics[width=\linewidth]{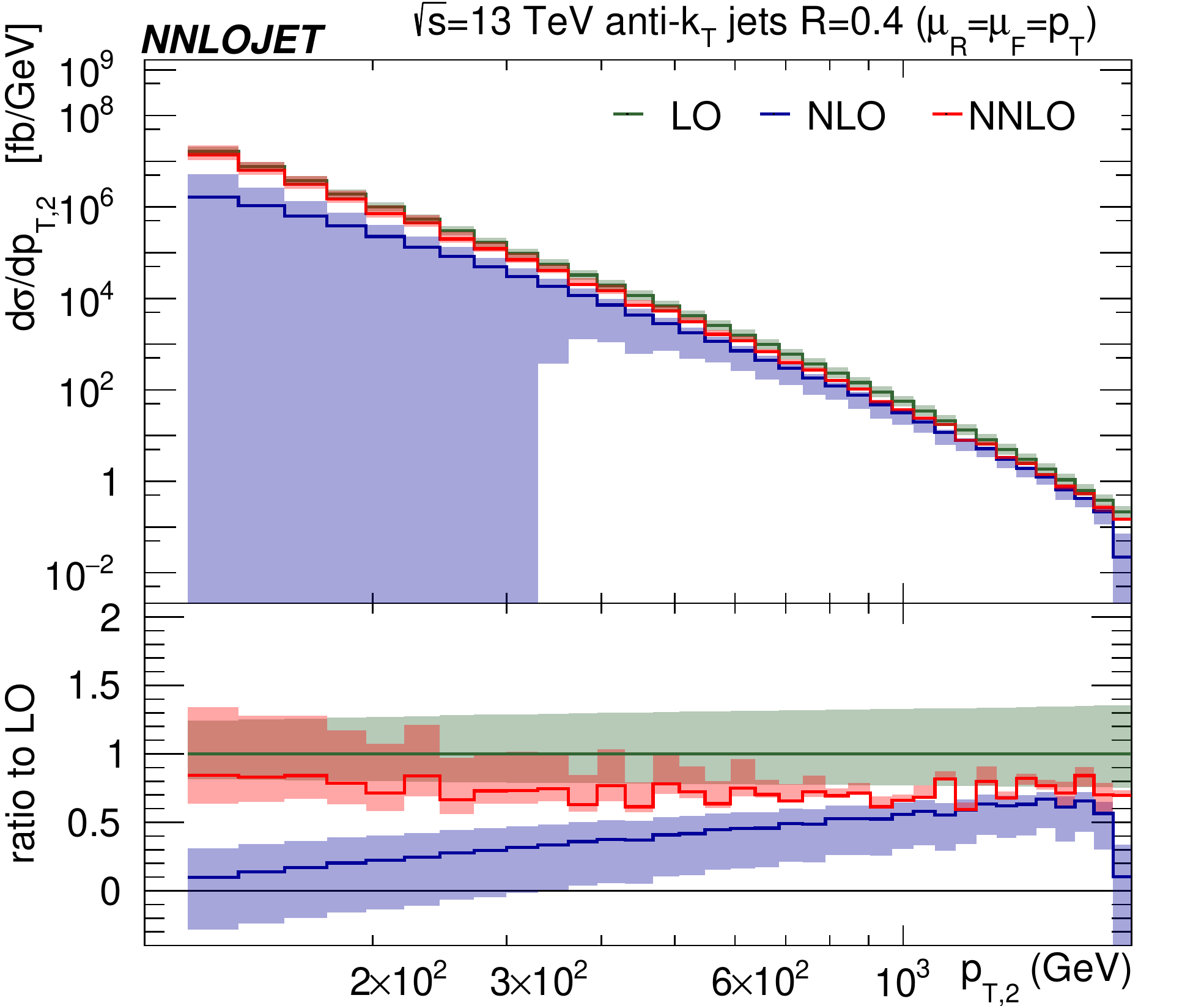}
  \caption{}
  \label{fig:PT2muPTR04}
\end{subfigure}
\caption{Perturbative corrections to the transverse momentum  distribution of the second jet 
at 13 TeV (CMS cuts, $|y| < 4.7$, $R=0.4$),  integrated over rapidity and normalised to 
the LO prediction. Central scale choice: (a) $\mu=\ptj1$ and (b) $\mu=\pt$. 
Shaded bands represent the theory uncertainty due to the variation of the factorization and renormalization scales.}
\label{fig:PT2R04}
\end{figure}

Given its potential impact on the scale uncertainty of the NNLO single jet inclusive cross section, 
we now focus our attention on the second jet transverse momentum distribution.
Fig.~\ref{fig:PT2R07} shows the perturbative expansion of the second jet $\pt$ distribution for the jet cone size of $R=0.7$ with the scale choice $\mu=\ptj1$ (left) and  $\mu=\pt$ (right). 
For the two scale choices, we observe that this distribution
is subject to very large perturbative corrections indicating potentially IR-sensitive effects. In particular we identify the presence of very large negative NLO corrections and large positive NNLO corrections generating an alternating series expansion with large coefficients. It is reassuring that the results at NNLO for the two scale choices are still largely identical despite this effect.
We can nonetheless discern a significantly improved behaviour in the perturbative expansion when the scale  $\mu=\pt$ is used. Both NLO and NNLO $K$-factors are significantly
reduced and the NLO and NNLO scale uncertainty bands are also closer to each other for the $\mu=\pt$ case.

The same behaviour can be observed for the smaller jet cone size of $R=0.4$ in Fig.~\ref{fig:PT2R04}  where the sensitivity to IR effects is even more pronounced. In this case, we find a negative
NLO cross section for the scale choice $\mu=\ptj1$ which is clearly exhibiting a pathological behaviour. 
The NNLO corrections fix this unphysical behaviour even when the
scale $\mu=\ptj1$ is used, but similarly to the $R=0.7$ case, we see a significantly better convergence of the 
perturbative series using $\mu=\pt$ as the central scale choice.  

Interestingly enough we observe also for both cone sizes, that in this contribution the NNLO scale band (in red) is larger than the LO scale band (in green). As explained in Section~\ref{subsec:indiv}, 
the notion of leading and subleading jet is not well defined at leading order ($\ptj1 = \ptj2$ at LO) and for this reason the NLO result is the first non-trivial contribution sensitive to the difference
in $\pt$ between the leading and subleading jet. As such, the single jet inclusive observable is decomposed into IR-sensitive leading and subleading jet contributions
and the functional form of the scale can have an impact on the final result, when the kinematics of the scale choice affects the IR cancellations between the different contributions. 

\begin{figure}
\centering
\begin{subfigure}{.5\textwidth}
  \centering
  \includegraphics[width=\linewidth]{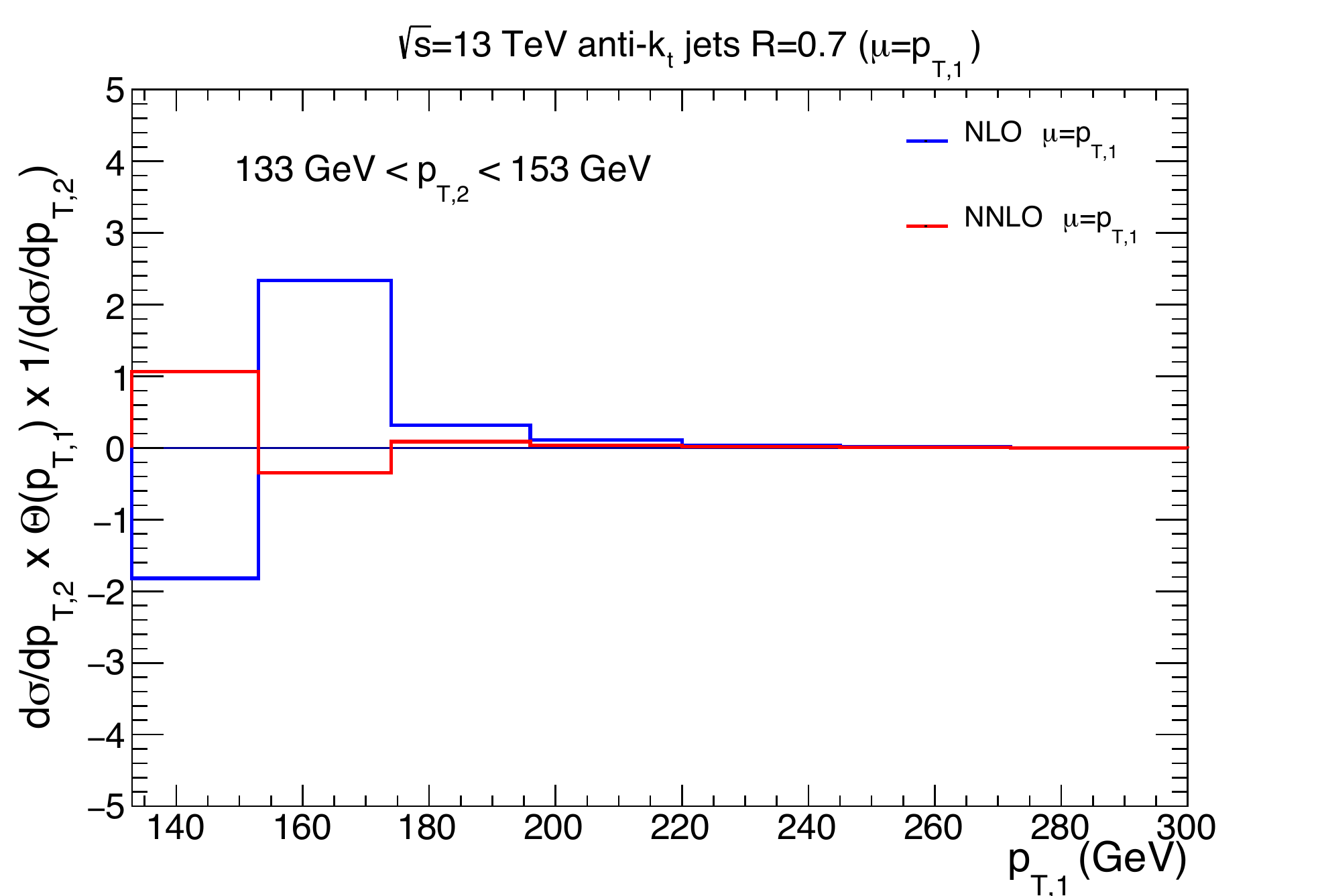}
  \caption{}
\end{subfigure}%
\begin{subfigure}{.5\textwidth}
  \centering
  \includegraphics[width=\linewidth]{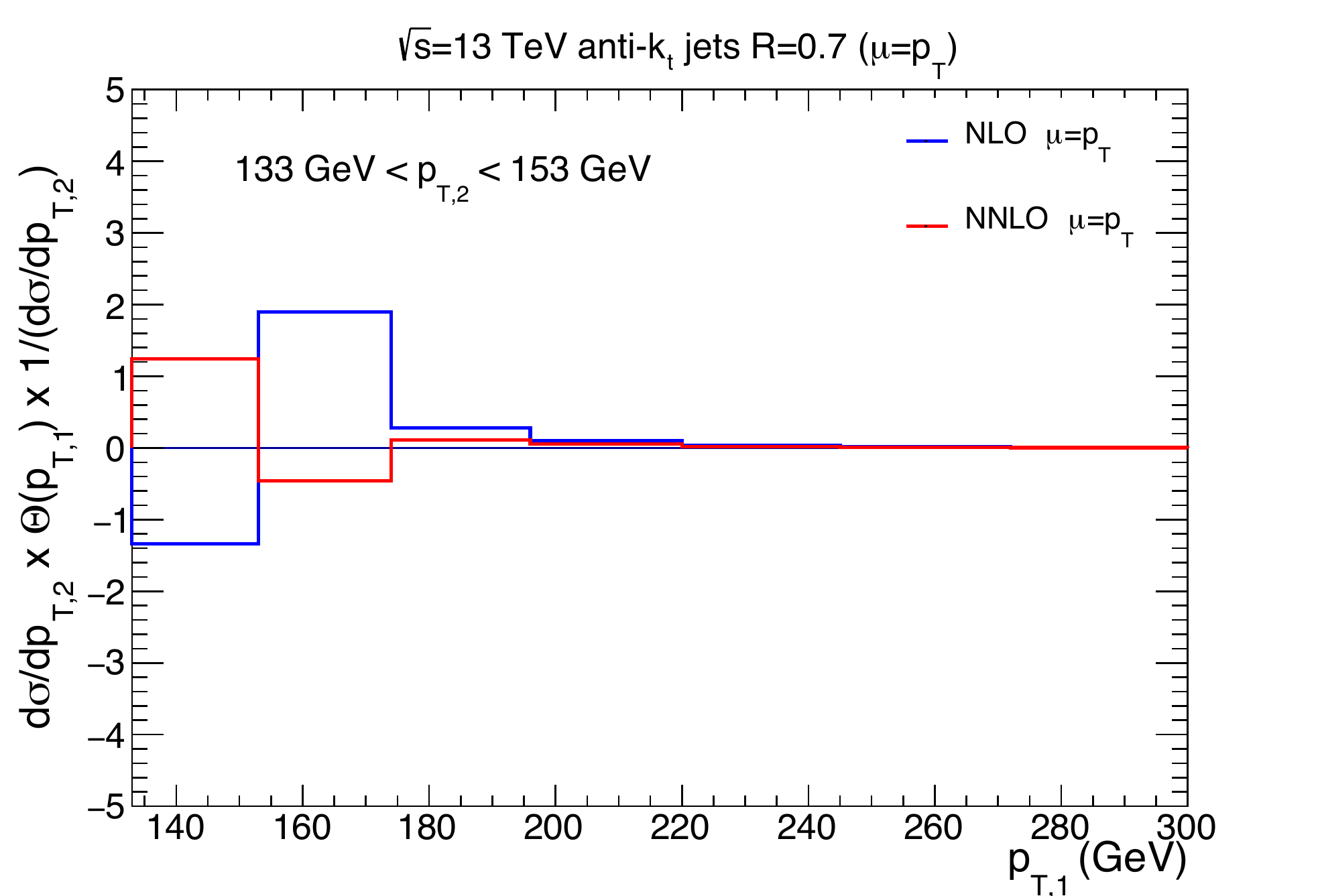}
  \caption{}
\end{subfigure}
\begin{subfigure}{.5\textwidth}
  \centering
  \includegraphics[width=\linewidth]{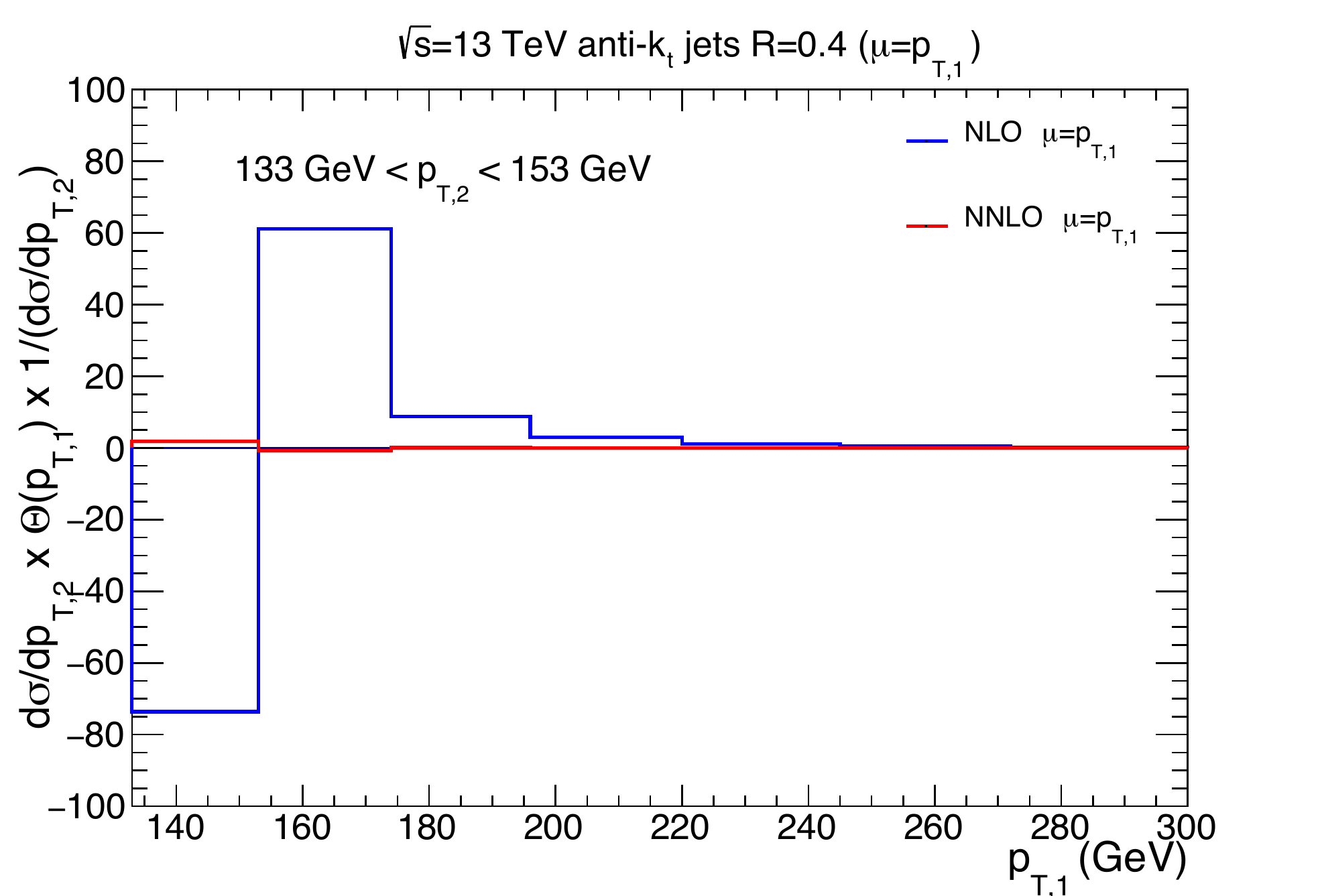}
  \caption{}
\end{subfigure}%
\begin{subfigure}{.5\textwidth}
  \centering
  \includegraphics[width=\linewidth]{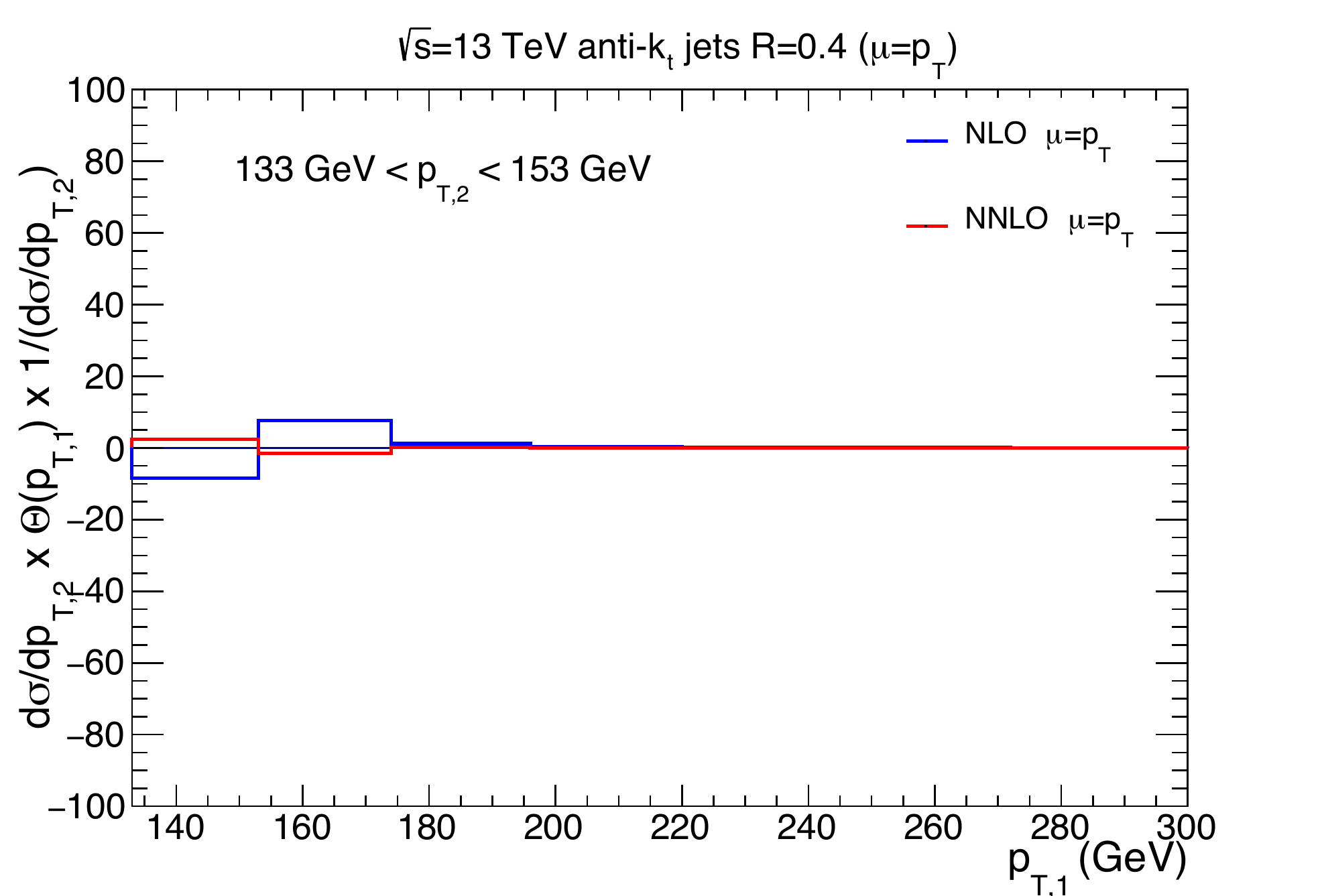}
  \caption{}
\end{subfigure}
\caption{Decomposition of events contributing to a single bin in $\ptj2$ according to the transverse momentum 
of the leading jet in the event $\ptj1$. CMS cuts at 13 TeV with jet resolution $R=0.7$ and scale choice (a) 
$\mu=\ptj1$; (b) $\mu = \pt$;  jet resolution $R=0.4$ and scale choice (c) $\mu=\ptj1$; (d) $\mu = \pt$;.
\label{fig:PTimb}}
\end{figure}

In order to understand the source of the IR sensitivity in the second-jet contribution Fig.~\ref{fig:PTimb} shows the fractional contribution 
to the second jet $\pt$ distribution in a given $\ptj2$ interval (133 GeV $< \ptj2 <$ 153 GeV) for particular $\ptj1$ slices plotted along the $x$-axis, for either  $\mu=\ptj1$ (left frames) or $\mu=\pt$ (right frames), 
and using $R=0.7$ (upper frames) and $R=0.4$ (lower frames). The bin content 
is constrained to sum to unity by construction. We observe 
that this is achieved from a large cancellation (for both scale choices) 
between the first bin of the distribution (where $\ptj1=\ptj2$) and the adjacent bin where ($\ptj1\gtrsim \ptj2$). In particular at NLO (in blue) the entire second bin
content is filled from the NLO real emission (where $\ptj1$ can be larger than $\ptj2$ for the first time) while the virtual correction contributes to the first bin only.
When comparing the behaviour of the two scale choices we note that for $\mu=\ptj1$ the scale is increasing along the $x$-axis.
On the other hand for $\mu=\pt$, the scale is fixed to be equal to $\ptj2$ for all contributions and the cancellation between the large positive real emission and large negative virtual
correction is improved (as shown by the height of the bins). This effect is even more pronounced for the $R=0.4$ jet size as shown in the two lower frames of Fig.~\ref{fig:PTimb}.

We can therefore infer that we observe an instability at higher order in the second jet $\pt$ distribution when additional radiation is not recombined into the outgoing jet and generates an imbalance between $\ptj1$ and $\ptj2$. 
In this case, relatively soft emissions do not outbalance fully with virtual corrections and large logarithms appear.%
\footnote{
  The same instability is present in the inclusive dijet cross section. In that case, the use of asymmetric $\pt$ cuts on the first and second jet's $\pt$ increases the phase space available for soft emissions, suppressing the appearance of large effects from soft gluon emission.
} 
This effect has been observed to be particularly relevant for the smaller jet cone size distributions. After employing the $\mu=\pt$ scale choice we see an improved convergence for the second jet $\pt$ distribution.
The observed stabilisation for the jet-based scale $\mu=\pt$ as opposed to the event-based scale $\mu=\ptj1$ is 
at first sight counter-intuitive, as one should expect an event-based scale to lead to an improved infrared 
stability~\cite{Dasgupta:2016bnd}, since 
all contributions from a single 
parton-level event are  evaluated at the same scale.
The situation is somewhat different for the jet inclusive $\pt$ distribution, since its infrared sensitivity stems only from 
the contribution from the second jet, which has implicit restrictions on its 
allowed phase space. If the second-jet cross section
in a fixed kinematical bin is 
broken down according to the event properties that contribute to it, then the jet-based 
$\mu=\pt$ is a fixed scale, while the event-based $\mu=\ptj1$ 
becomes a dynamical scale. 

We conclude that by employing the scale $\mu=\pt$ we improve the stability of the second jet transverse momentum distribution with respect to $\mu=\ptj1$ by improving
the cancellation at fixed order between the real and virtual corrections. 
Since the leading jet $\ptj1$ contribution is identical with either $\mu=\pt$ and $\mu=\ptj1$, the single jet inclusive cross section is potentially more stable 
when using the jet based scale $\mu=\pt$.

\section{Comparison of different scale choices}
\label{Sec:kfactors13TeV}

The renormalization and factorization scales are arbitrary dimensionful parameters and any scale is \emph{a priori} an equally 
valid choice.
Moreover, any ambiguity induced by different choices of the scales should ideally reduce as higher order terms in the perturbative expansion  are included.
As was shown in the previous section, however, the inclusive $\pt$ distribution suffers from an infrared sensitivity that exhibits a strong dependence on the scale that is used and a suboptimal choice can introduce pathological behaviours in the predictions.

It is the aim of this section to go beyond the two scale choices $\mu=\pt$ and $\mu=\ptj1$ of the previous section and to study predictions for single jet inclusive production based on the comprehensive set of functional forms introduced in Section~\ref{subsec:scalechoice}. In particular, we will study the scale $\mu=\htp$ and the appropriate scaling factor in front of the central scale choice.
To this end, we introduce a set of criteria that define desirable properties for a suitable scale choice:
\begin{enumerate}[(a)]
  \item \emph{perturbative convergence:} 
  We require that the size of the corrections reduces at each successive order in the perturbative expansion.
  \item \emph{scale uncertainty as error estimate:} 
  In order to have a reliable estimate of theory uncertainties due to missing higher-order corrections, we require overlapping scale-uncertainty bands between the last two orders, i.e.\ between the NLO and NNLO predictions. 
  Ideally, the central prediction with the highest accuracy should lie within the scale variation of the order that precedes it.
  \item \emph{perturbative convergence of the individual jet spectra:}
  Based on the observation of the previous section, where the \pt spectra of the individual jets receive large corrections with cancellations in the inclusive distribution, we further demand the convergence of the corrections 
  to the individual \ptj1 and \ptj2 distributions.
  \item \emph{stability of the second jet distribution:}
  The comparison between the scales $\mu=\pt$ and $\ptj1$ has exposed the second jet distribution to be especially sensitive to the scale choice, sometimes even exhibiting unphysical behaviour where the scale variation predicts negative cross section.
  We therefore introduce an additional criterion based on the second jet distribution and its associated scale uncertainty and require the predictions to provide physical, positive cross sections.
\end{enumerate}
In this way, a careful assessment of the behaviour of each scale can be made purely based on the 
behaviour of the 
predictions in perturbation theory, prior to any comparisons with experimental data (which are deferred to Section~\ref{Sec:numerics13TeV}).

Section~\ref{Sec:delpt1pt2} is devoted to a comparison of the different scales on the basis of the criteria defined above and identifying the choices that satisfy our requirements 
on transverse momentum distributions integrated over rapidity. It is the aim of this section to arrive at a sensible scale choice for single jet inclusive production. 
In Section~\ref{Sec:delpt1pt2diff}, we validate the optimal scale choices we made by further looking at the inclusive jet $\pt$ distribution differentially in rapidity.

\subsection{Assessment of the convergence criteria}
\label{Sec:delpt1pt2}

In order to test the convergence criteria~(a--d) defined in the introduction of this section on a more quantitative level, we define the following correction factors for the individual jet \pt spectra
\begin{align}
  \delta k_i^\text{NLO} &= \frac{\rd \sigma^\text{NLO} / \rd \ptj{i} }{\rd \sigma^\text{LO} / \rd \pt} - 1 \;, 
  \nonumber \\
 \delta k_i^\text{NNLO} &= \frac{\rd \sigma^\text{NNLO} / \rd \ptj{i} }{\rd \sigma^\text{NLO} / \rd \pt} - 1 \; ,
 \end{align}
where \ptj{i} denotes the $i$-th leading jet in the event.
The $K$-factor for the inclusive jet distribution can be expressed in terms of the $\delta k_i$ as follows,
\begin{align}
  K^\text{NLO} &= \frac{\rd \sigma^\text{NLO} / \rd \pt }{\rd \sigma^\text{LO} / \rd \pt}
  = 1 + \delta k_\Sigma^\text{NLO} \;, &
  \delta k_\Sigma^\text{NLO} &= \sum_{i\in\text{jets}} \delta k_i^\text{NLO} \;, \nonumber \\
  K^\text{NNLO} &= \frac{\rd \sigma^\text{NNLO} / \rd \pt }{\rd \sigma^\text{NLO} / \rd \pt}
  = 1 + \delta k_\Sigma^\text{NNLO} \;, &
  \delta k_\Sigma^\text{NNLO} &= \sum_{i\in\text{jets}} \delta k_i^\text{NNLO} \;,
\end{align}
and conditions (a,c) are then given by
\begin{align*}
  \text{(a)}\quad& 
  \left\lvert \delta k_\Sigma^\text{NNLO} \right\rvert < \left\lvert \delta k_\Sigma^\text{NLO} \right\rvert \;, \\
  \text{(c)}\quad& 
  \left\lvert \delta k_i^\text{NNLO} \right\rvert < \left\lvert \delta k_i^\text{NLO} \right\rvert \quad\forall i \;.
\end{align*} 
Given that the measured single jet inclusive sample receives contributions predominantly from the two leading jets in the event, it is sufficient to test condition~(c) only for $i=1,\,2$.

\begin{figure}[ht!]
\centering
\begin{subfigure}{.5\textwidth}
  \centering
  \includegraphics[width=1.07\linewidth]{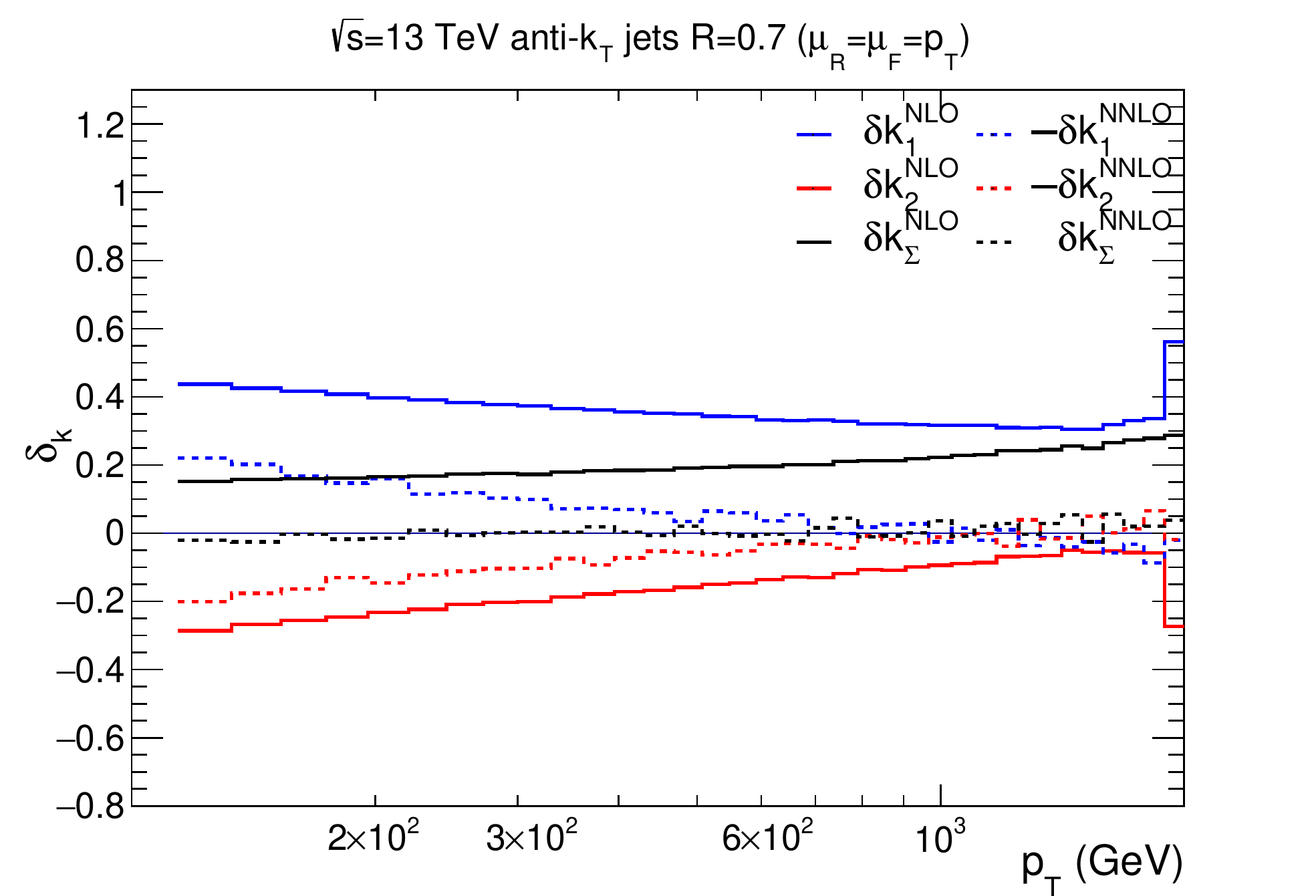}
  \caption{}
\end{subfigure}%
\begin{subfigure}{.5\textwidth}
  \centering
  \includegraphics[width=1.07\linewidth]{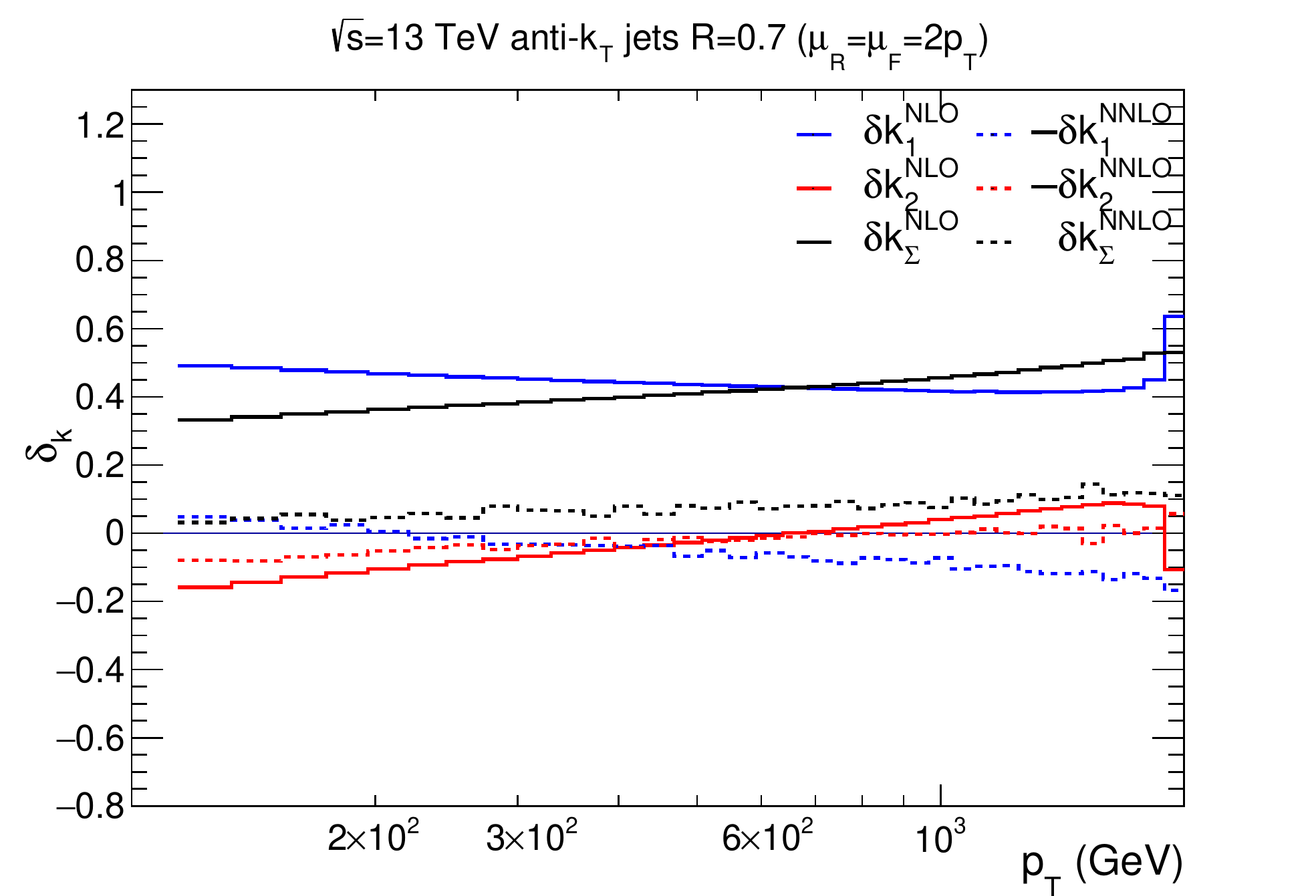}
  \caption{}
\end{subfigure}
\begin{subfigure}{.5\textwidth}
  \centering
  \includegraphics[width=1.07\linewidth]{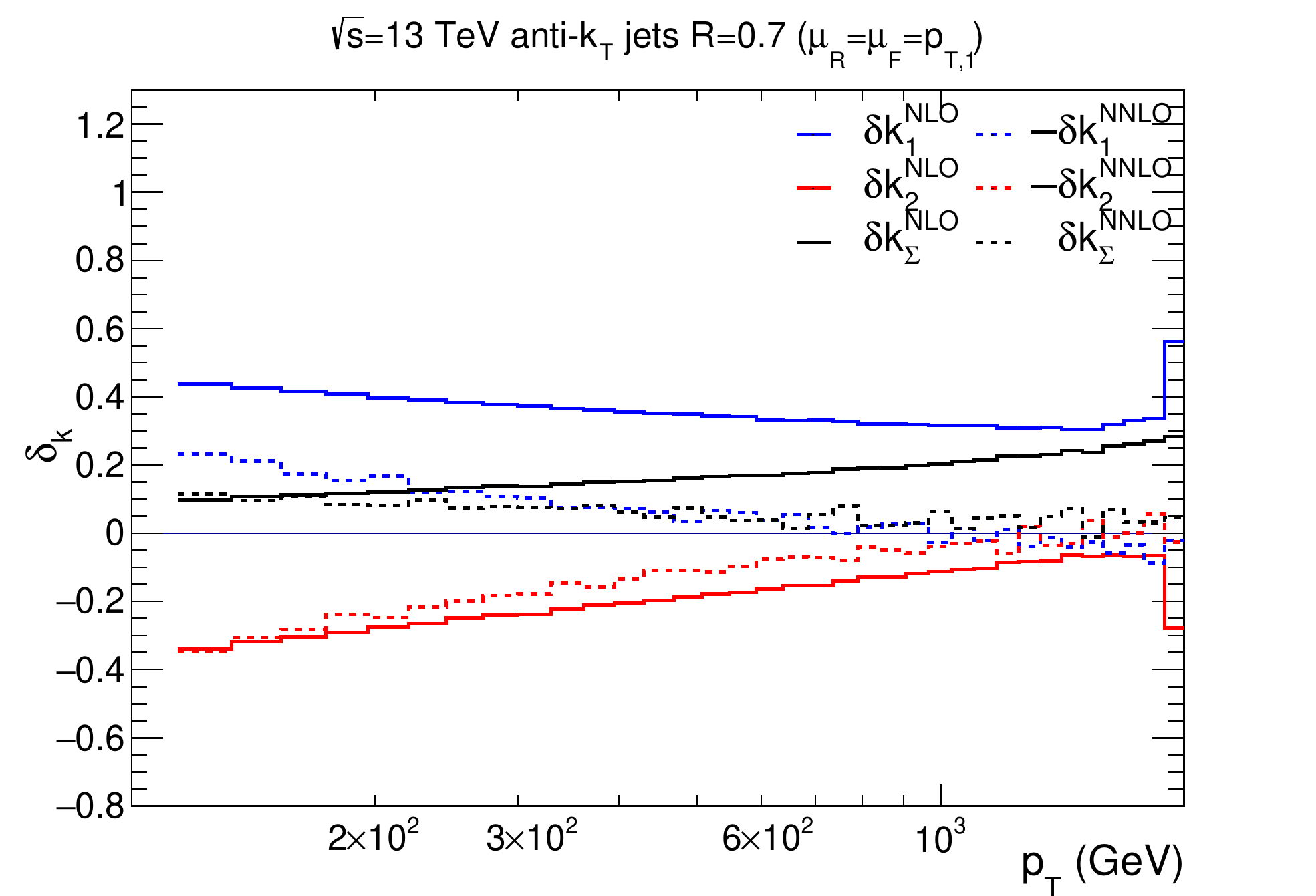}
  \caption{}
\end{subfigure}%
\begin{subfigure}{.5\textwidth}
  \centering
  \includegraphics[width=1.07\linewidth]{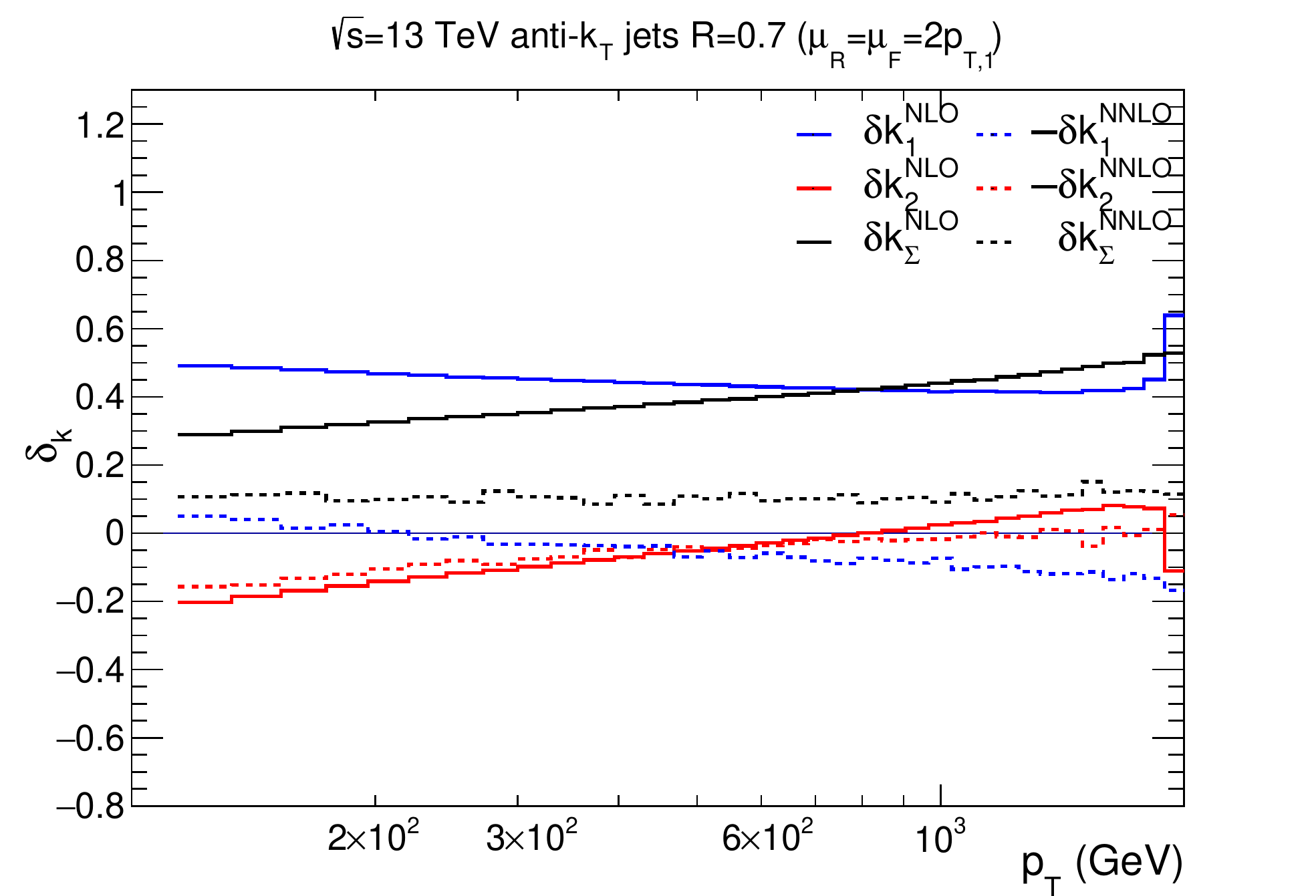}
  \caption{}
\end{subfigure}
\begin{subfigure}{.5\textwidth}
  \centering
  \includegraphics[width=1.07\linewidth]{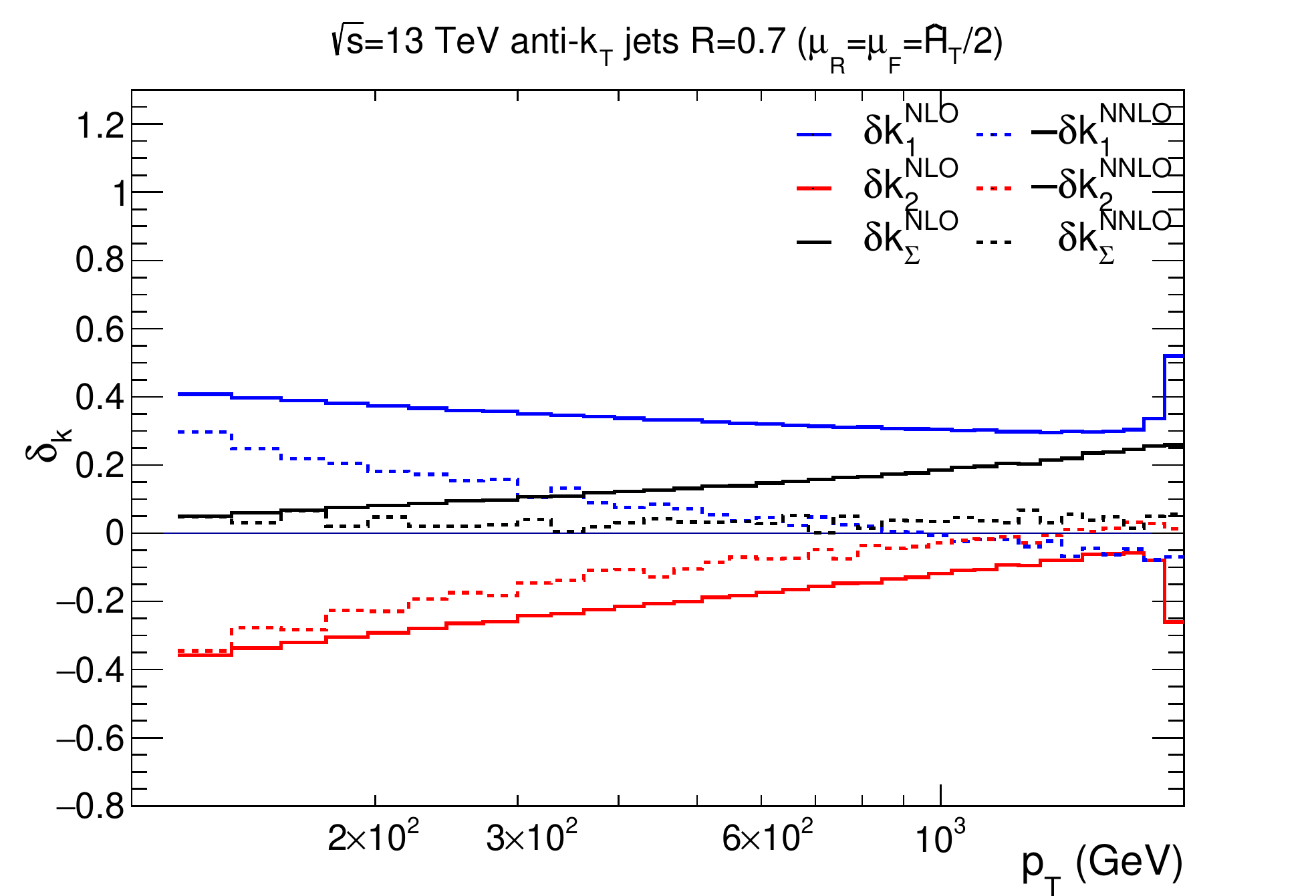}
  \caption{}
\end{subfigure}%
\begin{subfigure}{.5\textwidth}
  \centering
  \includegraphics[width=1.07\linewidth]{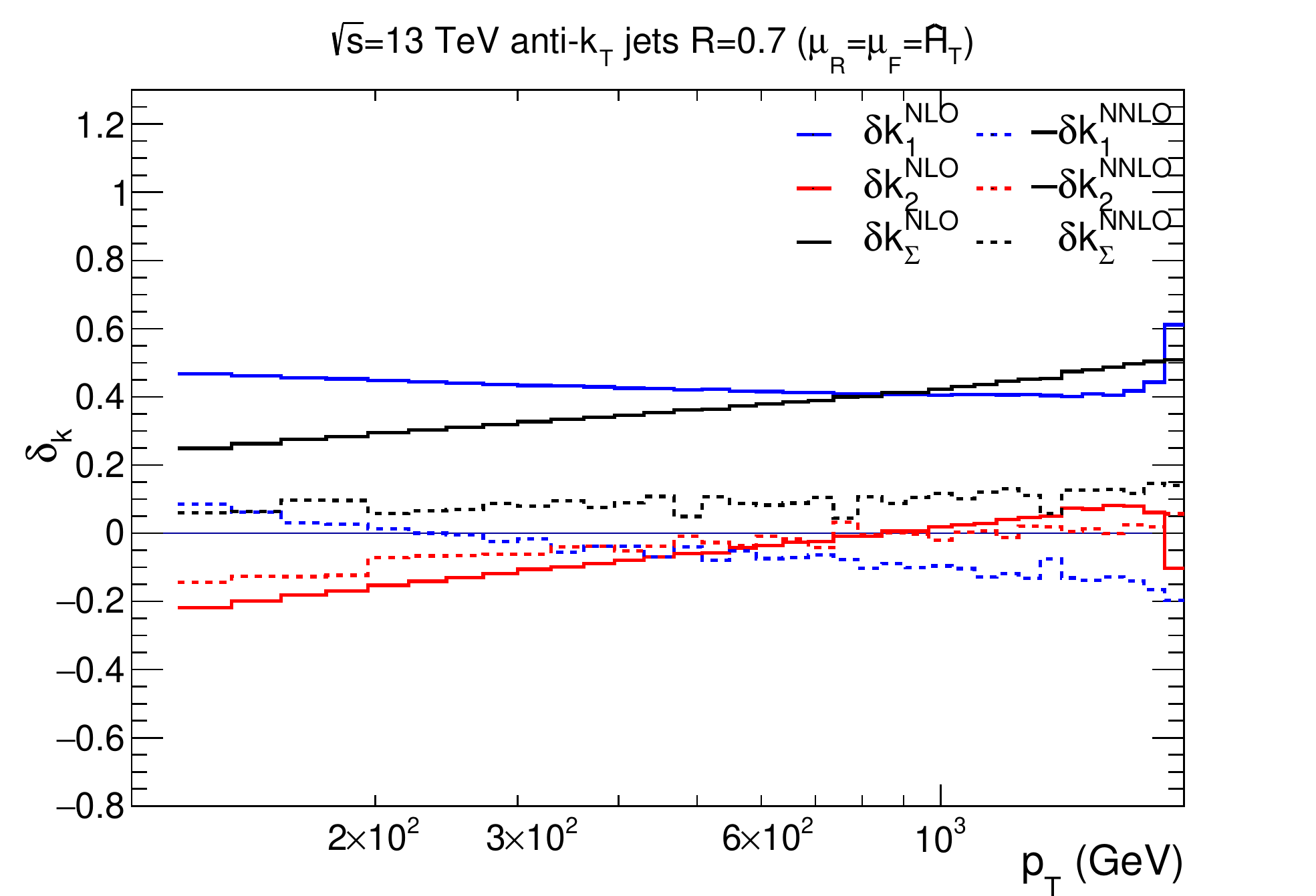}
  \caption{}
\end{subfigure}
\caption{Differential correction factors at $\sqrt{s}=13~\TeV$ for the leading jet ($\delta k_1^\text{N${}^k$LO}$, blue), subleading jet ($\delta k_2^\text{N${}^k$LO}$, red) and the inclusive jet distribution ($\delta k_\Sigma^\text{N${}^k$LO}$, black) with $R=0.7$ and integrated over rapidity for the scale choices (a) $\mu=\pt$, (b) $\mu=2\,\pt$, (c) $\mu=\ptj1$, (d) $\mu=2\,\ptj1$, (e) $\mu=\htp/2 $, (f) $\mu=\htp$.  
\label{fig:pt1pt2AR07}
}
\end{figure}

\begin{figure}[ht!]
\centering
\begin{subfigure}{.5\textwidth}
  \centering
  \includegraphics[width=1.07\linewidth]{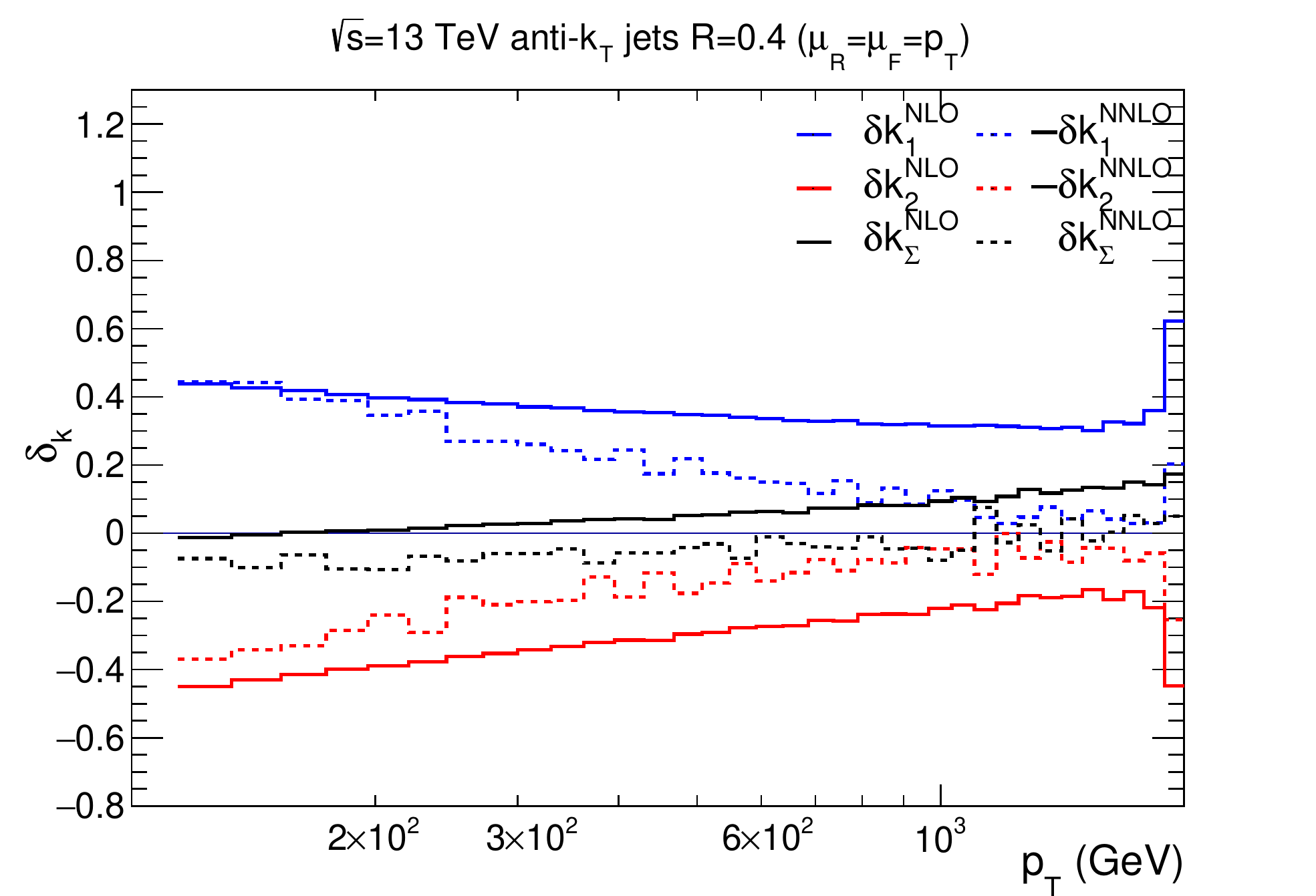}
  \caption{}
\end{subfigure}%
\begin{subfigure}{.5\textwidth}
  \centering
  \includegraphics[width=1.07\linewidth]{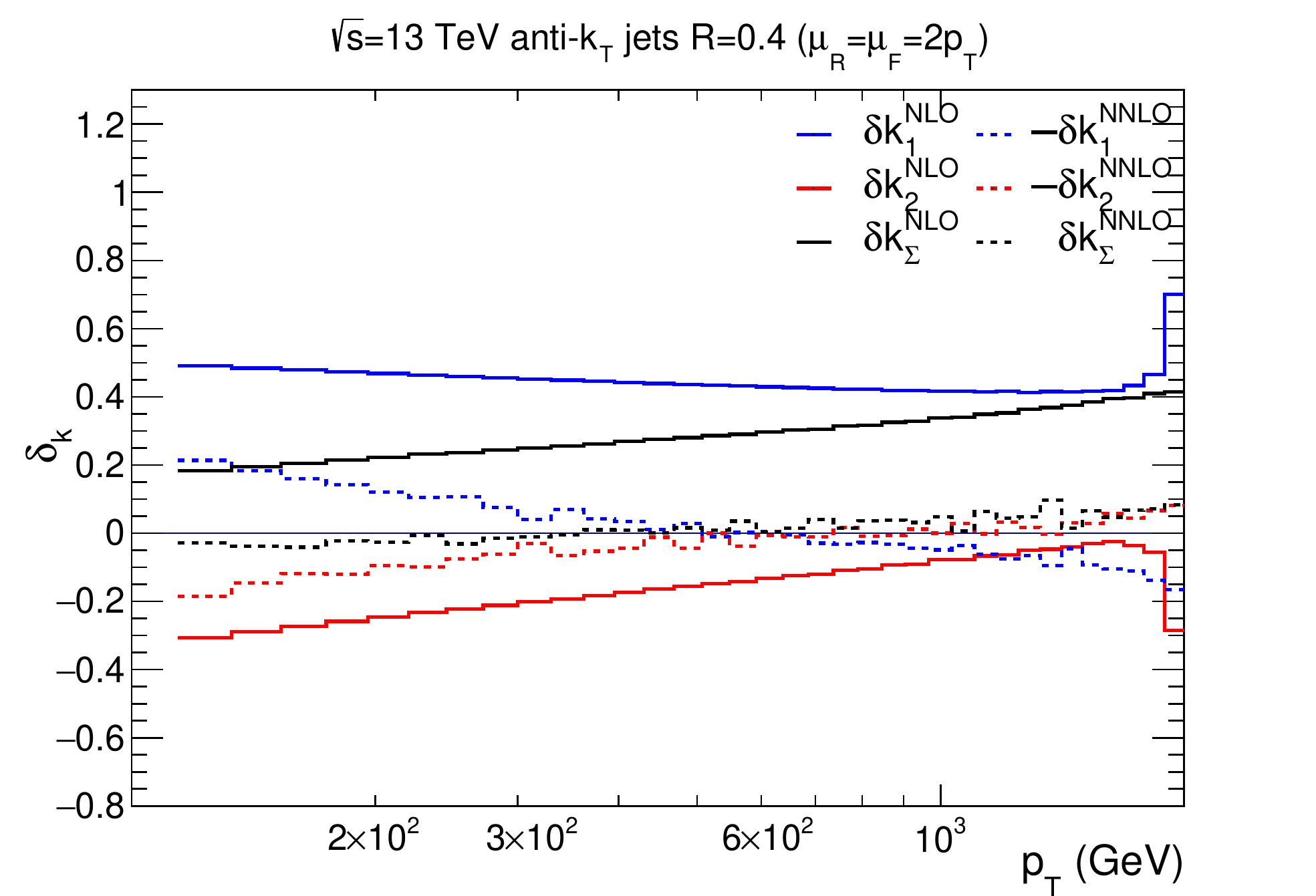}
  \caption{}
\end{subfigure}
\begin{subfigure}{.5\textwidth}
  \centering
  \includegraphics[width=1.07\linewidth]{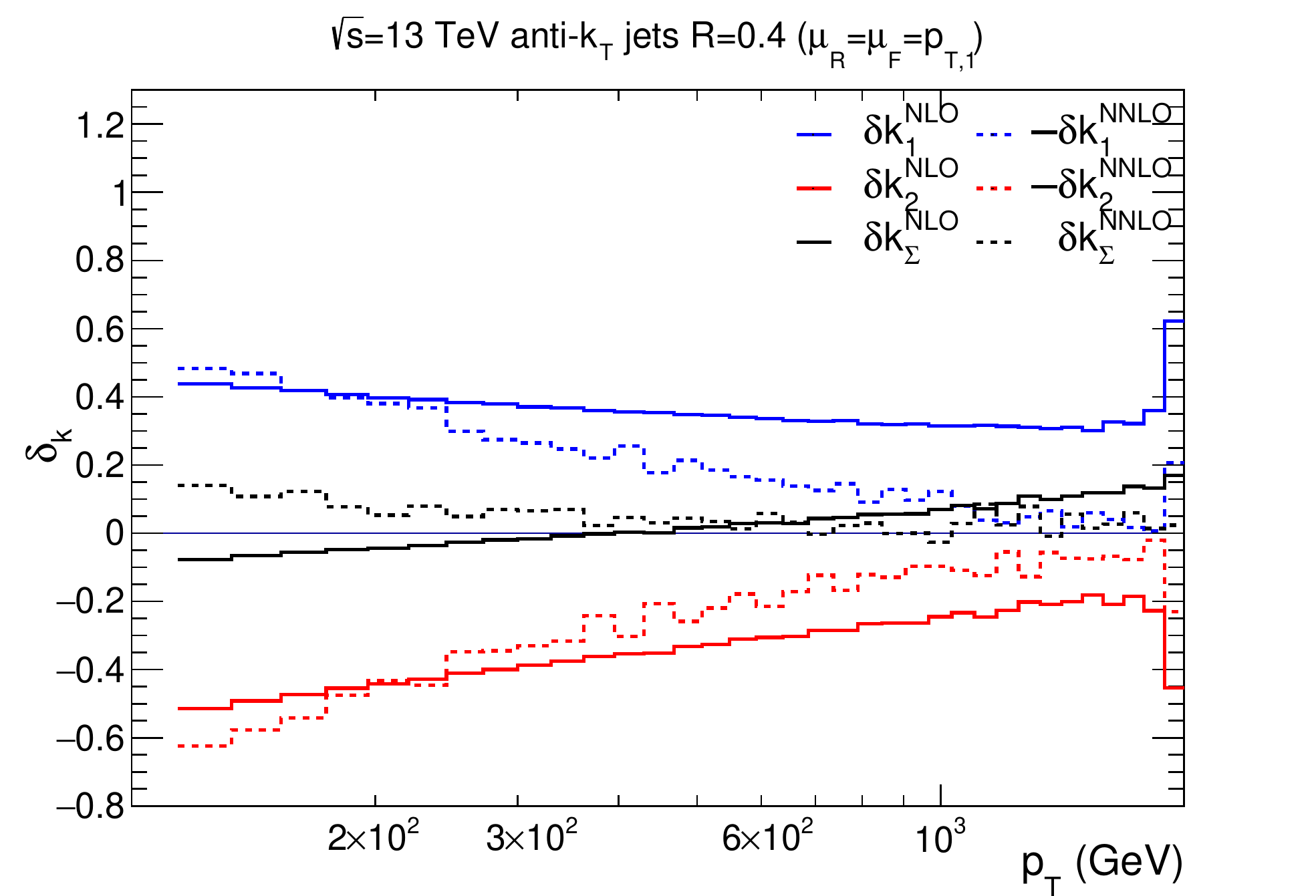}
  \caption{}
\end{subfigure}%
\begin{subfigure}{.5\textwidth}
  \centering
  \includegraphics[width=1.07\linewidth]{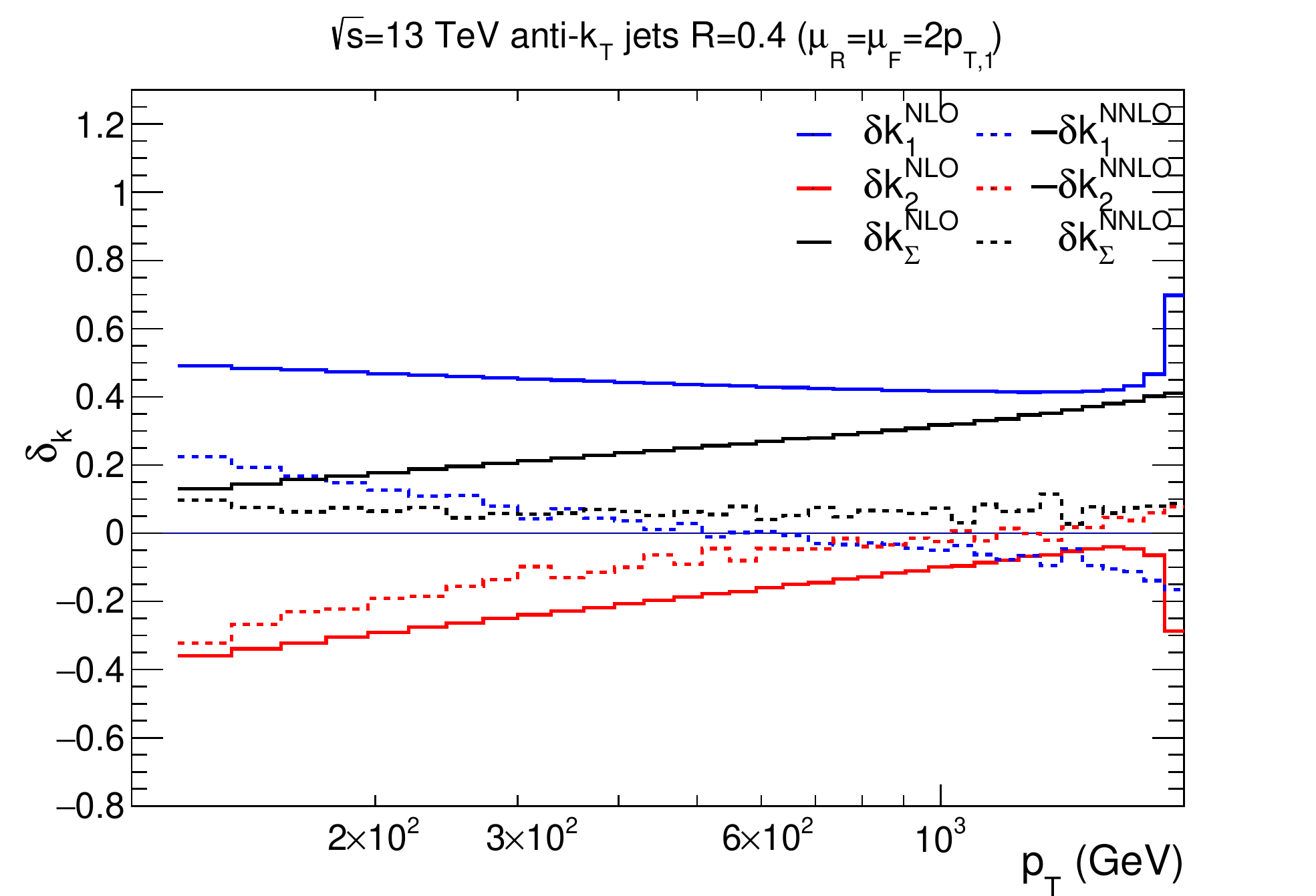}
  \caption{}
\end{subfigure}
\begin{subfigure}{.5\textwidth}
  \centering
  \includegraphics[width=1.07\linewidth]{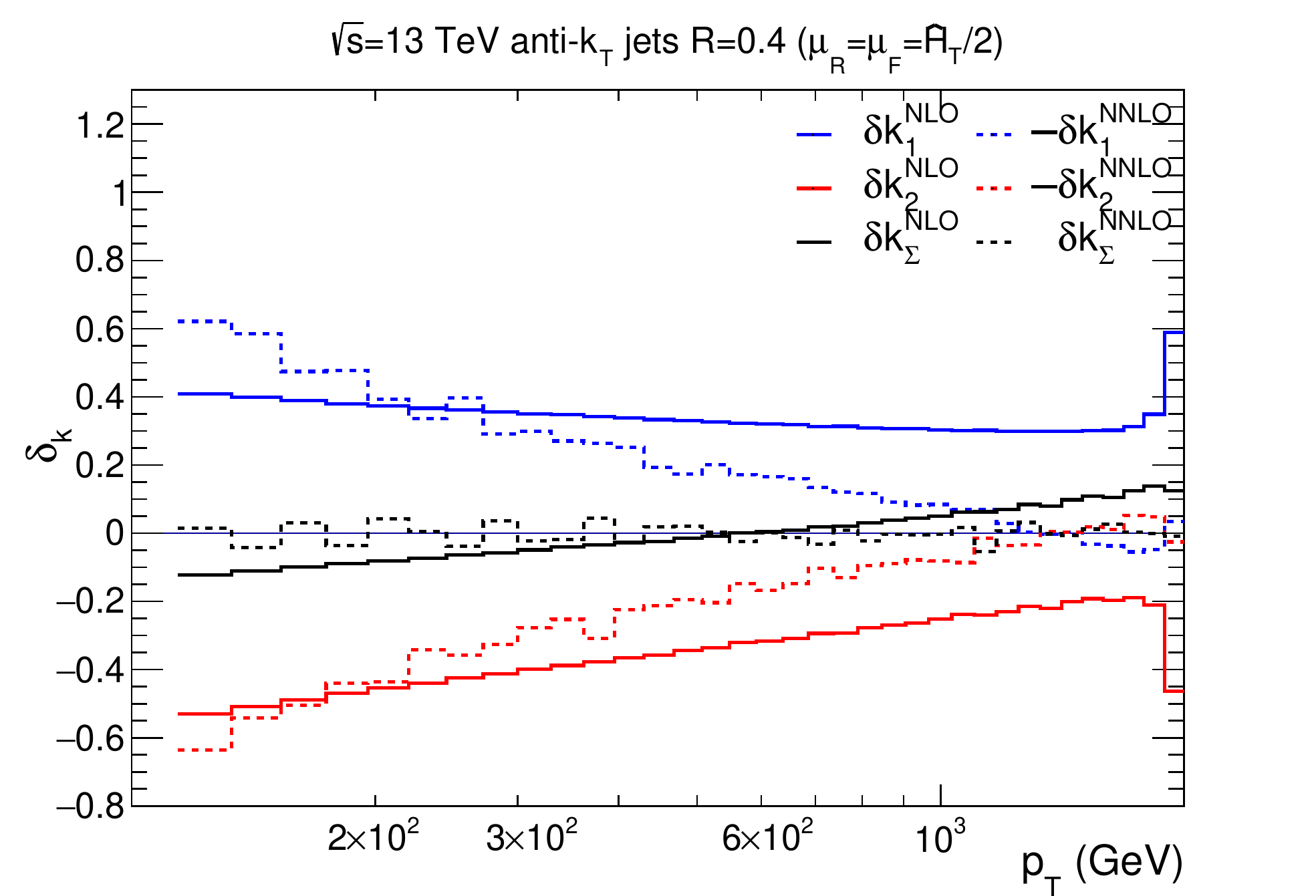}
  \caption{}
\end{subfigure}%
\begin{subfigure}{.5\textwidth}
  \centering
  \includegraphics[width=1.07\linewidth]{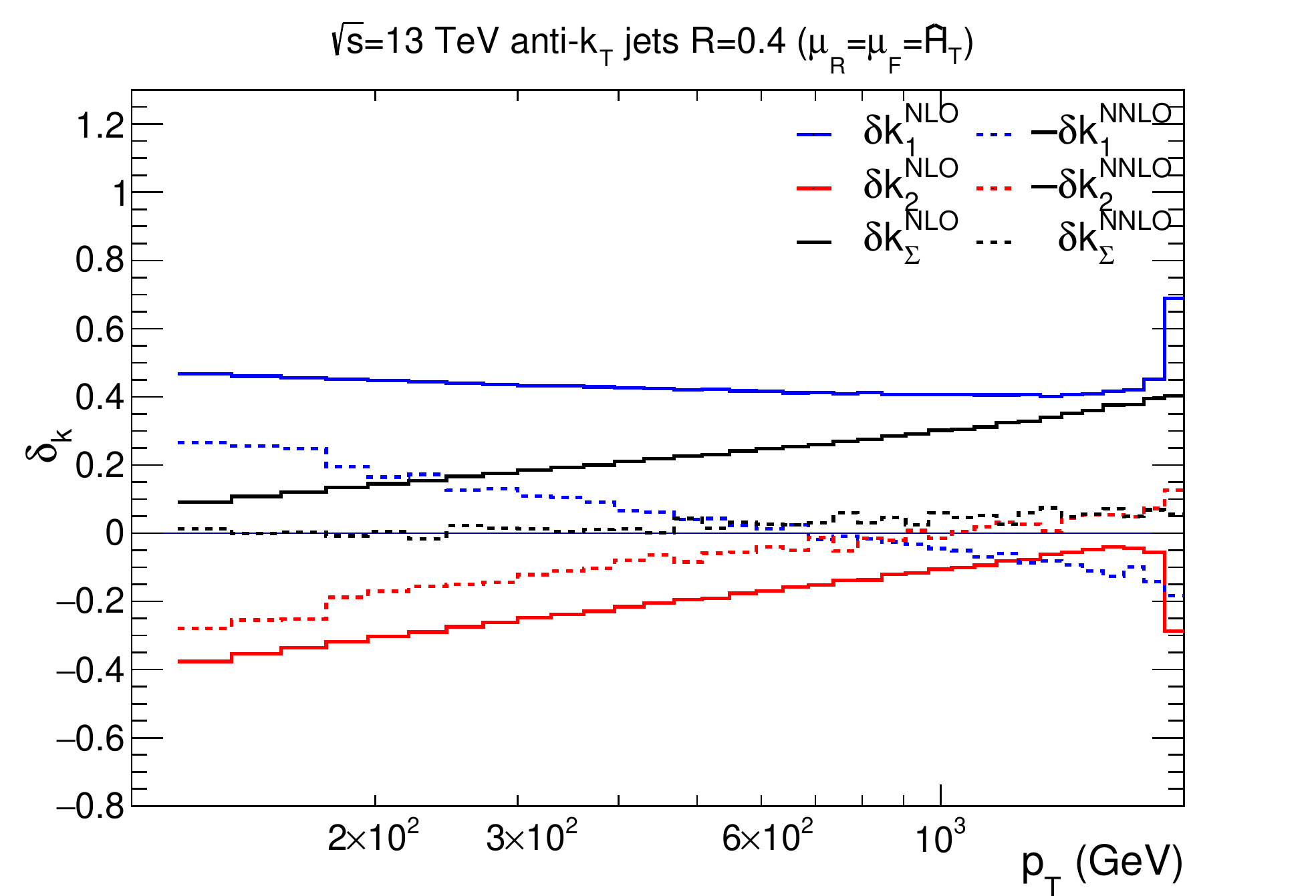}
  \caption{}
\end{subfigure}
\caption{Differential correction factors at $\sqrt{s}=13~\TeV$ for the leading jet ($\delta k_1^\text{N${}^k$LO}$, blue), subleading jet ($\delta k_2^\text{N${}^k$LO}$, red) and the inclusive jet distribution ($\delta k_\Sigma^\text{N${}^k$LO}$, black) with $R=0.4$ and integrated over rapidity for the scale choices (a) $\mu=\pt$, (b) $\mu=2\,\pt$, (c) $\mu=\ptj1$, (d) $\mu=2\,\ptj1$, (e) $\mu=\htp/2 $, (f) $\mu=\htp$.  
\label{fig:pt1pt2AR04}
}
\end{figure}

Figures~\ref{fig:pt1pt2AR07} and \ref{fig:pt1pt2AR04} show the correction factors $\delta k_1$, $\delta k_2$, and $\delta k_\Sigma$ for the set of scale choices of Sect.~\ref{subsec:scalechoice} at NLO (solid lines) 
and NNLO (dashed lines) for the cone sizes $R=0.7$ and $R=0.4$, respectively.
As anticipated, we find large cancellations between the leading (blue) and the subleading jet contributions (red) at each order in the perturbative expansion for any scale choice.

In particular, we can observe a very large negative (positive) NLO coefficient for the second (first) jet contributions in solid red (blue) respectively. This effect explains why the second jet contribution
to the inclusive jet $\pt$ sample at NLO is significantly reduced and the first jet fractions dominates (as shown in Figs.~\ref{fig:jetfracMUpt1},~\ref{fig:jetfracMUpt}). At the next order, the sign of the
NNLO coefficient is reversed for the leading and subleading jet (dashed blue and red respectively), resulting in the leading and subleading fractions to become similar over the
whole $\pt$ range at NNLO. 

Given that the aforementioned feature is common for all scale choices, we can now apply the criteria (a,c) to assess which scale choices show the most stable behaviour in the
perturbative expansion. Criterion~(c) is concerned with the \emph{spread} of the blue and red curves, associated with $\delta k_1$ and $\delta k_2$, respectively.
Going from NLO to NNLO, we require the size of the corrections to the individual jet \pt to become smaller and therefore that the dashed curves exhibit a smaller spread than the corresponding solid ones.
We observe that for the scale choices  $\mu=\pt$, $\mu=\ptj1$, and $\mu=\htp/2$ this condition is not fulfilled, specially at low $\pt$ and in particular for the smaller jet cone size $R=0.4$.

\begin{figure}[t!]
\begin{subfigure}{\textwidth}
\centering
\includegraphics[width=\linewidth]{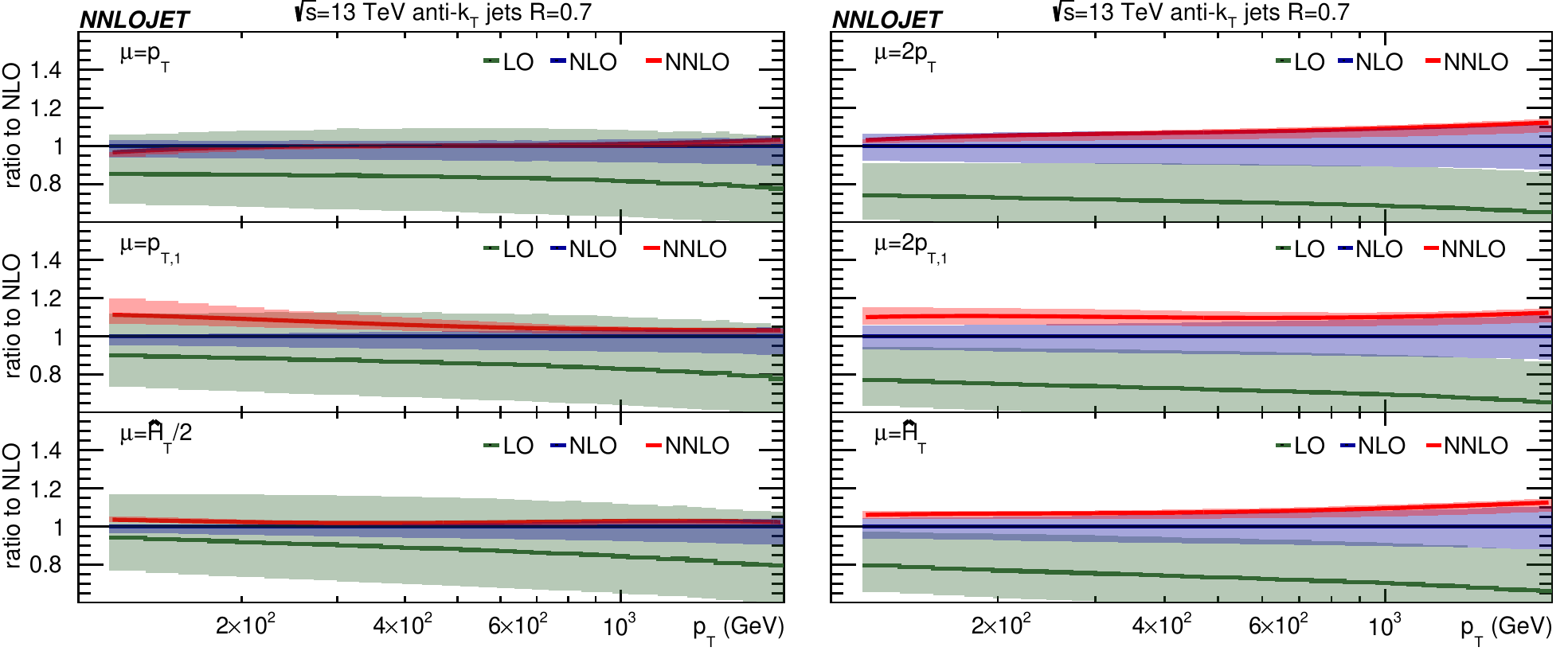}  
\caption{}
\end{subfigure}
\begin{subfigure}{\textwidth}
\vspace{0.7cm}
\centering
\includegraphics[width=\linewidth]{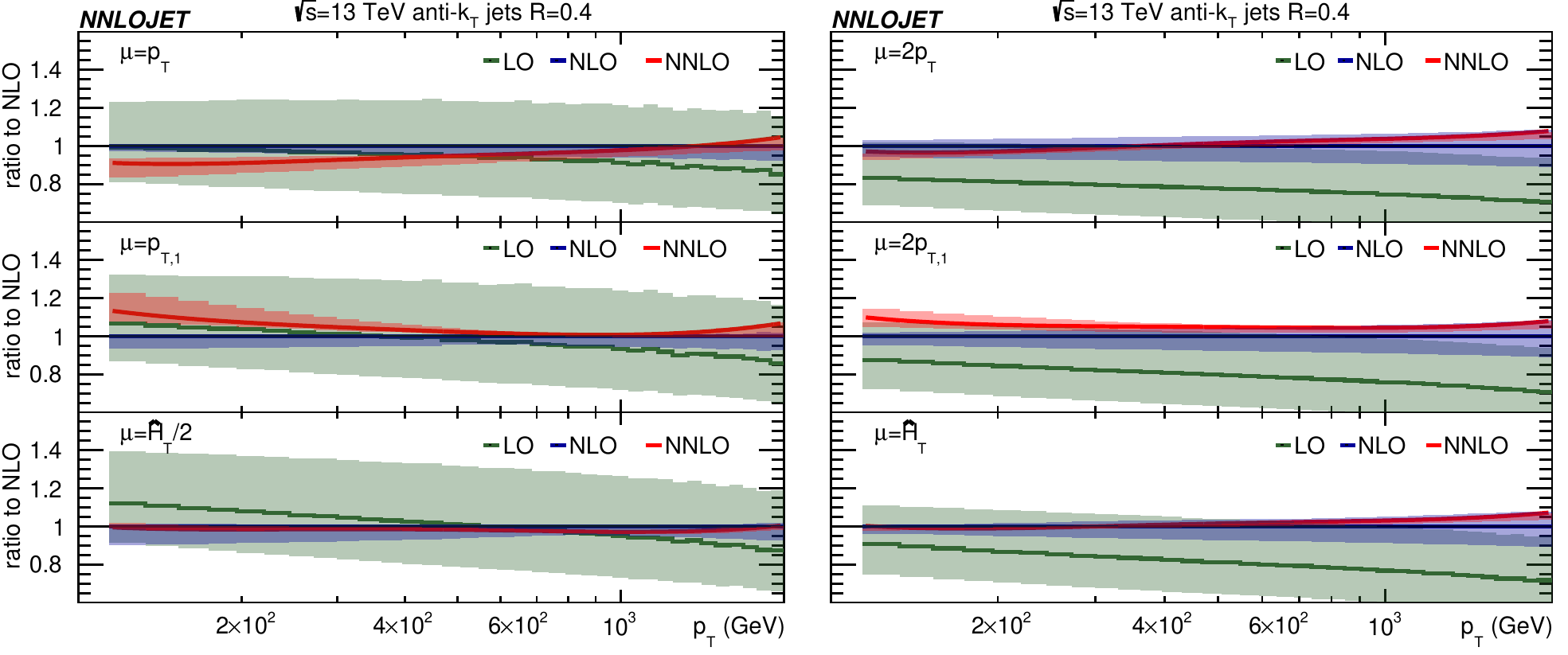}  
\caption{}
\end{subfigure}
\caption{Inclusive jet $\pt$ spectrum integrated over rapidity at LO (green), NLO (blue) and NNLO (red) normalised to the NLO prediction as a function of the central scale choice for (a) cone size $R=0.7$ and (b) cone size $R=0.4$.}
\label{fig:criterion-b}
\end{figure}

The net effect on the inclusive $\pt$ spectrum is given by the correction factors $\delta k_\Sigma$, shown as the black lines.
With criterion~(a), we require the $K$-factor at NNLO to be smaller than at NLO, i.e.\ the dashed black lines to be closer to zero than the solid ones.
Here, we observe that the scales $\pt$ and $\ptj1$ give rise to sizeable NNLO corrections that are larger in magnitude than the corresponding corrections at NLO for $R=0.4$,
while for the bigger cone size of $R=0.7$ the same effect is observed to be true again for the $\ptj1$ scale choice.
The remaining scales $2\,\ptj1$, $2\,\pt$, $\htp$, and $\htp/2$ fulfil criterion~(a), with NNLO corrections at the level of $5$--$10\%$.

In Fig.~\ref{fig:criterion-b} we examine criterion~(b) on the theory error estimate by plotting the predictions at a given order with their respective scale uncertainty bands normalised to the NLO prediction.
Given the potentially large impact of the cone size, we present results for both $R=0.7$ (top) and $R=0.4$ (bottom).
For both cone sizes we observe that the scale choices $\ptj1$ and $2\,\ptj1$ give rise to scale uncertainty bands at NLO and NNLO that do not overlap in the low-$\pt$ region.
For the scale $\pt$, the conclusion depends strongly on the cone size where we observe overlapping bands for $R=0.7$ but not for $R=0.4$.
The remaining three scales $2\,\pt$, $\htp$, and $\htp/2$, on the other hand, exhibit good convergence with overlapping scale uncertainty bands independently on the cone size. 

\begin{figure}[t!]
\centering
\begin{subfigure}{.5\textwidth}
  \centering
  \includegraphics[width=.96\linewidth]{figures/PT2KFmuPT-R07.pdf}
  \caption{}
\end{subfigure}%
\begin{subfigure}{.5\textwidth}
  \centering
  \includegraphics[width=.96\linewidth]{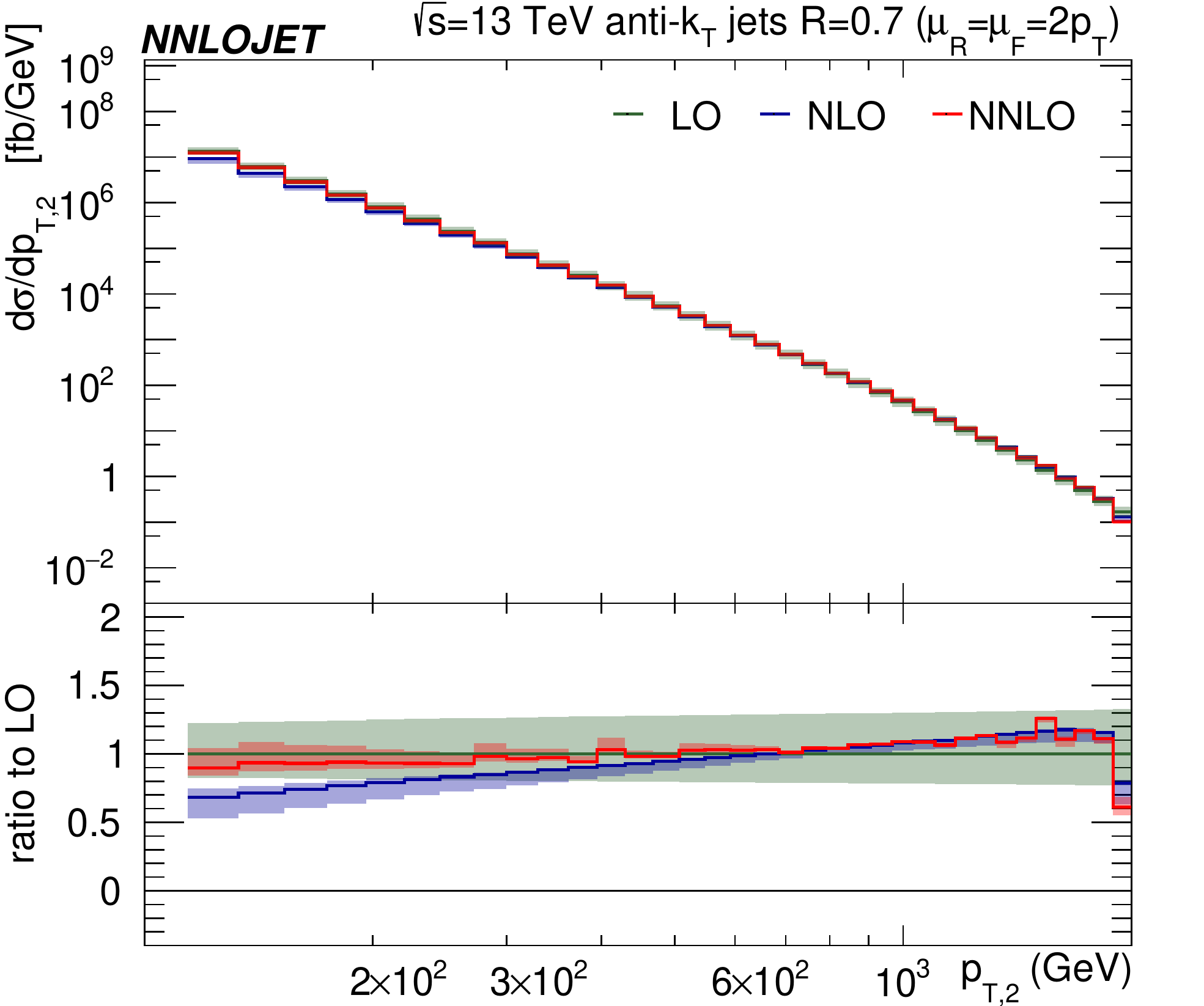}
  \caption{}
\end{subfigure}
\begin{subfigure}{.5\textwidth}
  \centering
  \includegraphics[width=.96\linewidth]{figures/PT2KFmuPT1-R07.pdf}
  \caption{}
\end{subfigure}%
\begin{subfigure}{.5\textwidth}
  \centering
  \includegraphics[width=.96\linewidth]{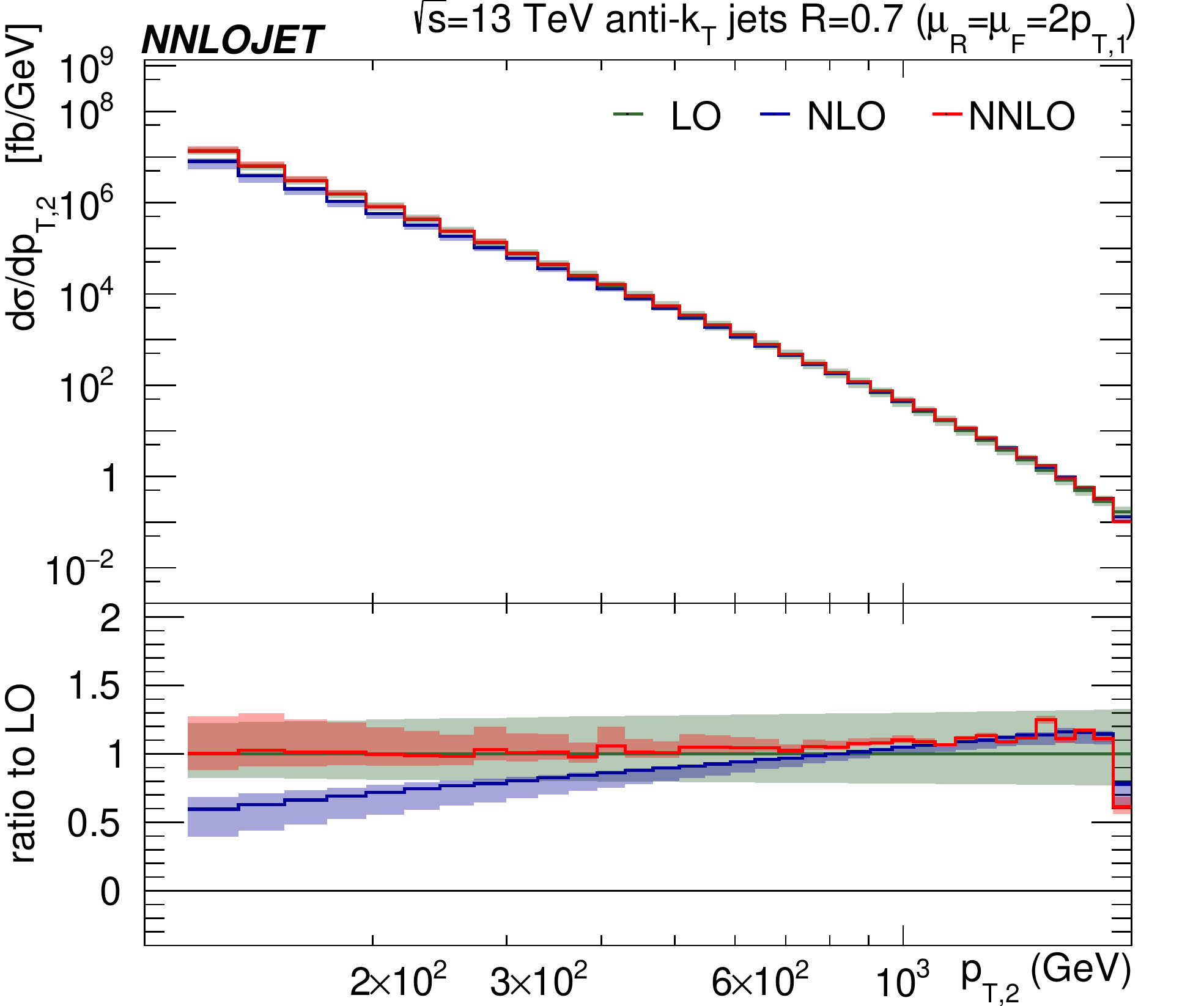}
  \caption{}
\end{subfigure}
\begin{subfigure}{.5\textwidth}
  \centering
  \includegraphics[width=.96\linewidth]{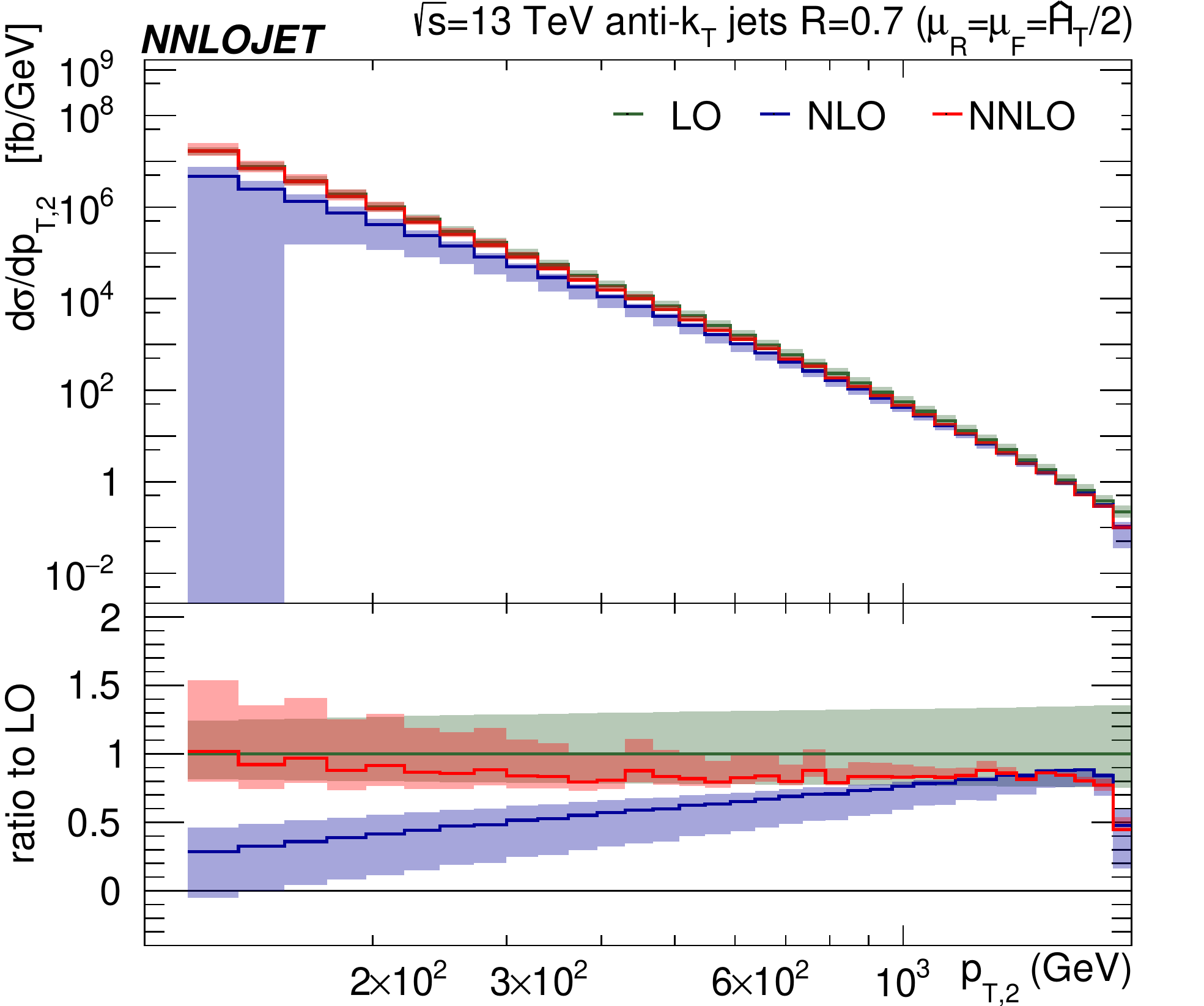}
  \caption{}
\end{subfigure}%
\begin{subfigure}{.5\textwidth}
  \centering
  \includegraphics[width=.96\linewidth]{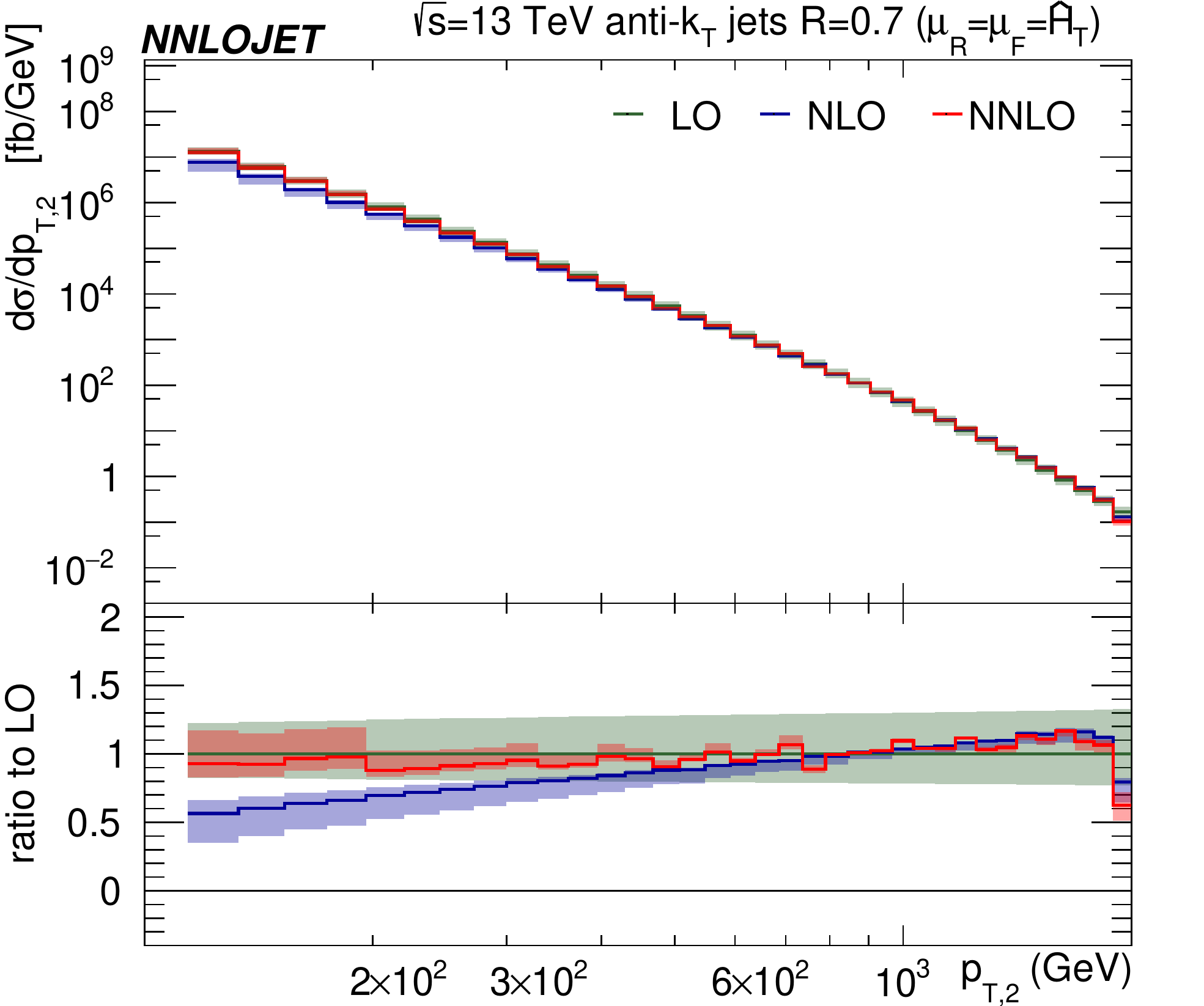}
  \caption{}
\end{subfigure}
\caption{The transverse-momentum distribution of the subleading jet with $R=0.7$ for (a) $\mu=\pt$, (b) $\mu=2\,\pt$, (c) $\mu=\ptj1$, (d) $\mu=2\,\ptj1$, (e) $\mu=\htp/2 $, (f) $\mu=\htp$.\label{fig:criterion-d-R07}}
\end{figure}

\begin{figure}[t!]
\centering
\begin{subfigure}{.5\textwidth}
  \centering
  \includegraphics[width=.96\linewidth]{figures/PT2KFmuPT-R04.pdf}
  \caption{}
\end{subfigure}%
\begin{subfigure}{.5\textwidth}
  \centering
  \includegraphics[width=.96\linewidth]{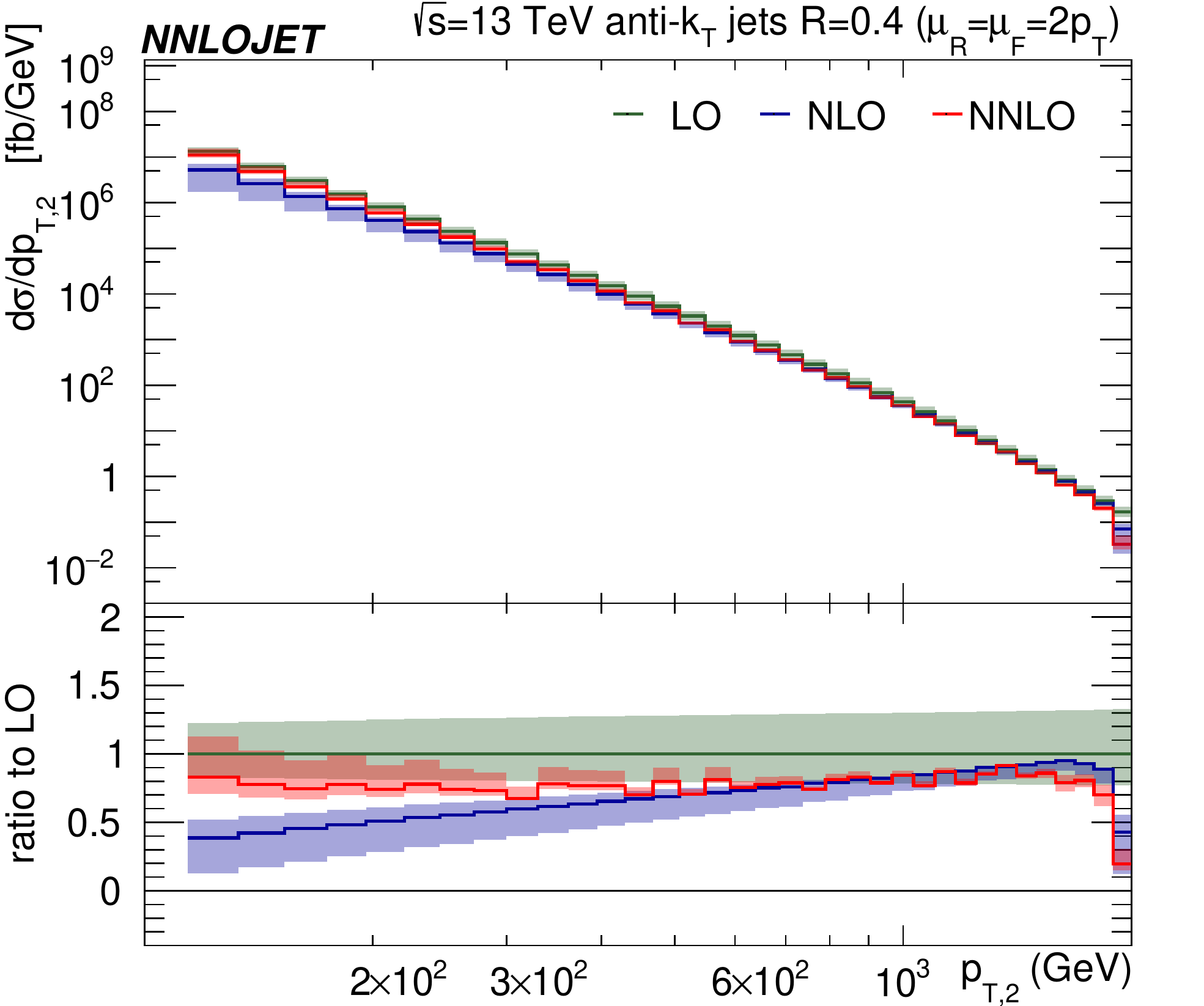}
  \caption{}
\end{subfigure}
\begin{subfigure}{.5\textwidth}
  \centering
  \includegraphics[width=.96\linewidth]{figures/PT2KFmuPT1-R04.pdf}
  \caption{}
\end{subfigure}%
\begin{subfigure}{.5\textwidth}
  \centering
  \includegraphics[width=.96\linewidth]{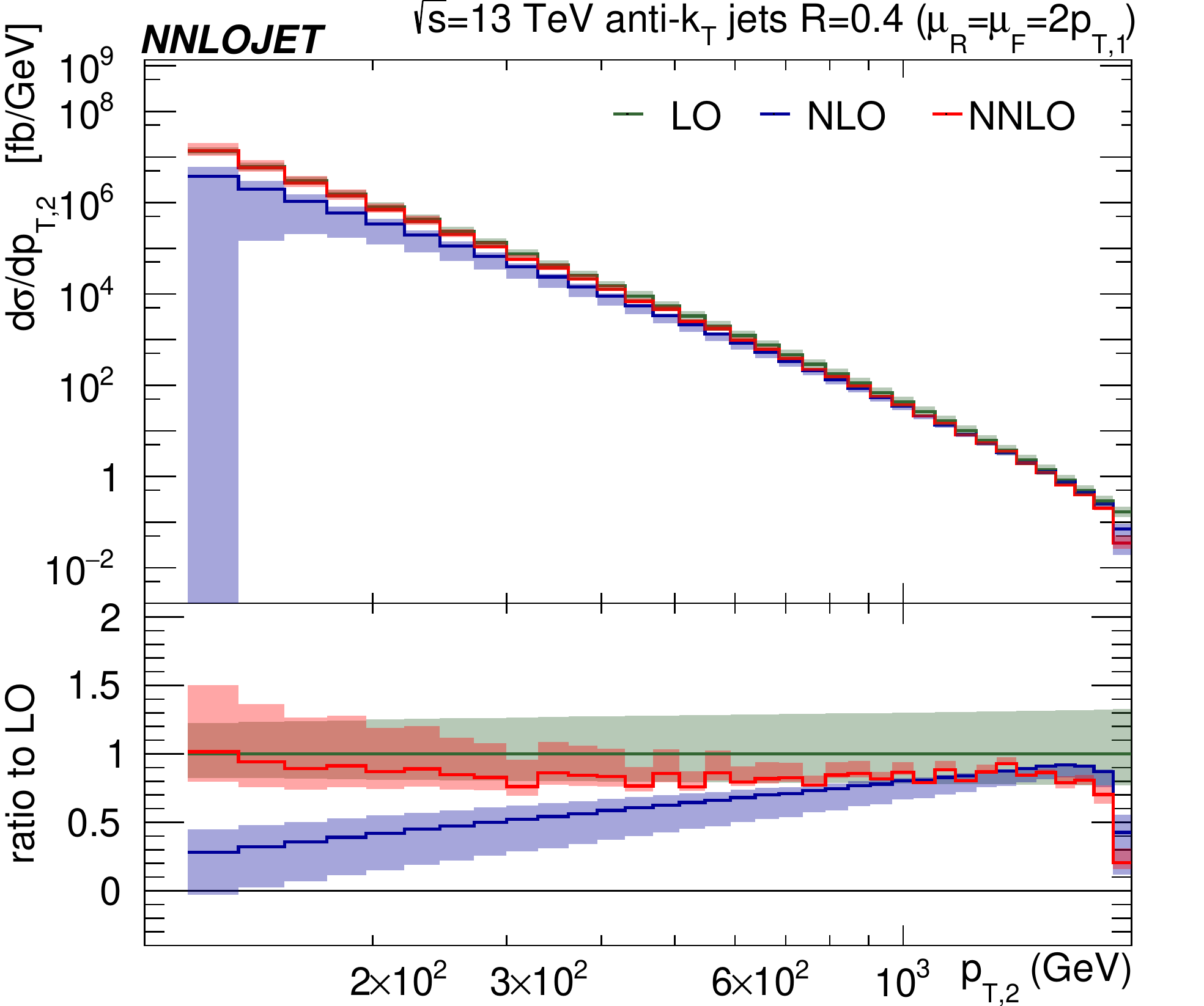}
  \caption{}
\end{subfigure}
\begin{subfigure}{.5\textwidth}
  \centering
  \includegraphics[width=.96\linewidth]{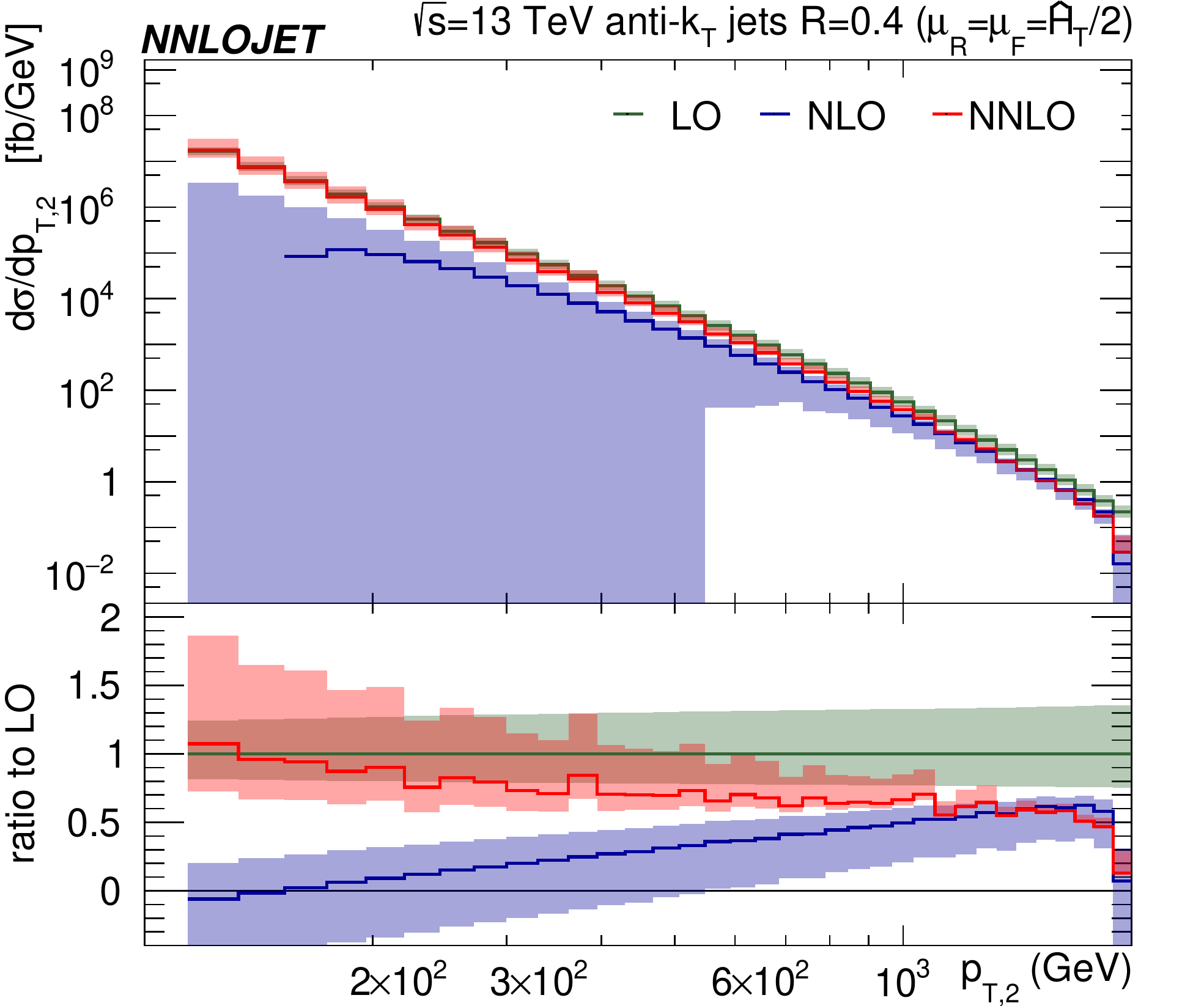}
  \caption{}
\end{subfigure}%
\begin{subfigure}{.5\textwidth}
  \centering
  \includegraphics[width=.96\linewidth]{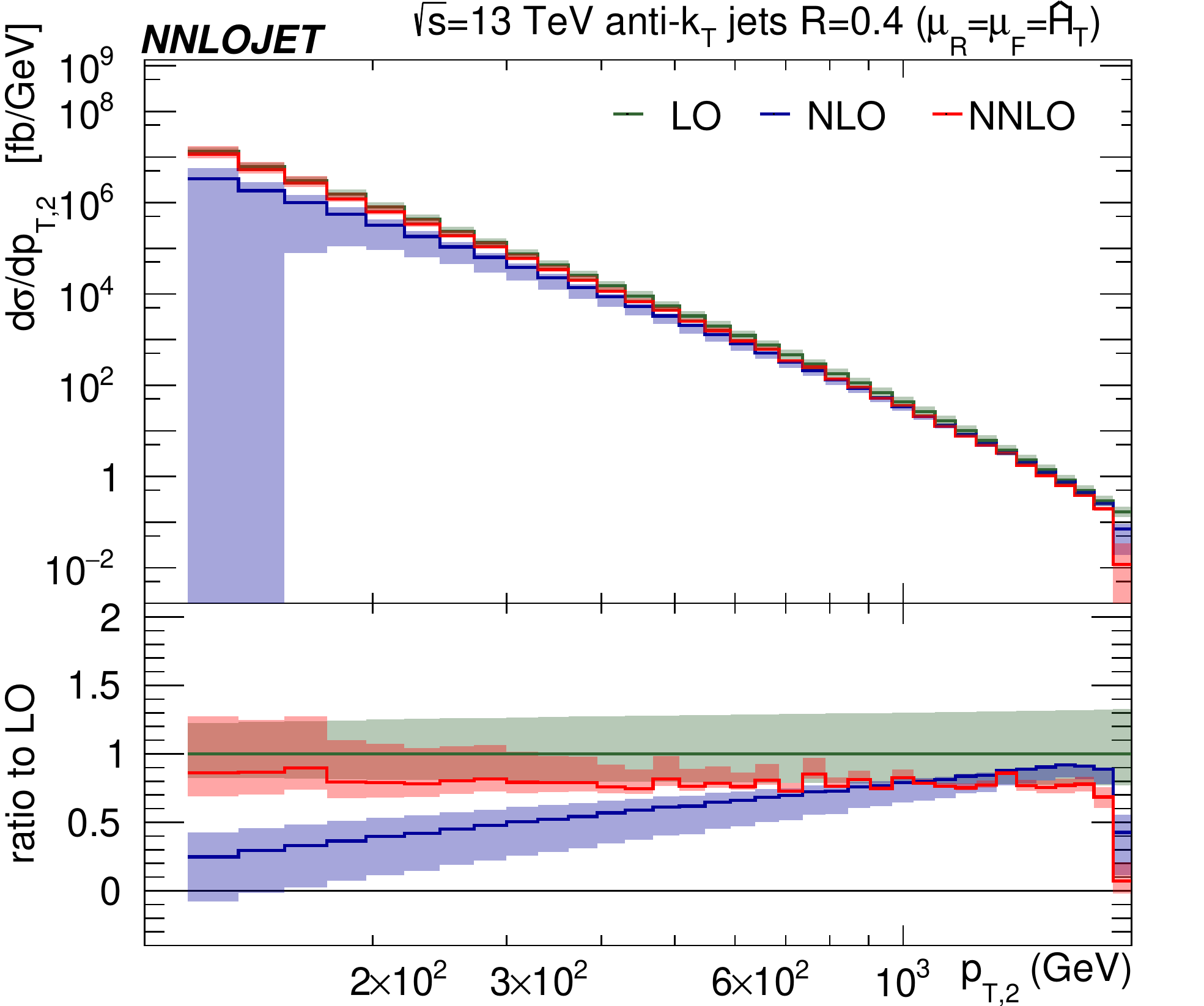}
  \caption{}
\end{subfigure}
\caption{The transverse-momentum distribution of the subleading jet with $R=0.4$ for (a) $\mu=\pt$, (b) $\mu=2\,\pt$, (c) $\mu=\ptj1$, (d) $\mu=2\,\ptj1$, (e) $\mu=\htp/2 $, (f) $\mu=\htp$.\label{fig:criterion-d-R04}}
\end{figure}

Finally, we study criterion~(d) by investigating the perturbative behaviour of the second jet distribution and its associated scale uncertainties.
In Figs.~\ref{fig:criterion-d-R07} and \ref{fig:criterion-d-R04} we show the corrections to the \pt distribution of the second jet for the cone sizes $R=0.7$ and $0.4$, respectively.
As was already mentioned in the study of criterion~(c), we clearly observe an improved perturbative behaviour with smaller higher-order corrections and scale uncertainties for the three harder scale choices $2\,\pt$, $2\,\ptj1$, and $\htp$ compared to their respective counterparts that are smaller by a factor of a half.
For $R=0.4$, we find that only the scale $\mu=2\,\pt$ is able to predict positive NLO cross sections across the entire $\pt$ range, both for the central value as well as for its variation.
Although the scale choices $2\,\ptj1$ and $\htp$ give rise to NLO scale uncertainties that extend to negative cross section values, this behaviour is less critical as it only occurs in the very first bin(s) below $\pt\lesssim150~\GeV$ and the central predictions remain positive.
The situation is much more severe for the remaining three scales $\pt$, $\ptj1$, and $\htp/2$, where the NLO prediction exhibits the unphysical behaviour of negative cross sections already starting from $\pt\sim400$--$600~\GeV$. 
In the case of $\mu=\ptj1$ and $\htp/2$, even then central prediction turns negative in the lowest \pt bin(s) below $\sim150~\GeV$. 
For the larger cone size of $R=0.7$, on the other hand, the issue of negative cross sections at NLO is largely alleviated, where only the choice $\mu=\htp/2$ exhibits this unphysical behaviour. 

\begin{table}[t!]
\begin{subtable}{.9\textwidth}
  \centering
  \begin{tabular}{l c c c c}
    \toprule
    & \multicolumn{4}{c}{criterion} \\
    scale & (a) & (b) & (c) & (d) \\
    \midrule
    $\ptj1$    & --         & --         & \checkmark & \checkmark \\
    $2\,\ptj1$ & \checkmark & --         & \checkmark & \checkmark \\
    $\pt$      & --         & \checkmark & \checkmark & \checkmark \\
    $2\,\pt$   & \checkmark & \checkmark & \checkmark & \checkmark \\
    $\htp/2$   & \checkmark & \checkmark & \checkmark & --         \\
    $\htp$     & \checkmark & \checkmark & \checkmark & \checkmark \\
    \bottomrule
  \end{tabular}
  \caption{$R=0.7$}
  \label{tab:scale-criteria-r07}
\end{subtable}
\begin{subtable}{.9\textwidth}
\vspace{0.5cm}
  \centering
  \begin{tabular}{l c c c c}
    \toprule
    & \multicolumn{3}{c}{criterion} \\
    scale & (a) & (b) & (c) & (d) \\
    \midrule
    $\ptj1$    & --         & --         & --         & --           \\
    $2\,\ptj1$ & \checkmark & --         & \checkmark & (\checkmark) \\
    $\pt$      & --         & --         & --         & --           \\
    $2\,\pt$   & \checkmark & \checkmark & \checkmark & \checkmark   \\
    $\htp/2$   & \checkmark & \checkmark & --         & --           \\
    $\htp$     & \checkmark & \checkmark & \checkmark & (\checkmark) \\
    \bottomrule
  \end{tabular}
  \caption{$R=0.4$}
  \label{tab:scale-criteria-r04}
\end{subtable}  
  \caption{Summary of scales vs.\ criteria for (a) $R$=0.7 and (b) $R$=0.4 cone sizes.}
\end{table}

We summarise the findings of this section in Tables~\ref{tab:scale-criteria-r07},~\ref{tab:scale-criteria-r04} for cone sizes of $R=0.7$ and $R=0.4$ respectively. 
By comparing the two tables we see that, as expected, the various scale choices behave in a much more similar way for the larger cone size than for $R=0.4$.
It is interesting to note that the two most commonly used scales $\mu=\pt$ and $\ptj1$ perform by far the worst among the set of scale choices considered here.
In particular, they are not able to fulfil any of the criteria for the smaller cone size of $R=0.4$.
On the other hand, the scale $\mu=2\,\pt$ fulfils all the requirements we identified at the beginning of this section, while the scale $\mu=\htp$ satisfies all of the criteria for $\pt > 150~\GeV$.
We therefore identify $\mu=2\,\pt$ and $\mu=\htp$ as the two theoretically best-motivated scale choices for single jet inclusive production, noting that the former belongs to the class of jet-based scales and the latter is an event-based scale.

\subsection{Results for central and forward rapidity slices}
\label{Sec:delpt1pt2diff}

\begin{figure}[t!]
  \begin{subfigure}{\textwidth}
  \centering
  \includegraphics[width=\linewidth]{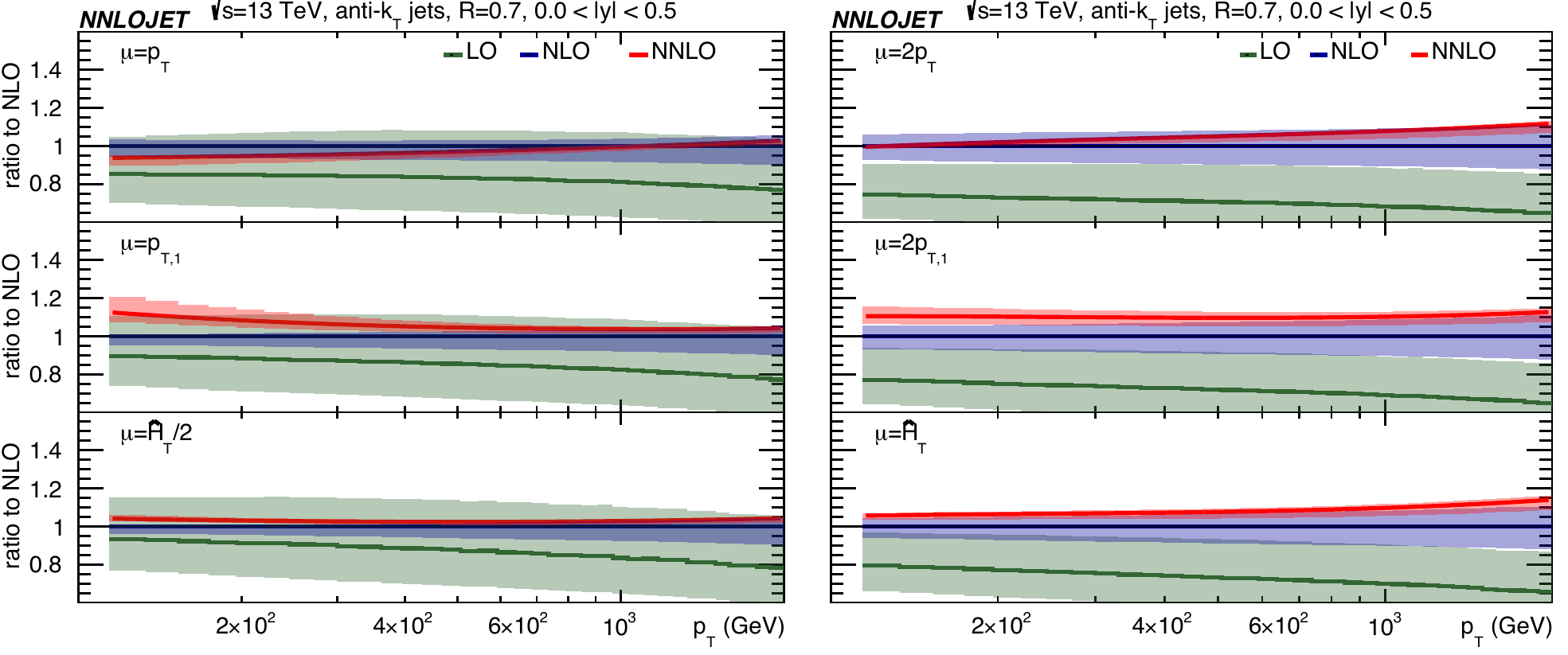}
  \caption{}
  \end{subfigure}%
  \vspace{0.7cm}
  \begin{subfigure}{\textwidth}
  \centering
  \includegraphics[width=\linewidth]{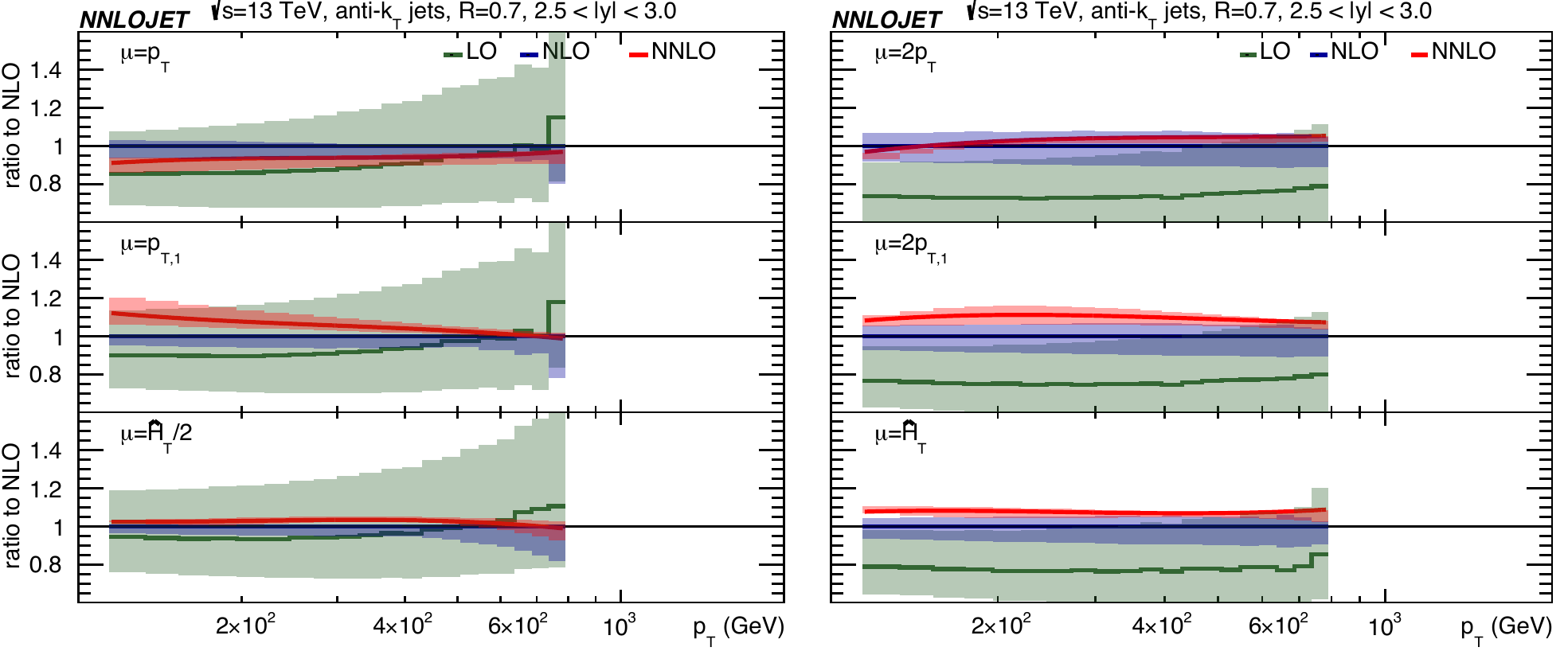}
  \caption{}
  \end{subfigure}
\caption{LO/NLO (green), NLO/NLO (blue) and NNLO/NLO (red) $K$-factors at $\sqrt{s}=$13 TeV for (a) central rapidity, $0.0 < |y| < 0.5$, and (b) forward rapidity, $2.5 < |y| < 3.0$, as
a function of the central scale choice for $R=0.7$ and CMS cuts.}
\label{fig:KFr07a}
\end{figure}

\begin{figure}[t!]
\centering
\begin{subfigure}{.5\textwidth}
  \centering
  \includegraphics[width=\linewidth]{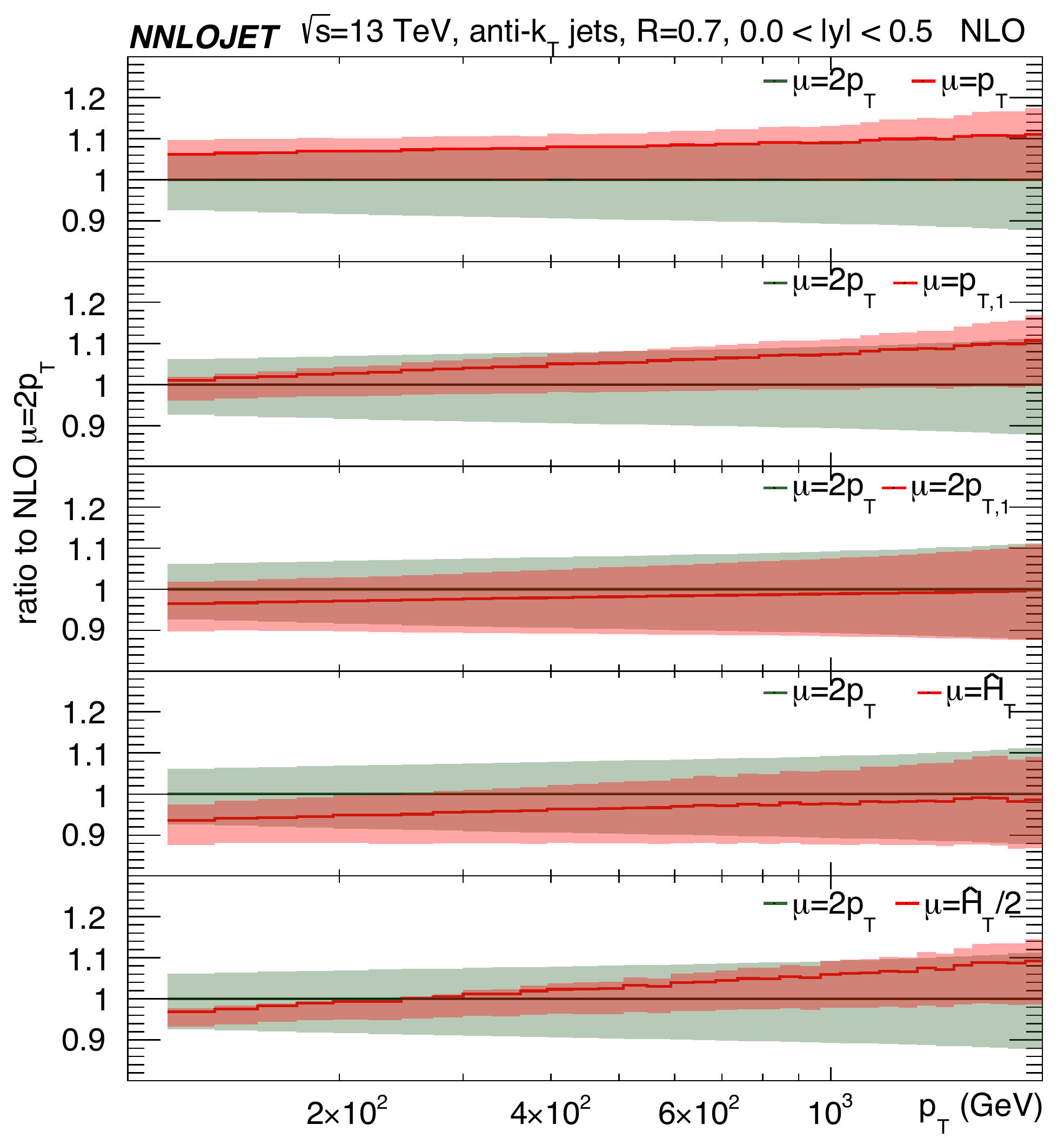}
  \caption{}
\end{subfigure}%
\begin{subfigure}{.5\textwidth}
  \centering
  \includegraphics[width=\linewidth]{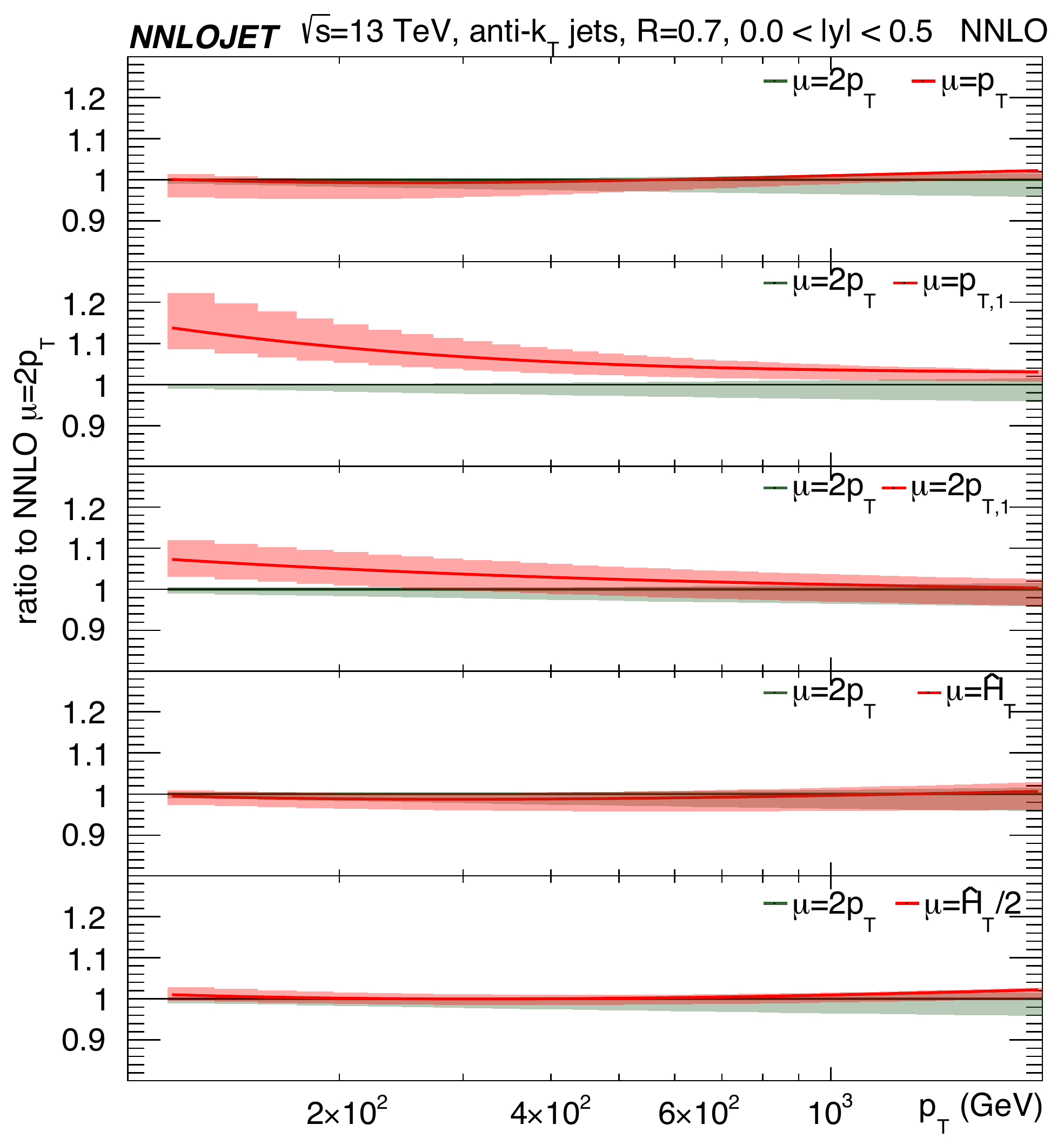}
  \caption{}
\end{subfigure}
\caption{Ratio of 13 TeV single jet inclusive cross sections to the $\mu=2\,\pt$ scale choice at (a) NLO and (b) NNLO with $R=0.7$ and CMS cuts.}
\label{fig:ratiotoPT107}
\end{figure}

Having discussed at length the behaviour of the leading and subleading jet contributions as a function of the scale choice integrated over rapidity,  for the remainder of this section, we will focus on the single jet inclusive observable for the different rapidity bin intervals used by the CMS collaboration~\cite{Khachatryan:2016wdh}.

Figure~\ref{fig:KFr07a} shows the perturbative corrections for the single jet inclusive cross section at NLO and at NNLO for the same six scale choices discussed earlier: $\mu=\ptj1$, $\mu=\pt$, $\mu=2\,\ptj1$, $\mu=2\,\pt$, $\mu=\htp$, and $\mu=\htp /2$ for a jet cone size of $R=0.7$ and for jets produced at (a) central rapidity ($|y| < 0.5$) and (b) forward rapidity ($2.5 < |y| < 3.0$).
The shaded bands represent the scale variation around the respective central scale choice.

Focussing first on the central rapidity region shown in Fig.~\ref{fig:KFr07a}(a), we see that the shape and size of the LO/NLO $K$-factor (green) for the $\mu=\ptj1$, $\mu=\pt$ and $\mu=\htp /2$ scales are fairly similar. However, we observe larger NLO radiative corrections when these central choices are rescaled by a factor of 2.  

Inspection of the NNLO/NLO $K$-factor (red) reveals that the size and shape of the NNLO corrections are generally smaller than the NLO ones, but that there is some dependence on the functional form of the scale choice. While the NNLO/NLO $K$-factor is never more than $\pm 20\%$ for any of the scale choices, the dependence on $\pt$ is quite varied.  For $\mu=\pt$ ($\mu=2\,\pt$), the corrections grow from $-10\%$ ($0\%$) at low $\pt$ to a few percent ($10\%$) at large $\pt$, while for $\mu=\ptj1$ ($\mu=2\,\ptj1$), the corrections fall from $+15\%$ ($12\%$) at low $\pt$ to a few percent ($10\%$) at large $\pt$. For $\mu=\htp $, the corrections are always positive, growing from a few percent at low $\pt$ to $12\%$ at large $\pt$. In the case of $\mu=\htp /2$, the NNLO/NLO $K$-factor is always small.
The same qualitative behaviour can be observed in the predictions for jet production at forward rapidity ($2.5 < |y| < 3.0$), shown in Fig.~\ref{fig:KFr07a}(b).

Because of the significantly different behaviour of the perturbative expansion for each scale choice, it is instructive to compare the respective absolute cross sections in the central rapidity region with a fixed normalisation. Fig.~\ref{fig:ratiotoPT107}(a) shows the NLO results for all six scales normalised to one common NLO prediction, namely that for $\mu=2\,\pt$. For $R$=0.7 we see at NLO, that the scale uncertainty bands of the various NLO predictions (red band) are largely overlapping with the scale uncertainty for $\mu=2\,\pt$ (green band) indicating little scale choice ambiguity in the NLO predictions.  In other words, the change in functional form of the scale choice is largely captured by the scale uncertainty of the NLO result. 

Performing the same comparison for the different scale choices at NNLO in Fig.~\ref{fig:ratiotoPT107}(b) and normalising to the NNLO prediction with $\mu=2\,\pt$, we observe the anticipated dramatic reduction in the scale variation with respect to NLO (as indicated by the reduction in the thickness of the red and green bands compared to Fig.~\ref{fig:ratiotoPT107}(a)).  We also conclude that the NNLO predictions are generally all in good agreement, particularly at high $\pt$, and independently of the scale choice. However, at low-$\pt$ we do observe larger differences, where the scales $\mu=\ptj1$, $\mu=2\,\ptj1$ tend to look similar and predict a larger NNLO cross section of approximately 10\% with respect to 
the scales $\mu=\pt$, $\mu=2\,\pt$, $\mu=\htp $, $\mu=\htp /2$.
The size of this effect combined with a significant reduction in the scale uncertainty of the NNLO prediction introduces an ambiguity because the scale variation of the NNLO cross section no longer captures the predictions of different functional forms for the central scale choice.  This has an important interplay with PDF extractions~\cite{Harland-Lang:2017ytb} using jet data and NNLO predictions, and also a significant impact when comparing the NNLO predictions with jet data.

\begin{figure}[t!]
  \begin{subfigure}{\textwidth}
  \centering
  \includegraphics[width=\linewidth]{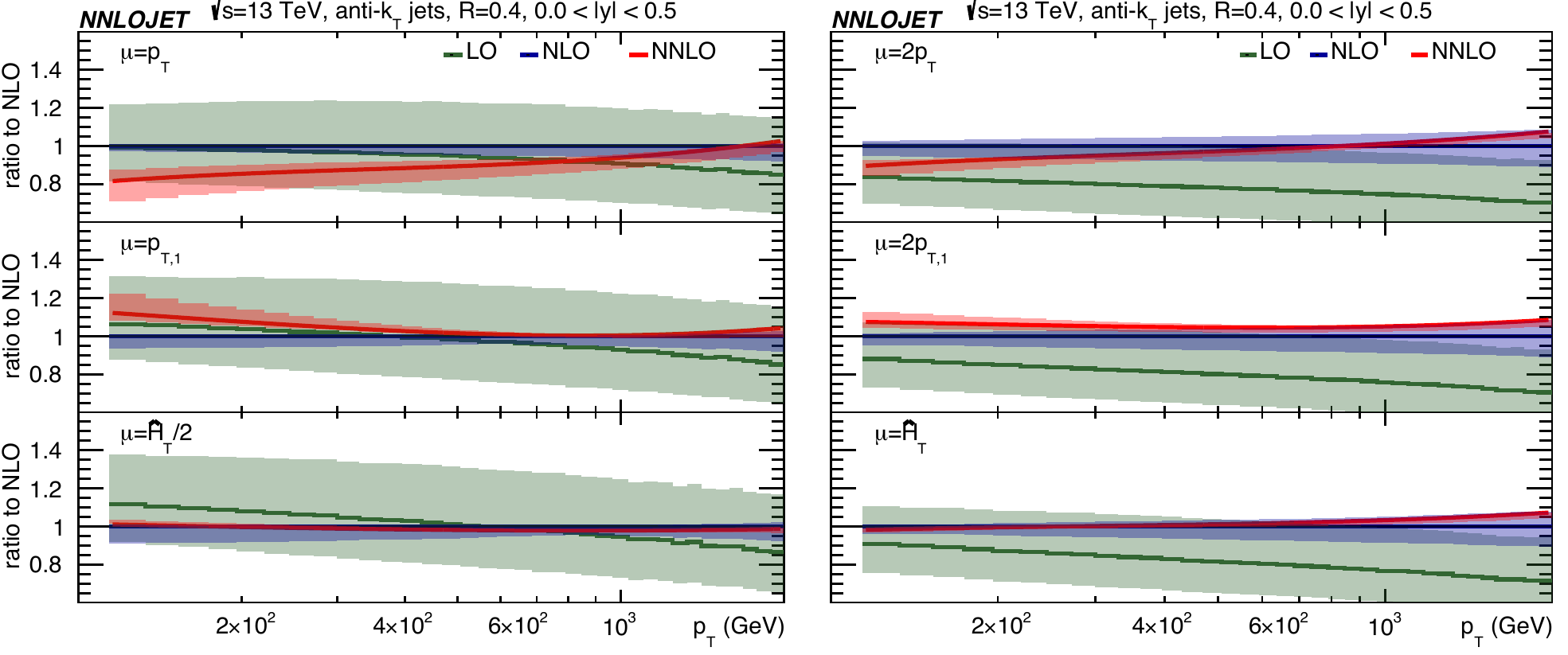}
  \caption{}
  \end{subfigure}%
  \vspace{0.7cm}
  \begin{subfigure}{\textwidth}
  \centering
  \includegraphics[width=\linewidth]{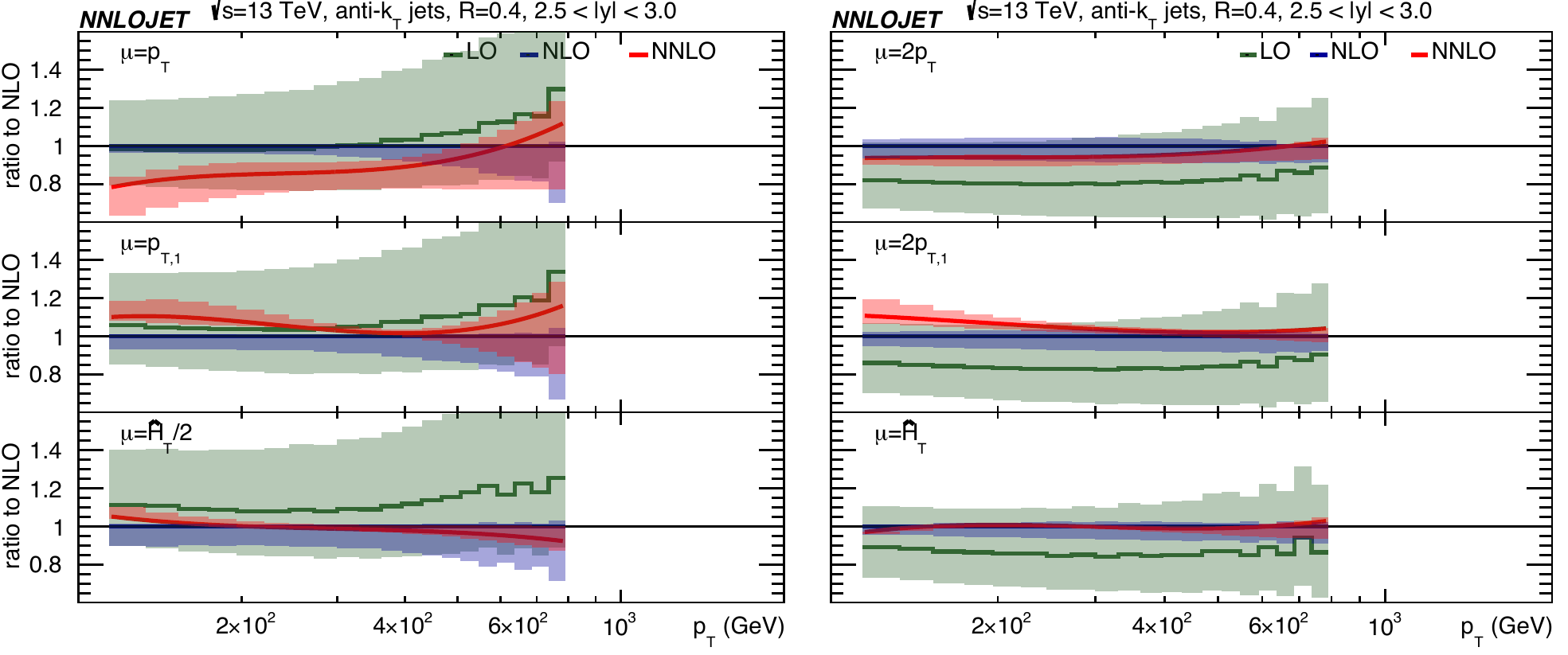}
  \caption{}
  \end{subfigure}
\caption{LO/NLO (green), NLO/NLO (blue) and NNLO/NLO (red) $K$-factors at $\sqrt{s}=$13 TeV for (a) central rapidity, $0.0 < |y| < 0.5$, and (b) forward rapidity, $2.5 < |y| < 3.0$, as
a function of the central scale choice for $R=0.4$ and CMS cuts.}
\label{fig:KFr04a}
\end{figure}

\begin{figure}[t!]
\begin{subfigure}{.5\textwidth}
  \centering
  \includegraphics[width=\linewidth]{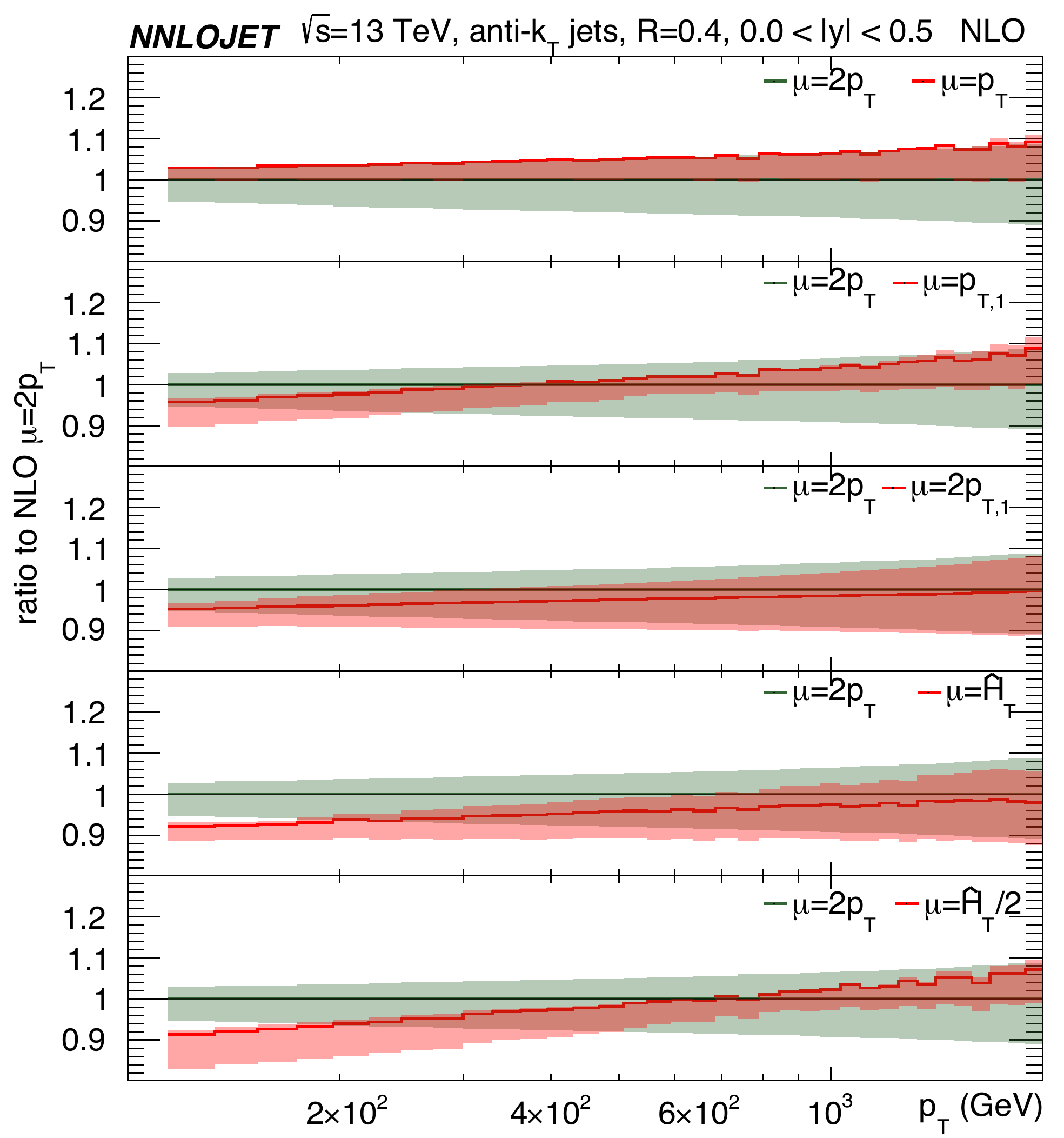}
  \caption{}
\end{subfigure}%
\begin{subfigure}{.5\textwidth}
  \centering
  \includegraphics[width=\linewidth]{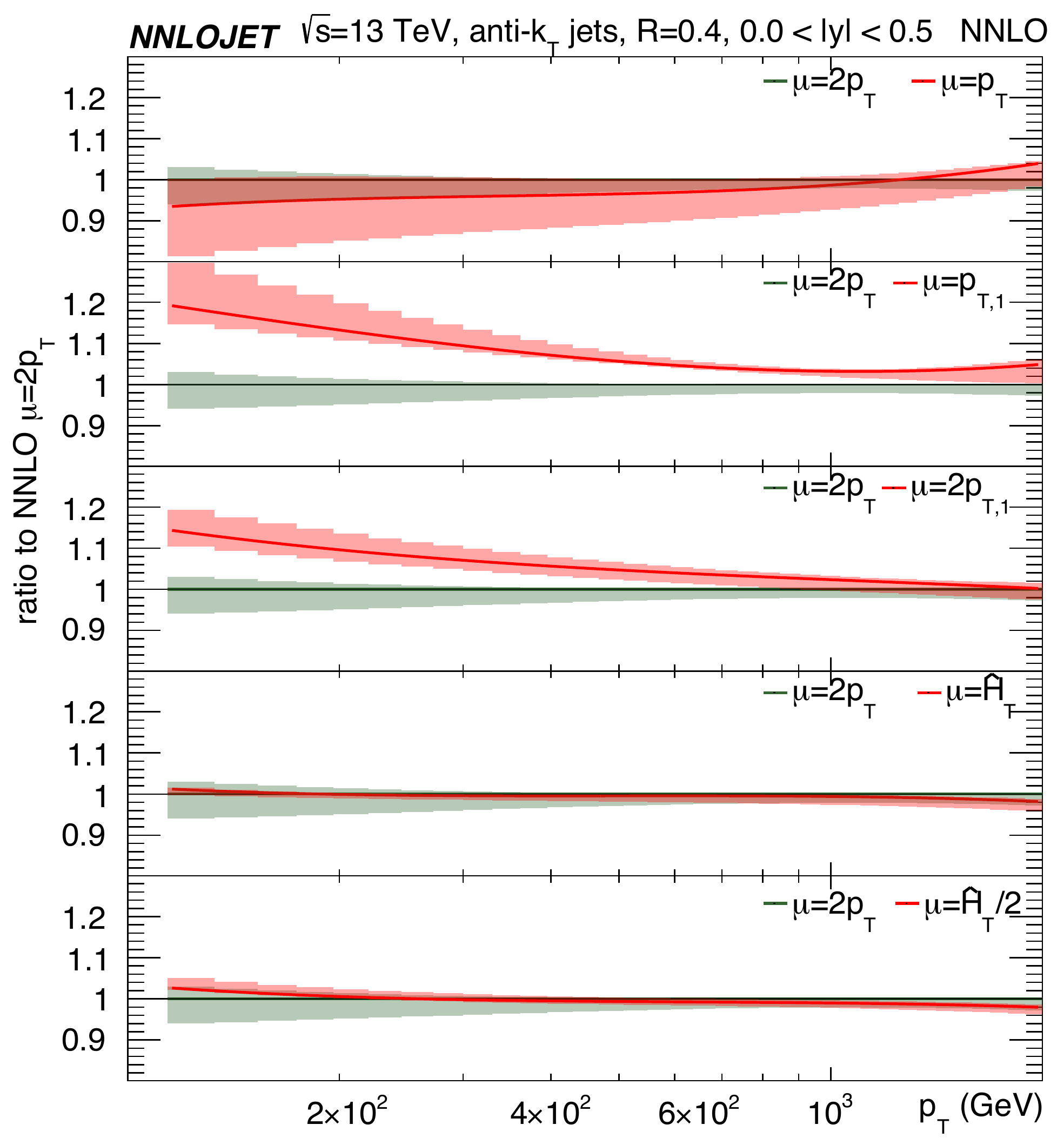}
  \caption{}
\end{subfigure}
\caption{Ratio of 13 TeV single jet inclusive cross sections to the $\mu=2\,\pt$ scale choice at (a) NLO and (b) NNLO with $R=0.4$ and CMS cuts.}
\label{fig:ratiotoPT104}
\end{figure}

We present the study of the perturbative corrections for the smaller jet cone size $R=0.4$ in Fig.~\ref{fig:KFr04a}. Compared to the results with $R$=0.7 (Fig.~\ref{fig:KFr07a}) we observe smaller NLO scale uncertainty bands and smaller NLO corrections. In particular at low-$\pt$, for $R$=0.4
the central scale choices tend to sit at the upper edge of the band, and accidentally minimise the scale uncertainty. At NNLO there is a more symmetric scale variation, however the 
NNLO/NLO $K$-factors behave rather differently. The effect of the NNLO radiative corrections is positive for the scales $\mu=\ptj1$,  $\mu=2\,\ptj1$, negligible for the scales $\mu=\htp $,  $\mu=\htp /2$, and negative for the scales $\mu=\pt$,  $\mu=2\,\pt$. As expected, the magnitude of the ambiguity in the scale choice for inclusive jet production is more severe for the smaller jet cone sizes. 

The respective absolute cross sections at NLO and NNLO for all six scale choices compared to the prediction at the same order computed with $\mu=2\,\pt$ for the jet size of $R$=0.4 are shown in Fig.~\ref{fig:ratiotoPT104}(a) and (b), respectively. As in Fig.~\ref{fig:ratiotoPT107}, the red band reflects the scale uncertainty for the various scale choices, while the green band shows the scale uncertainty for $\mu=2\,\pt$.
As with the larger cone size, $R=0.7$, Fig.~\ref{fig:ratiotoPT104}(a) shows scale variations at NLO whose uncertainty largely captures the effects of changing the functional form of the scale.
That is to say that the red and green bands largely overlap.

The same comparison is shown at NNLO in Fig.~\ref{fig:ratiotoPT104}(b) where, as for $R=0.7$,  we observe a dramatic reduction in the scale variation with respect to NLO (as indicated by the reduction in the thickness of the red and green bands compared to Fig.~\ref{fig:ratiotoPT104}(a)), except at low-$p_{T}$ for the scale choices $\mu=\pt$, $\mu=\ptj1$ and $\mu=2\,\ptj1$. For this reason, we can conclude that the NNLO predictions are generally in very good agreement at high $\pt$, independently of the scale choice. At low  $\pt$ we find larger differences, where in particular
the scales $\mu=\ptj1$, $\mu=2\,\ptj1$ tend to look similar and predict a larger NNLO cross section of approximately 15\%-20\% with respect to $\mu=\pt$, $\mu=2\,\pt$, $\mu=\htp $, $\mu=\htp /2$.

The instability of the single jet inclusive cross section at low $\pt$ has been thoroughly discussed in Sections~\ref{Sec:scalecomp} and \ref{Sec:delpt1pt2}. Due to implicit restrictions on its kinematics, it was found that the contribution from the second jet in the event is particularly sensitive to higher order effects, and that the perturbative stability of the predictions can be improved to some extent by adopting sensible scale choice criteria. 
Moreover, the largest difference in cross section and NNLO scale uncertainty is associated to using either $\mu=\pt$, $\mu=\ptj1$ or $\mu=2\,\ptj1$
as central scale choices. As documented in Tables~\ref{tab:scale-criteria-r07},~\ref{tab:scale-criteria-r04}, these scale choices introduce pathological behaviours in the 
perturbative expansion of the single jet inclusive observable. Since the spread in the NNLO predictions including these scale choices is larger in size than the NNLO scale variation, their inclusion
(and associated pathological behaviours)
is therefore overestimating the residual scale uncertainty at NNLO. It is therefore sensible to adopt well-motivated criteria for fixing the scale choice that best maximise the impact of the knowledge 
of the higher order QCD corrections to the observable, to the extent that pathological behaviours are avoided. We have observed that the best perturbative stability can be obtained for  
 $\mu=2\,\pt$ or $\mu=\htp$, where the perturbative convergence of the individual jet contributions is vastly 
improved with respect to the other functional forms of the scale choice. It is therefore not surprising that these scales tend to show smaller NNLO corrections and lead to smaller residual NNLO scale uncertainties. 

In the remainder of this paper we will employ these two functional forms of the central scale choice to compare our predictions with jet data from the CMS dataset at $\sqrt{s}=13$ TeV for the first time.

\section{Comparison with CMS jet measurements at \texorpdfstring{$\sqrt{s}=13~\TeV$}{13 TeV}}
\label{Sec:numerics13TeV}

\begin{figure}[t!]
  \begin{subfigure}{.5\textwidth}
  \includegraphics[width=\linewidth]{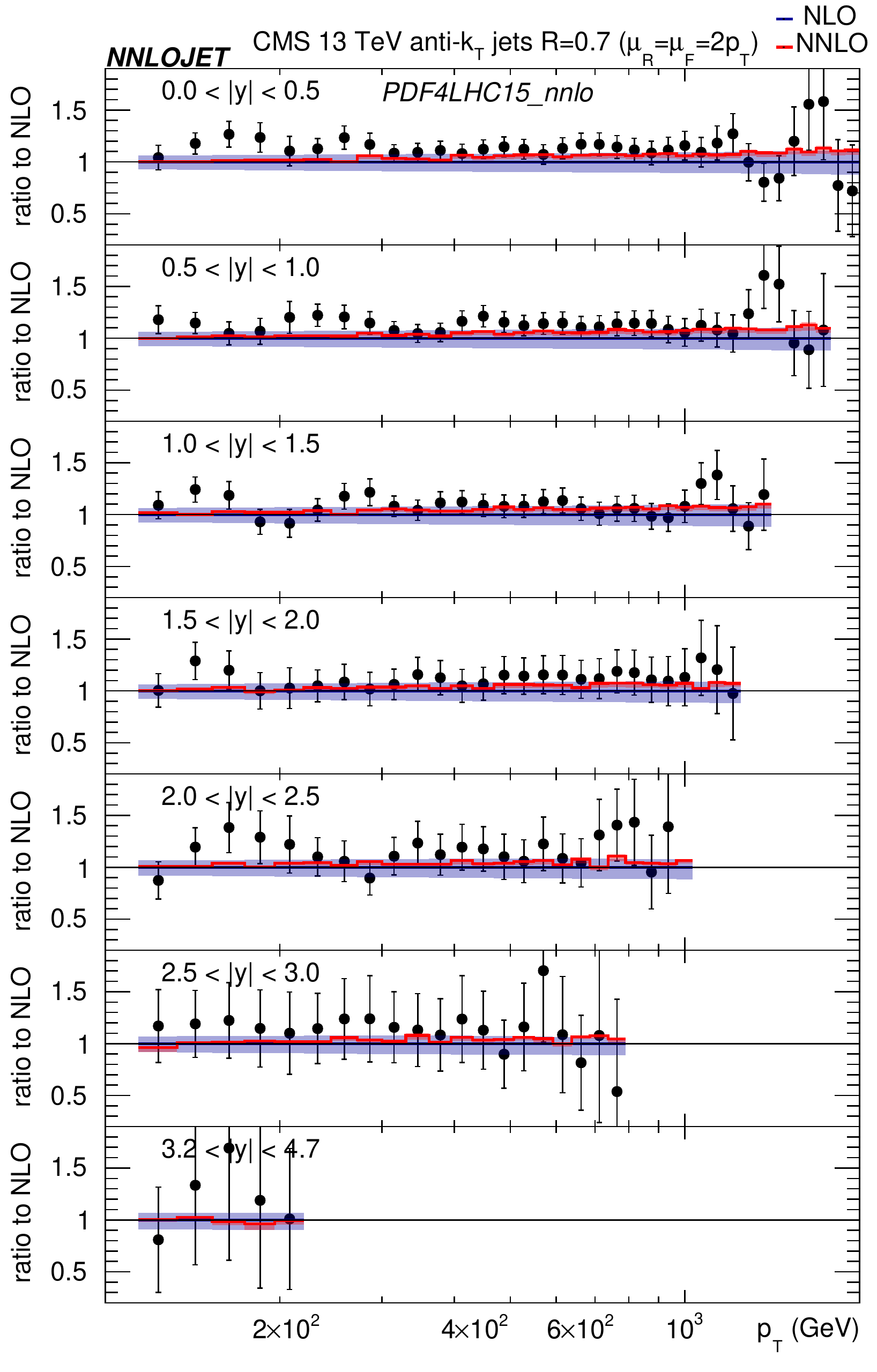}
  \caption{}
  \end{subfigure}%
  \begin{subfigure}{.5\textwidth}
  \includegraphics[width=\linewidth]{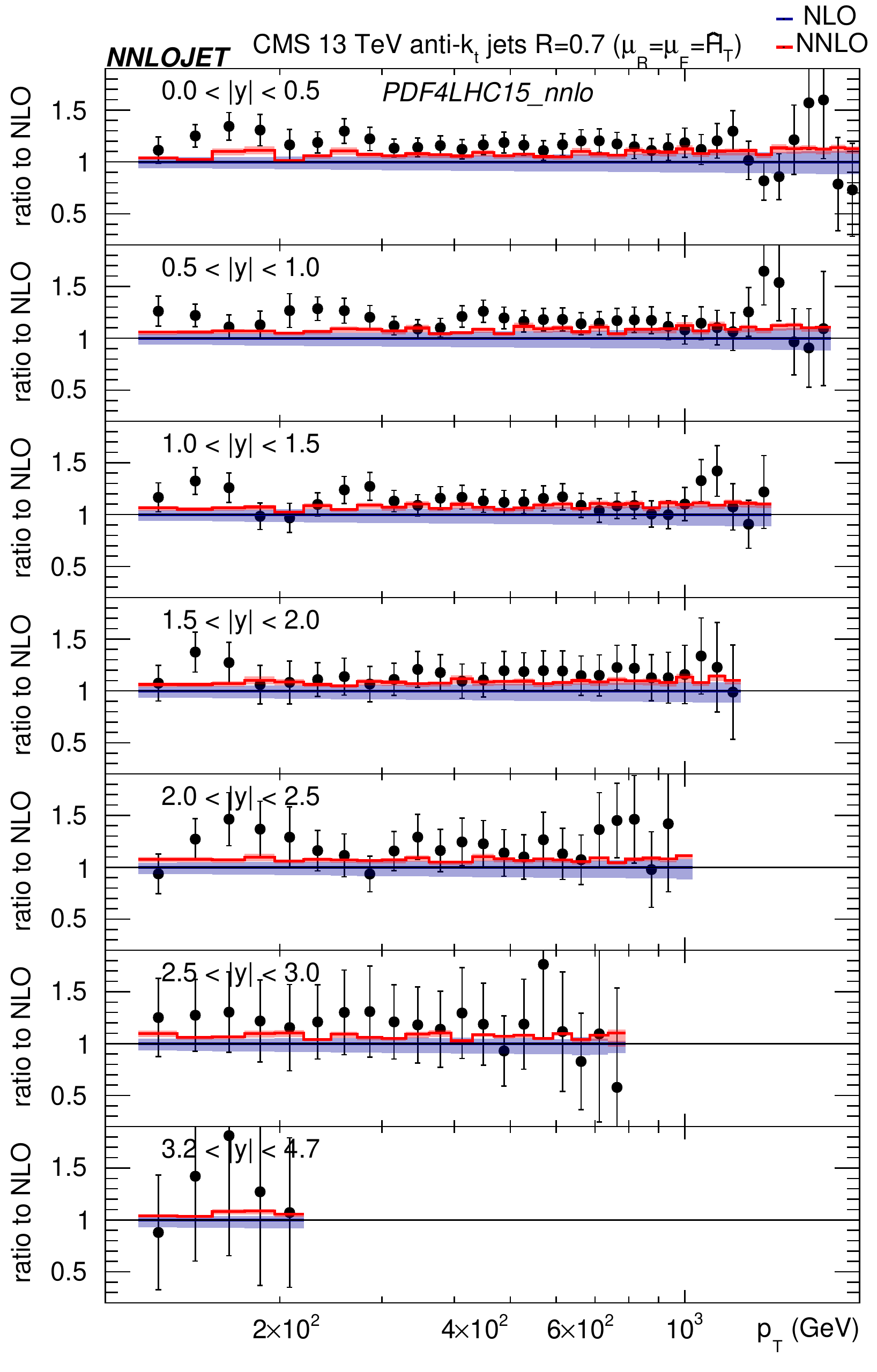}
  \caption{}
  \end{subfigure}  
\caption{Double-differential single jet inclusive cross-sections measurement by CMS~\cite{Khachatryan:2016wdh} and NNLO perturbative QCD predictions as a function of the jet
$p_{T}$ in slices of rapidity, for anti-$k_{T}$ jets with $R=0.7$ normalised to the NLO result for (a) $\mu=2\,\pt$, (b) $\mu=\htp $ scales. 
The shaded bands represent the scale uncertainty.}
\label{fig:pTallscales07}
\end{figure}

\begin{figure}[t!]
  \begin{subfigure}{.5\textwidth}
  \includegraphics[width=\linewidth]{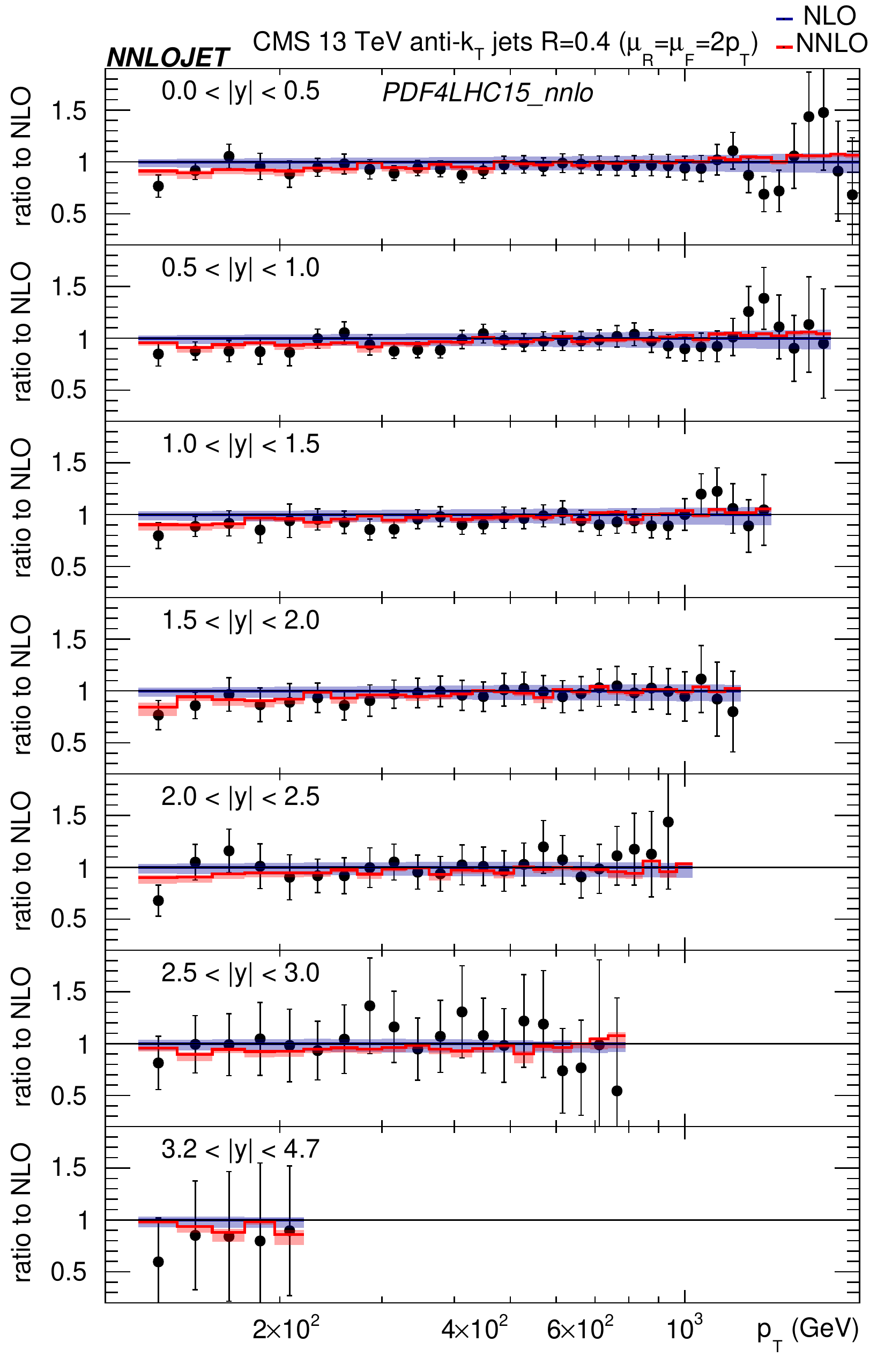}
  \caption{}
  \end{subfigure}%
  \begin{subfigure}{.5\textwidth}
  \includegraphics[width=\linewidth]{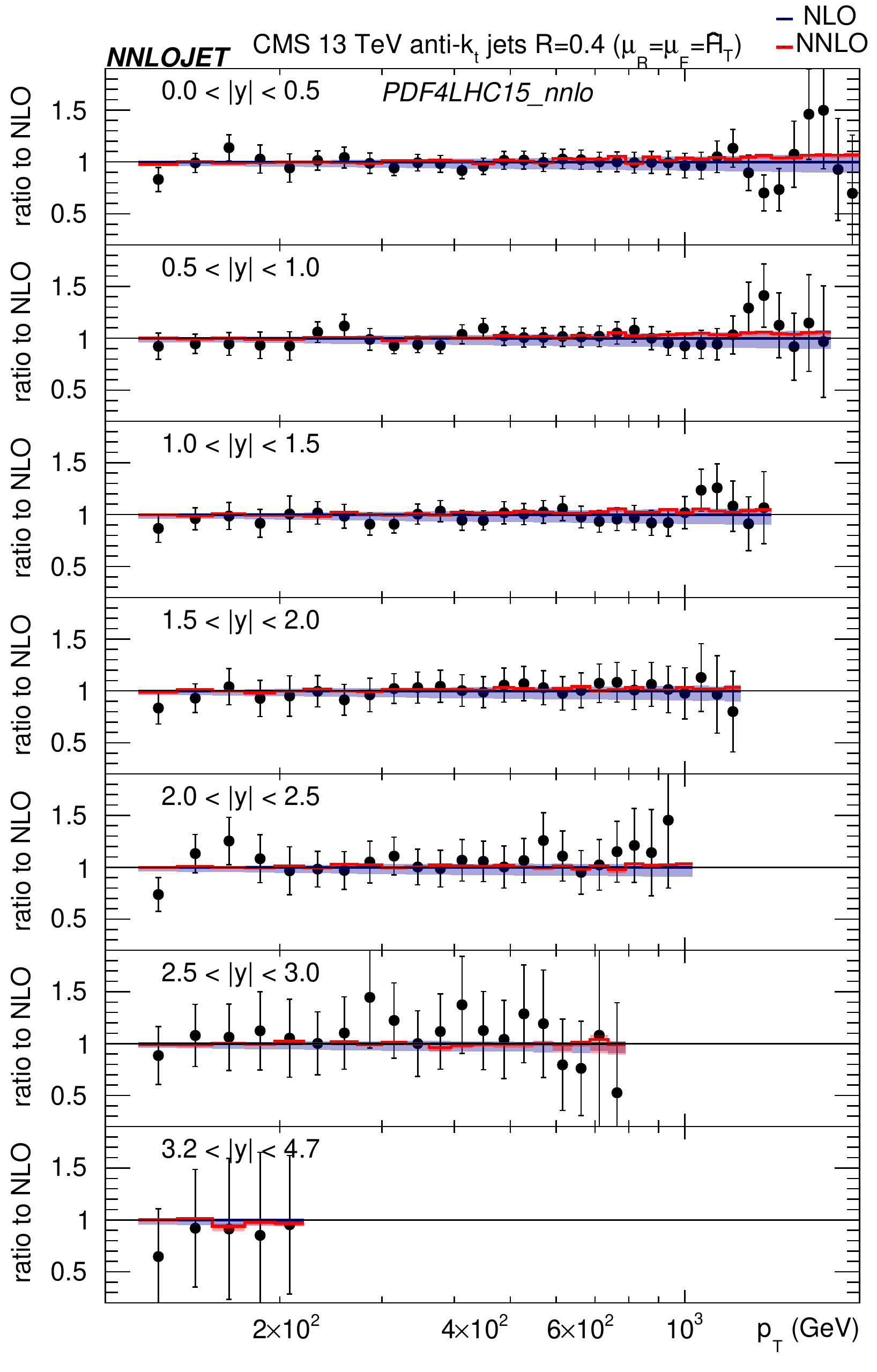}
  \caption{}
  \end{subfigure}  
\caption{Double-differential single jet inclusive cross-sections measurement by CMS~\cite{Khachatryan:2016wdh} and NNLO perturbative QCD predictions as a function of the jet
$p_{T}$ in slices of rapidity, for anti-$k_{T}$ jets with $R=0.4$ normalised to the NLO result for (a) $\mu=2\,\pt$, (b) $\mu=\htp $ scales. 
The shaded bands represent the scale uncertainty.}
\label{fig:pTallscales04}
\end{figure}

Having discussed how the jet kinematics at the LHC differently affects each of the event-based and jet-based scale choices,
in this section we present predictions for the double differential jet cross section at NLO and NNLO for the CMS measurement at $\sqrt{s}=13$ TeV~\cite{Khachatryan:2016wdh}. 
We use the same numerical setup as described in Section~\ref{subsec:setup} and do not include non-perturbative effects from underlying event and hadronization in our predictions. 
An assessment of the size of the non-perturbative contributions has been presented in~\cite{Khachatryan:2016wdh} and we note that these can vary significantly with the jet $\pt$ and the $R$ cone size.
In the study in~\cite{Khachatryan:2016wdh} the non-perturbative corrections are expected to be negligible for $R=0.4$ but can reach up to~10\%-15\% for $R=0.7$ at low-$p_{T}$.

Figure~\ref{fig:pTallscales07} displays the NLO and NNLO predictions for the jet-based scale choice $\mu=2\,\pt$,
as well as for the event-based scale choice $\htp$, compared to the CMS 13 TeV data~\cite{Khachatryan:2016wdh} with a jet cone size of $R=0.7$.
For both scale choices we observe small positive NNLO corrections across all rapidity slices,
that improve the agreement with the CMS data, as compared to the NLO prediction. 
In addition we identify a reduction in the scale uncertainty from NLO to NNLO across the entire $\pt$ range. 

Figure~\ref{fig:pTallscales04} shows the NLO and NNLO predictions for the smaller jet cone size of $R=0.4$ (where non-perturbative corrections are expected to be less important than for 
$R=0.7$~\cite{Khachatryan:2016wdh}). Similarly to the $R=0.7$ case, we see that both scale choices provide reasonable predictions and that the agreement with data is improved at NNLO. For the $\mu=2\,\pt$ scale choice this is 
achieved by having small negative NNLO corrections while for $\mu=\htp $ the NNLO corrections are flat leading to a smaller residual scale variation at NNLO
than for $\mu=2\,\pt$.

\section{Summary and Conclusions}
\label{Sec:conclusions}

In this paper we have studied single jet inclusive production at hadron colliders and the jet transverse momentum distribution obtained by adding up the contributions from all jets that are observed in an event. 
Our predictions include the most up-to-date second order NNLO corrections in the perturbative expansion of the observable. 

In detail we presented a breakdown of the inclusive jet-$\pt$ sample into leading and subleading jet contributions and found large radiative corrections to the first and second jet contributions
(that dominate the inclusive jet sample) that largely cancel each other. By investigating the second-jet transverse momentum distribution we identified large cancellations between
different kinematical event configurations, which are aggravated by certain types of scale choices. Since the notion of leading and subleading jet is not well defined at leading order ($\ptj1 = \ptj2$ at LO), 
the single jet inclusive observable is decomposed into IR-sensitive leading and subleading jet contributions
and the functional form of the scale can have an impact on the final result, when the kinematics of the scale choice affects the IR cancellations between the different contributions. We have found this
effect to be worse for the smaller jet cone size $R$=0.4 than for $R$=0.7.

The smaller cone size increases the contribution from events where relatively soft emissions are not recombined with outgoing jets. These do not cancel fully with virtual corrections, leading to 
an imbalance between $\ptj1$ and $\ptj2$. Since the second jet contribution to the inclusive jet sample is increased at NNLO with respect to NLO we have identified this effect
to be the cause in the mismatch between inclusive jet predictions at NNLO that employ $\mu=\pt$ or $\mu=\ptj1$ as central scale choices. By investigating the kinematical properties of events that contribute to a 
fixed bin in $\ptj2$ (as function of $\ptj1$ in the event), we found that the imbalance between real and virtual 
emissions is much more serious for 
 $\mu=\ptj1$ than for $\mu=\pt$, which can be understood from the fact that the former is changing event-by-event in 
 this distribution, while the latter remains constant. 
  
We have observed that the spread in the NNLO predictions that use different functional forms of the scale is larger in size than the NNLO scale variation 
in the low-$\pt$ region of the transverse momentum distribution. For this reason we have introduced a sensible set of criteria
that define desired properties for a suitable scale choice and that can maximise the impact of the knowledge of the higher order QCD corrections to the observable, to the extent that 
scale choices that introduce pathological behaviours can be identified and avoided.

We have identified $\mu=2\,\pt$ and $\mu=\htp$ as the two scales that fulfil all the criteria that we have defined, observing that they lead to important cancellations between the 
leading and subleading jet contributions which result in an improved perturbative convergence on the transverse momentum distributions, with overlapping scale uncertainty bands.

Subsequently we used these two functional forms of the central scale choice to compare our NNLO predictions with jet data from the CMS dataset at $\sqrt{s}=13$ TeV~\cite{Khachatryan:2016wdh} for the first time. 
We have observed that both recommended scale choices are stable and provide reasonable predictions for the two jet cone sizes employed in the measurement across the entire $\pt$ and
rapidity region where the observable is defined. In particular we find an improved agreement with data at NNLO with respect to NLO with a significant reduction
in scale uncertainty by roughly more than a factor of 2 in a wide range of $\pt$ and rapidity. We have refrained from comparing the measurement to predictions that employ
scale choices that contain pathological behaviours since these scale choices are not recommended on the grounds introduced in this study. 

The central
scale choices  $\mu=2\,\pt$ and $\mu=\htp$ are clearly found to be 
favoured in terms of stability and convergence of the predictions for single jet inclusive production. Both 
yield very similar predictions at NNLO. We expect that our 
findings will enable improved precision studies based on single jet inclusive production data, especially in using 
them as precision probes of the parton distributions in the proton and for a determination of QCD parameters. 

\acknowledgments 
The authors thank Xuan Chen, Juan Cruz-Martinez, Rhorry Gauld, Marius H\"ofer, Imre Majer, Tom Morgan, Jan Niehues, Duncan Walker and James Whitehead for useful discussions and their many contributions to the \textsc{NNLOjet} code.
This research was supported in part by the UK Science and Technology Facilities Council, by the Swiss National Science Foundation (SNF) under contracts 200020-175595 and 200021-172478, and CRSII2-160814, by the Research Executive Agency (REA) of the European Union through the ERC Advanced Grant MC@NNLO (340983) and by the Funda\c{c}\~{a}o para a Ci\^{e}ncia e Tecnologia (FCT-Portugal), project
UID/FIS/00777/2013.

\bibliography{refs}

\end{document}